\documentclass[longbibliography,amsmath,amssymb,aps,showpacs,twocolumn,floats,prx,groupaddress]{revtex4-2}

\usepackage{graphicx}
\usepackage{amssymb}
\usepackage{amsmath}
\usepackage{amstext}
\usepackage{amsfonts}
\usepackage{amsthm}
\usepackage{mathtools}
\usepackage{scalerel}
\usepackage{mathrsfs}
\usepackage{braket}
\usepackage{bm}
\usepackage[colorlinks,allcolors=blue,urlcolor=blue]{hyperref}
\usepackage{footnote}
\usepackage{color}
\usepackage{multirow}
\usepackage{enumerate}
\usepackage{soul}
\usepackage{lipsum}
\usepackage{verbatim}
\usepackage{upgreek}
\usepackage[dvipsnames]{xcolor}
\usepackage[normalem]{ulem}
\usepackage[FIGTOPCAP]{subfigure}
\pdfoutput=1 
\usepackage[shortlabels]{enumitem}\setlist[enumerate, 1]{1\textsuperscript{o}}

\newcommand{\tr}{{\mathrm{Tr}}}                     
\newcommand{\real}{\mathrm{Re}}                     
\newcommand{\imag}{\mathrm{Im}}                     
\newcommand{\comm}[2]{[{#1},{#2}]}                  
\newcommand{\acomm}[2]{\{{#1},{#2}\}}               
\newcommand{\ketbra}[2]{\ket{#1}\!\bra{#2}}         
\newcommand{\hrm}[1]{\hat{#1}^{\dagger}}            
\newcommand{\abs}[1]{\left\lvert{#1}\right\rvert}   
\newcommand{\expt}[1]{\langle #1 \rangle}           
\newcommand{\hc}{\mathrm{h.c.}}                     
\newcommand{\sgn}{\mathrm{sgn}}                     
\newcommand{\ext}{\mathrm{ext}}                     
\newcommand{\echo}{\mathrm{echo}}                   
\newcommand{\add}{\mathrm{rest}}                    
\newcommand{\ifo}{\protect\scaleobj{0.8}{\mathrm{IFO}}}     

\newcommand{\A}{\protect\scaleobj{0.8}{\mathrm{A}}}        
\newcommand{\B}{\protect\scaleobj{0.8}{\mathrm{B}}}        
\newcommand{\C}{\protect\scaleobj{0.8}{\mathrm{C}}}        
\newcommand{\K}{\protect\scaleobj{0.8}{\mathrm{K}}}        
\newcommand{\J}{\protect\scaleobj{0.8}{\mathrm{J}}}        
\newcommand{\Q}{\protect\scaleobj{0.8}{\mathrm{Q}}}        
\newcommand{\R}{\protect\scaleobj{0.8}{\mathrm{R}}}        
\newcommand{\q}{\protect\scaleobj{0.8}{\mathrm{Q}}}        

\newcommand{\lbld}{\mathcal{L}}                     
\newcommand{\diss}{\mathcal{D}}                     
\newcommand{\freq}[1]{\omega_{#1}}                  
\newcommand{\lw}[1]{\gamma_{#1}}                    
\newcommand{\dtn}[1]{\Delta_{#1}}                    
\newcommand{\disp}[1]{\chi_{#1}}                    
\newcommand{\coop}[1]{\mathcal{C}_{#1}}             
\newcommand{\rate}{\Gamma}                          
\newcommand{\mrate}{\Gamma_{\mathrm{meas}}}         
\newcommand{\drate}{\Gamma_{\mathrm{d}}}            
\newcommand{\mdrate}{\Gamma_{\mathrm{d,m}}}         
\newcommand{\pdrate}{\Gamma_{\mathrm{d,p}}}         
\newcommand{\nth}[1]{n^{\mathrm{th}}_{#1}}          

\newcommand{\symform}{\boldsymbol{\Omega}}
\newcommand{\symelem}{\Omega}
\newcommand{\smat}{\mathbf{S}}
\newcommand{\smelem}{\mathrm{S}}
\newcommand{\covar}{\boldsymbol{\Sigma}}
\newcommand{\celem}{\Sigma}
\newcommand{\mean}{\boldsymbol{\mu}}
\newcommand{\melem}{\mu}
\newcommand{\qvec}{\bm{r}}
\newcommand{\qelem}{r}
\newcommand{\fvec}{\bm{\xi}}
\newcommand{\felem}{\xi}
\newcommand{\charw}{w}
\newcommand{\wig}{W}

\newcommand{\pqelem}{\partial r}

\newcommand{\pfelem}{\partial \xi}
\newcommand{\lp}[1]{\overleftarrow{\partial #1}}
\newcommand{\rp}[1]{\overrightarrow{\partial #1}}

\newcommand{\HTwo}{\bm{H}}
\newcommand{\HTwoElem}{H}
\newcommand{\HOne}{\bm{h}}
\newcommand{\HOneElem}{h}
\newcommand{\HNull}{h}
\newcommand{\dissmat}{\boldsymbol{\Gamma}}
\newcommand{\idmat}{\bm{I}}
\newcommand{\nullmat}{\bm{0}}
\newcommand{\xmat}{\bm{X}}
\newcommand{\jmat}{\bm{J}}
\newcommand{\zmat}{\bm{Z}}

\newcommand{\defeq}{\vcentcolon=}

\newcommand{\meas}{\protect\scaleobj{0.8}{M}}
\newcommand{\SNR}{\mathrm{SNR}}
\newcommand{\nonrec}[2]{N^{(#1,#2)}}
\newcommand{\norm}[1]{|| {#1} ||}
\newcommand{\polyroot}{\mathrm{r}}
\newcommand{\inn}{\mathrm{in}}
\newcommand{\out}{\mathrm{out}}
\newcommand{\diag}{\mathrm{diag}}

\begin{document}

\title{Qubit measurement and backaction in a multimode nonreciprocal system}

\author{B. T. Miller}
\thanks{These authors contributed equally to this work.}
\author{F. Lecocq}
\email{florent.lecocq@nist.gov}
\affiliation{National Institute of Standards and Technology, 325 Broadway, Boulder, CO 80305, USA}
\affiliation{University of Colorado, 2000 Colorado Ave., Boulder, CO 80309, USA} 

\author{L. Orr$^{1,\color{blue}{*}}$  and A. Metelmann$^{1,2,3,}$}
\email{anja.metelmann@kit.edu}
\affiliation{${}^1$ Institute for Theoretical Condensed Matter Physics, Karlsruhe Institute of Technology, 76131 Karlsruhe, Germany}
\affiliation{${}^2$Institute for Quantum Materials and Technology, Karlsruhe Institute of Technology, 76344 Eggenstein-Leopoldshafen, Germany}
\affiliation{${}^3$Institut de Science et d’Ingénierie Supramoléculaires (ISIS, UMR7006), University of Strasbourg and CNRS, 67000 Strasbourg, France}

\date{\today}

\begin{abstract}
High fidelity qubit readout is a cornerstone for quantum information protocols. In traditional superconducting qubit readout, a chain of microwave amplifiers and nonreciprocal components aid in detecting the qubit's state with tolerable added noise and backaction. However, the loss, size, and magnetic field of standard nonreciprocal components have sparked a decades-long search for more efficient and scalable alternatives. One prominent approach employs networks of parametrically coupled modes to achieve nonreciprocity. While this class of devices can be directly integrated with the qubit's readout cavity, current understanding of the resulting single quantum system is substantially lacking. Here we provide a first-principles theoretical tool to understand and design networks of linear modes integrated with embedded qubits. We utilize this theory to inform and analyze the experimental implementation of a qubit readout with an integrated three-mode nonreciprocal system. In doing so, we achieve excellent agreement between the experimental and theoretical qubit measurement and dephasing rates. We then theoretically analyze the same system operated as an integrated nonreciprocal amplifier, predicting high efficiency for reasonable experimental parameters.
\end{abstract}

 
\maketitle


\section{Introduction}
\label{sec:intro}
Qubit measurement is a fundamental requirement of quantum computation and sensing protocols. In superconducting systems, qubit information is typically encoded in a weak microwave field that is sent through a series of amplifiers and circulators. The amplifiers aid in preserving the measurement signal-to-noise ratio (SNR) throughout the readout chain, while the circulators directionally route signals and suppress amplifier backaction. However, the commercially-available ferrite circulators commonly used have a large physical footprint and operate with a strong magnetic field. This hinders their integrability with superconducting circuits, and thus the scalability of superconducting quantum processors and sensors. In addition, their intrinsic and associated wiring losses are currently the dominant sources of measurement inefficiency.

These issues have inspired a decades-long search for low-loss, compact alternatives to ferrite circulators which operate with small magnetic fields. These alternatives have utilized mixing and delay lines \cite{kamal_noiseless_2011, chapman_widely_2017, rosenthal_breaking_2017}, on-chip switches \cite{chapman_design_2019}, loops of DC voltage-biased Josephson junctions \cite{navarathna_passive_2023,fedorov_nonreciprocity_2024}, pumped nonlinear transmission lines \cite{ranzani_wideband_2017}, and networks of parametrically coupled modes \cite{estep_magnetic-free_2014, ranzani_graph-based_2015, metelmann_nonreciprocal_2015, kerckhoff_-chip_2015, sliwa_reconfigurable_2015,
peterson_strong_2019, naaman_synthesis_2022, kwende_josephson_2023, bello_high-fidelity_2026}. Several of these strategies have been extended to create a parametric amplifier with reverse isolation \cite{Lecocq2017, Lecocq2020, malnou_travelling-wave_2025, ranadive_travelling-wave_2025, kow_traveling-wave_2025}, facilitating high-fidelity, high-efficiency, and low-backaction qubit measurement without ferrite circulators at the amplifier input \cite{Abdo2014, abdo_high-fidelity_2021, Rosenthal2021, Lecocq2021A}. Among the various schemes to achieve this integrated readout, networks of parametrically coupled linear modes currently exhibit the lowest backaction while maintaining record efficiency. However, the quantitative understanding of qubit measurement and backaction in these networks is rudimentary, presenting a hurdle to the optimization of their design.

\begin{figure}[t]
    \includegraphics[scale=1]{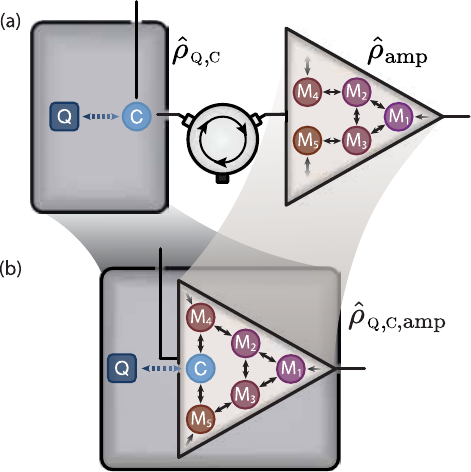}
    \caption{\textbf{Integrating the readout network}. \textbf{(a)} Conceptual diagram of a dispersive superconducting qubit measurement with a traditional readout chain consisting of a qubit (Q) and cavity (C) subsystem $\hat{\rho}_{\Q,\C}$, as well as an arbitrary resonant parametric amplifier $\hat{\rho}_{\mathrm{amp}}$, separated by an isolator. The amplifier is comprised of a network of coupled modes labeled $\mathrm{M}_{n}$, $n \in \mathbb{N}$. \textbf{(b)} When the intermediate isolator is removed, the cavity mode must be included in the network of modes comprising the amplifier. The qubit and readout network therefore combine into a single quantum system described by the inseparable density matrix $\hat{\rho}_{\Q,\C,\mathrm{amp}}$. }
    \label{fig:arb_qubit_meas}
\end{figure}

Historically, the presence of ferrite circulators between the qubit's readout cavity and parametric amplifier have allowed for their independent design and optimization. However, in the absence of intermediate isolation, these two components become a single quantum system, conceptually represented in Fig.~\ref{fig:arb_qubit_meas}. As a consequence, the amplifier becomes qubit-state dependent, and conversely, the qubit's evolution is directly affected by the state of the amplifier \cite{Eddins2019}. In contrast with previous analyses of multimode networks, we present here a full theory that captures this interdependence. We leverage this theory to design and experimentally operate a system consisting of a qubit coupled to a three-mode readout network. 

This paper is organized as follows. In Sec.~\ref{sec:theory}, we introduce a general method for computing the measurement efficiency for an arbitrary linear readout network, emphasizing the phase-space method to compute the qubit's steady-state dephasing rate. In Sec.~\ref{sec:building_readout_network}, we apply general principles from the theory to motivate our experimental implementation of a three-mode readout network. In Sec.~\ref{sec:qubit_meas}, we utilize the theory to experimentally extract the thermal occupancy of each mode comprising the network. We then experimentally demonstrate the nonreciprocal routing of signals and noise in the readout network, extracting qubit measurement and dephasing rates that quantitatively agree with theory. Lastly, we theoretically examine the device's operation as an embedded amplifier by including parametric gain. We thereby show that this device serves not only as a test-bed for the theory presented in this work, but also as a promising step toward higher efficiency qubit measurement.

\section{General Theory}
\label{sec:theory}
To quantitatively understand the detection of a qubit state, the measurement rate can be related to the network's scattering matrix, which can be computed with existing methods \cite{ranzani_graph-based_2015, Lecocq2017, Lecocq2020}. In contrast, to accurately compute qubit backaction, one must model all components of the readout network's internal state which couple to the qubit. The presence of multiple modes and gain requires a high-dimensional truncated Hilbert space to represent the system state when numerically solving the full master equation, resulting in an expensive computation. To render the calculation of the qubit backaction tractable, we expand on a phase-space method akin to those presented in Refs.~\cite{Gambetta2006, Clerk2007,Eddins2019}. The key to this method is the assumption that all states within the network remain Gaussian, which greatly reduces the degrees of freedom in the model. This allows one to quantify the backaction of the readout network on the qubit with a quick numerical, or in some cases, analytical computation. 

\subsection{Background}
A quantum-nondemolition (QND) measurement \cite{braginsky_quantum_1980} of the state of a qubit involves the coherent coupling of the qubit observable $\hat \sigma_{z} = \ketbra{e}{e} - \ketbra{g}{g}$ to a meter. 
Ideally, the only source of incoherent measurement backaction are from fluctuations in the meter, which result in fluctuations of the qubit transition frequency, and hence qubit decoherence in the form of dephasing. We will start with a review of how this manifests in the evolution of the single qubit state, and then connect this to the well-known case of a dispersive measurement with a single linear mode. First, we expand the density operator for the total system in the basis of the measured qubit,
    \begin{equation}
        \hat{\rho} = \hat{\rho}_{ee} \ket{e}\!\bra{e} + \hat{\rho}_{gg} \ket{g}\!\bra{g} + \hat{\rho}_{eg} \ket{e}\!\bra{g} + \hat{\rho}_{ge} \ket{g}\!\bra{e},
    \end{equation}
where the operators $\hat{\rho}_{jk}$ act on the Hilbert space of the meter system. The loss of qubit coherence through dephasing will result in the decay of the magnitude of $\hat{\rho}_{eg}$, thereby reducing the purity of $\hat{\rho}$ over time. Modeling the backaction on the qubit state therefore requires solving the dynamics of the meter, and then tracing over the meter's degrees of freedom, $\tr_{\R}[\hat{\rho}_{eg}(t)]$.

The master equation governing the dynamics of the total system is described using a superoperator $\lbld$. We assume that the meter is only coherently coupled to the qubit transition operator $\hat{\sigma}_z$, and that $\lbld$ corresponds to a Lindblad master equation. Consequently, taking the trace with respect to the meter, $\tr_{\R}[\lbld(\hat{\rho})]$, yields the following reduced master equation \cite{Gambetta2008} for the qubit density operator, $\hat{\rho}_{\q} = \tr_{\R}[\hat{\rho}]$,
    \begin{align}\label{eq:reduced_qubit_lindbladian}
        \frac{d}{dt} \hat{\rho}_{\q} =& \, -i \frac{\tilde{\omega}_{\q} + B(t)}{2} \comm{\hat{\sigma}_z}{\hat{\rho}_{\q}} +
        \frac{\rate_\varphi + \drate(t)}{2} \diss[\hat{\sigma}_z](\hat{\rho}_{\q}) \nonumber \\[-0.5em]
        & \, + \rate_- \diss[\hat{\sigma}_-](\hat{\rho}_{\q}) + \rate_+ \diss[\hat{\sigma}_+](\hat{\rho}_{\q}),
    \end{align}
where $\diss[\hat{o}](\hat{\rho}) = \hat{o} \hat{\rho} \hrm{o} - \frac{1}{2} \acomm{\hrm{o} \hat{o}}{\hat{\rho}}$ is the dissipation super-operator. Here, we also account for the coupling between the qubit and the environment, leading to the Lamb-shifted qubit transition frequency $\tilde{\omega}_{\q}$, pure qubit dephasing with rate $\rate_\varphi$, and the energy relaxation and excitation rates $\rate_{\mp}$.

The backaction on the qubit from the meter is encapsulated in the induced dephasing rate $\drate(t)$ and the frequency shift $B(t)$. Given that $\tr_{\R}[\hat{\rho}_{eg}(t)] = \braket{e | \hat{\rho}_{\q}(t) | g}$, a connection can be established between the state of the meter and the decoherence of the qubit state modeled by  the reduced master equation from above,
    \begin{align}\label{eq:resonator_trace}
        \tr_{\R}[\hat{\rho}_{eg}(t)]
        = & \, e^{-t\left(i \tilde{\omega}_{\q} + \rate_2\right)} \\[-0.8em] \nonumber
        & \times e^{- \int^t_0 \left(i B(\tau) + \drate(\tau)\right) d\tau} \,\tr_{\R}[\hat{\rho}_{eg}(0)],
    \end{align}
where $\rate_2 = \rate_{\varphi} + (\rate_+ + \rate_-)/2$ is the innate qubit decoherence rate. From this expression, it is evident that, in the presence of dephasing, $\abs{\tr_{\R}[\hat{\rho}_{eg}(t)]} \neq \abs{\tr_{\R}[\hat{\rho}_{eg}(0)]}$, and so the trace of $\hat{\rho}_{eg}$ is not conserved. 

Hence, in order to simulate the dynamics of $\drate(t)$ and $B(t)$ we must generate tractable equations of motion for $\tr_{\R}[\hat{\rho}_{eg}(t)]$. For various implementations of a single linear mode dispersively coupled to a qubit, this has been done using a phase space quasi-probability distribution (QPD) for $\hat{\rho}_{eg}(t)$ \cite{Gambetta2006,Serban2007,Clerk2007,Eddins2019}. These methods transform the master equation for $\hat{\rho}_{eg}(t)$ into a partial differential equation (PDE) for the associated QPD, whose dynamics are then solved. 

While the choice of representation is arbitrary, herein we will use the Wigner phase space QPD for $\hat{\rho}_{eg}(t)$, which will be denoted $\wig_{eg}(q,p;t)$ for a meter which is comprised of a single linear mode, where $q$ and $p$ are the mode's phase-space coordinates. The trace of $\hat{\rho}_{eg}(t)$ is then obtained by integrating the Wigner QPD over the entire phase space,
    \begin{equation}\label{eq:single_wigner_trace}
        \tr_{\R}[\hat{\rho}_{eg}(t)] \equiv \int_{\mathbb{R}^2} \wig_{eg}(q,p;t) \, dq \, dp = e^{-\nu(t)}.
    \end{equation}
The parameter $\exp[-\nu(t)]$ corresponds to the zeroth order moment of $\wig_{eg}(q,p;t)$. Using Eq.~\eqref{eq:resonator_trace}, we can infer that $\nu(t)$ contains all the information about the induced dephasing and frequency shift on the qubit:
    \begin{equation}\label{eq:deph_freqshift_defn}
        \drate(t) = \frac{d}{dt}\real\left[\nu(t)\right] \qquad
        B(t) = \frac{d}{dt} \imag\left[\nu(t)\right].
    \end{equation}
Here, the intrinsic qubit dynamics in the $e^{-t\left(i \tilde{\omega}_{\q} + \rate_2\right)}$ term has been removed from the evolution of $\wig_{eg}(q,p;t)$. In practice, the induced frequency shift on the qubit, $B(t)$, may be eliminated by a unitary transformation, and as a result, in this work we will focus solely on the dephasing from the coupled meter, $\drate(t)$ as it is responsible for irreversible loss of qubit information. Provided that qubit and single mode meter are only dispersively coupled, $\chi \hrm{a} \hat{a} \hat{\sigma}_z$ where the single mode operators obey $\comm{\hat{a}}{\hrm{a}}=1$, and the state of the linear mode is Gaussian, then $\dot{\nu}(t)$ is equal to \cite{Clerk2007}
    \begin{equation}
         \frac{d}{dt} \nu(t) \! = i \chi \left[ \celem_{qq}(t) + \celem_{pp}(t) + \melem_{q}^2(t) + \melem_{p}^2(t) -  1 \right] ,
    \end{equation}
where $\celem_{kk}(t)$ and $\melem_{k}(t)$ are the time-dependent covariances and means of $\wig_{eg}(q,p;t)$. 

The above expressions highlight how the dephasing may be obtained directly from the moments of $\wig_{eg}(q,p;t)$. The goal is then to systematically derive equations of motion for these moments, and hence, $\dot{\nu}(t)$ in a broad class of meters. In the next section we will move beyond single-mode dispersive readout, and provide a way to accomplish this for systems where the meter is a readout network comprised of an arbitrary number of linear modes, additionally incorporating other coupling mechanisms to the qubit.

\subsection{Generalization to multimode systems}
\label{sec:multimode_dephasing}
In this section, the phase-space procedure for calculating the dephasing is not only generalized to accommodate an arbitrary number of linear modes, but also systematized so that the system of dynamical equations may be immediately constructed from $\lbld$. We note that bold characters are used in this work to represent arrays, with upper-case letters for matrices and lower-case letters for vectors. Additionally, explicit time dependence is omitted from here on.

We again choose to use the Wigner phase space, and will require that the QPD be a Gaussian, and hence representable entirely by its first two moments. From this ansatz, the Wigner QPD for $\hat{\rho}_{eg}$ is
    \begin{align}\label{eq:wigner_ansatz}
        \wig_{eg}(\qvec) \, &= \frac{1}{(2\pi)^N} \int_{\mathbb{R}^{2N}} \hspace{-1mm} \tr_{\R} \! \left[\hat{\rho}_{eg} e^{i \fvec^T (\hat{\qvec} - \qvec)} \right] \! d \fvec \nonumber \\
        &= \frac{1}{(2\pi)^N \sqrt{\det[\covar]}} e^{- \frac{1}{2} (\qvec - \mean)^T \covar^{-1} (\qvec - \mean) - \nu}
    \end{align}
where $N$ is the number of linear modes, $\covar = \covar^T$ is the covariance matrix, and the elements of $\mean$ are the means. These first and second moments are complex-valued since $\hat{\rho}_{eg}$ is in general not Hermitian. As in Eq.~\eqref{eq:single_wigner_trace}, $\nu$ is defined such that $\tr_{\R}[\hat{\rho}_{eg}] = \exp[-\nu]$. 

The above Wigner QPD has been written in an arbitrary quadrature basis, $\qvec = (\qelem_1,\ldots,\qelem_{2N})$, where $\qelem_k$ is the phase-space coordinate associated with the quadrature operator $\hat{\qelem}_k$. The quadrature operators obey the usual canonical commutation relation, $\comm{\hat{\qelem}_j}{\hat{\qelem}_k} = i \symelem_{jk}$, where $\symelem_{jk}$ are the elements of the skew-symmetric symplectic form, $\symform = - \symform^T$. The choice of basis $\hat{\qvec}$ determines the structure of $\symform$, with the two most common being
    \begin{align}
        \hat{\qvec} = \begin{pmatrix} \hat{q}_1,\hat{p}_1,\ldots,\hat{q}_N,\hat{p}_N \end{pmatrix}, 
        \,\,\, & \,\,
        \symform = \bm{I}_N \otimes \begin{pmatrix} 0 & 1 \\ -1 & 0 \end{pmatrix}, \label{eq:quadrature_basis} \\
        \hat{\qvec} = \begin{pmatrix} \hat{q}_1,\ldots,\hat{q}_N,\hat{p}_1,\ldots,\hat{p}_N \end{pmatrix},
        \,\,\, & \,\,
        \symform = \begin{pmatrix} 0 & 1 \\ -1 & 0 \end{pmatrix} \otimes \bm{I}_N \label{eq:grouped_mode_basis},
    \end{align}
where $\bm{I}_N$ is the $N \times N$ identity matrix. The operators $\hat{q}_j = (\hrm{\jmath} + \hat{\jmath})/\sqrt{2}$ and $\hat{p}_k = i (\hrm{k} - \hat{k})/\sqrt{2}$, are the usual position and momentum quadratures, respectively, which satisfy $\comm{\hat{q}_j}{\hat{p}_k} = i \delta_{jk}$, and where the creation and annihilation operators obey $\comm{\hat{\jmath}}{\hrm{k}} = \delta_{jk}$. By avoiding a specific choice for $\hat{\qvec}$, the expressions in this section can be kept as simple and generic as possible.

Having specified what form $\hat{\rho}_{eg}$ must take, we can now proceed to describe the restrictions which must be placed on the superoperator $\lbld$ to ensure that $\hat{\rho}_{eg}$ is always Gaussian. In short, $\lbld$ must satisfy two simplified conditions. The first of these requires that all linear modes must evolve under linear dynamics. As a consequence, the Hamiltonian acting on the linear mode Hilbert space must be at most a bilinear function of the quadrature operators, while only collapse operators which are linear combinations of quadrature operators are permitted.

The second condition demands that the qubit dynamics cannot couple $\hat{\rho}_{eg}$ to any other component of $\hat{\rho}$, so that the dynamical equation for $\hat{\rho}_{eg}$ may be extracted from the following:
    \begin{equation}\label{eq:dynamics_offdiagonal}
        \frac{d}{dt} \left(\hat{\rho}_{eg} \ketbra{e}{g}\right) = \lbld \! \left(\hat{\rho}_{eg} \ketbra{e}{g}\right).
    \end{equation}
As a consequence, Jaynes-Cummings and Rabi interactions outside the dispersive regime cannot be handled by this method due to the presence of coherent qubit-state transitions. This condition is stricter for systems with multiple qubits, in which case $\hat{\sigma}_z$ must be a QND-observable of the measurement, $\big\langle\lbld^\dagger (\hat{\sigma}_z)\big\rangle = 0$.

In order to more easily apply these conditions to $\lbld$, we first express it as
    \begin{equation}\label{eq:total_lindbladian}
        \lbld (\hat{\rho}) 
        = \big(\lbld_{\R} + \lbld_{\q\R} + \lbld_{\q} \big)(\hat{\rho}),
    \end{equation}
where $\lbld_{\R}$ contains processes acting solely on the linear modes, those in $\lbld_{\q}$ act on the qubit, and $\lbld_{\q\R}$ on the combined system. We can now write general forms for the individual components of $\lbld$, assuming that this represents a Markovian Lindblad master equation. We start with the Lindbladian $\lbld_{\R}$, which may written in the quadrature basis as:
    \begin{align}\label{eq:cavity_lindblad}
        &\lbld_{\R}(\hat{\rho}) = -i \comm{\hat{H}_{\R}}{\hat{\rho}} + \!\! \sum_{j,k=1}^{2N} \!\! \rate_{jk} \! \left(\! \hat{\qelem}_k \hat{\rho} \hat{\qelem}_j - \frac{1}{2} \acomm{\hat{\qelem}_j \hat{\qelem}_k}{\hat{\rho}} \! \right) \!, \\[-0.5em]
        &\text{where} \quad \hat{H}_{\R} = \frac{1}{2} \hat{\qvec}^T \HTwo_{\R} \hat{\qvec} + \hat{\qvec}^T \HOne_{\R}. \nonumber
    \end{align}
Here, $\rate_{jk}$ are the elements of the matrix $\dissmat \in \mathbb{C}^{2N \times 2N}$, the eigenvalues of which are real with dimension of inverse time and so correspond to the decay rates of the system, while the eigenvectors combine the elements $\hat{\qvec}$ to yield the jump operators. Since the decay rates must be real and non-negative, $\dissmat$ must be a positive semidefinite Hermitian matrix, $\dissmat^\dagger = \dissmat \geq 0$. The matrix $\HTwo_{\R}$ contains the coefficients of the bilinear terms from the Hamiltonian $\hat{H}_{\R}$, while the vector $\HOne_{\R}$ corresponds to the linear coefficients. $\HTwo_{\R}$ therefore contains information about any detunings, squeezing, and beam splitter interactions, whereas the linear term $\HOne_{\R}$ represents drive terms of the form $(\varepsilon \hrm{a} + \varepsilon^* \hat{a})$. In order for $\hat{H}_{\R}$ to be Hermitian, it is required that $(\HTwo_{\R})^T = \HTwo_{\R} \in \mathbb{R}^{2N \times 2N}$ and $\HOne_{\R} \in \mathbb{R}^{2N}$.

For the Lindbladian $\lbld_{\q\R}$, this method is able to accommodate certain dressed-dissipation terms which may appear in the dispersive regime \cite{Boissonneault2009}. Despite this, only coherent couplings between the qubit and linear modes are considered here for simplicity:
    \begin{align}\label{eq:qubit_cavity_lindblad}
        &\lbld_{\q\R}(\hat{\rho}) = -i \comm{\hat{\sigma}_z \hat{H}_{\q\R}}{\hat{\rho}} ,\\
        &\text{where} \quad \hat{H}_{\q\R} = \frac{1}{2} \hat{\qvec}^T \HTwo_{\q\R} \hat{\qvec} + \hat{\qvec}^T \HOne_{\q\R} + \HNull_{\q\R}. \nonumber
    \end{align}
As above, $\HTwo_{\q\R}$ and $\HOne_{\q\R}$ correspond to the coefficients of the bilinear and linear terms from the Hamiltonian $\hat{H}_{\q\R}$, respectively. A constant term $\HNull_{\q\R}$, equivalent to an energy shift in the Hamiltonian, is also included, though is only of relevance when calculating the induced qubit frequency shift. Again, to ensure that $\hat{H}_{\q\R}$ is Hermitian, it is necessary that $(\HTwo_{\q\R})^T = \HTwo_{\q\R} \in \mathbb{R}^{2N \times 2N}$, $\HOne_{\q\R} \in \mathbb{R}^{2N}$, and $\HNull_{\q\R} \in \mathbb{R}$. Written in this form, dispersive interactions, $\hat{\sigma}_z \hrm{a} \hat{a}$, are contained within $\HTwo_{\q\R}$, while $\HOne_{\q\R}$ corresponds to longitudinal interactions, such as $\hat{\sigma}_z (\hat{a} + \hrm{a})$. As a result, $\lbld_{\q\R}$ encompasses all possible dispersive and longitudinal measurement processes.

Lastly, the Lindbladian $\lbld_{\q}$, which acts solely on the qubit, takes the following form,
    \begin{align}\label{eq:qubit_lindblad}
        \lbld_{\q}(\hat{\rho}) = & -i \frac{\tilde{\omega}_{\q}}{2} \comm{\hat{\sigma}_z}{\hat{\rho}} +
        \frac{\rate_\varphi}{2} \diss[\hat{\sigma}_z](\hat{\rho}) \nonumber \\[-0.5em]
        &+ \rate_- \diss[\hat{\sigma}_-](\hat{\rho}) + 
        \rate_+ \diss[\hat{\sigma}_+](\hat{\rho}).
    \end{align}
Upon tracing over the linear modes, the Lindbladian $\lbld$ reduces to the qubit master equation from Eq.~\eqref{eq:reduced_qubit_lindbladian}, as desired. As in the single mode case, the contribution from the intrinsic qubit processes in $\lbld_{\q}$ to the dynamics of $\hat{\rho}_{eg}$ may be eliminated by applying a shift, $\hat{\rho}_{eg} \rightarrow e^{-t\left(i \tilde{\omega}_{\q} + \rate_2\right)} \hat{\rho}_{eg}$, and so this contribution to Eq.~\eqref{eq:total_lindbladian} will be omitted.

Using Eq.~\eqref{eq:dynamics_offdiagonal}, we can now write a dynamical equation for $\hat{\rho}_{eg}$, which is not in Lindblad form:
    \begin{equation}\label{eq:eff_master_offdiagonal}
         \frac{d}{dt} \hat{\rho}_{eg} = \lbld_{\R}(\hat{\rho}_{eg}) - i \acomm{\hat{H}_{\q\R}}{\hat{\rho}_{eg}}.
    \end{equation}
From this equation, we can now generate differential equations for the moments along with $\nu(t)$, which according to Eq.~\eqref{eq:deph_freqshift_defn}, will contain all information about the induced dephasing and frequency shift on the qubit. This is accomplished by converting Eq.~\eqref{eq:eff_master_offdiagonal} into a PDE for $\wig_{eg}(\qvec)$, which is further transformed into another PDE for the characteristic function of $\wig_{eg}(\qvec)$, $\charw_{eg}(\fvec) = \tr_{\R} [\hat{\rho}_{eg} \exp[i \fvec^T \hat{\qvec}]]$, by performing a Fourier transform over all phase-space coordinates. Based on our ansatz for the Wigner function in Eq.~\eqref{eq:wigner_ansatz}, this characteristic function may be written as
   \begin{equation}\label{eq:char_func_ansatz}
         \charw_{eg}(\fvec) = e^{- \frac{1}{2} \fvec^T \covar \fvec + i \mean^T \fvec - \nu},
    \end{equation}
where $\fvec = (\felem_1,\ldots,\felem_{2N})$ are the coordinates of the reciprocal Fourier space. The characteristic function is more natural to work with since the moments of the QPD may be generated through differentiation about the origin, $\fvec = \nullmat$. The PDE for $\charw_{eg}(\fvec)$ is therefore converted into the following system of differential equations for $\covar$, $\mean$, and $\nu$:
    \begin{align}\label{eq:moment_odes}
        \frac{d}{dt} \covar &= \bm{A} \covar + \covar \bm{A}^T - \covar \bm{B} \covar + \bm{C} ,\nonumber \\[0.3em]
        \frac{d}{dt} \mean &= (\bm{A} - \covar \bm{B}) \mean + \bm{d} - \covar \bm{f} ,\nonumber \\[0.1em]
        \frac{d}{dt} \nu &= g + \bm{f}^T \mean + \frac{1}{2} \mean^T \bm{B} \mean + \frac{1}{2} \tr \left[\bm{B} \covar\right],
    \end{align}
where we use the Lindbladians in Eqs.~\eqref{eq:cavity_lindblad} and \eqref{eq:qubit_cavity_lindblad} to define:
    \begin{align}\label{eq:array_definitions}
        \bm{A} &= \symform \! \left(\HTwo_{\R} + \imag[\dissmat]\right), &
        \bm{d} &= \symform \HOne_{\R}, \nonumber \\
        \bm{B} &= 2 i \HTwo_{\q\R} ,&
        \bm{f} &= 2 i \HOne_{\q\R} ,\nonumber \\[-0.5em]
        \bm{C} &= -\symform \! \left(\frac{i}{2} \HTwo_{\q\R} + \real[\dissmat] \right) \! \symform ,&
        g &= 2 i \HNull_{\q\R}.
    \end{align}
Supporting details for this derivation are given in Appendix~\ref{apdx:theory}. With this result, we may now write explicit solutions for the dephasing in terms of the moments of $\wig_{eg}(\qvec)$ and the coupling between the qubit and linear modes. Using Eq.~\eqref{eq:deph_freqshift_defn} and the above expressions, we find that $\drate$ may be decomposed into a sum of two components,
    \begin{equation}\label{eq:drate}
        \drate = \pdrate + \mdrate,
    \end{equation}
which may be expressed as
    \begin{align}
        &\pdrate = - \tr \! \left[\imag[\covar] \HTwo_{\q\R} \right] ,\label{eq:pdrate}\\
        &\mdrate = - 2 \Big(\imag[\mean]^T \HOne_{\q\R} + \imag[\mean]^T \HTwo_{\q\R} \real[\mean] \Big)\label{eq:mdrate}.
    \end{align}
The first component, $\pdrate$, is termed the \textit{parasitic dephasing} herein, and represents backaction which occurs when the variance of the linear modes coupled to the qubit are above vacuum. A non-zero $\pdrate$ may therefore arise due to thermal fluctuations \cite{Clerk2007} or the presence of amplification \cite{Eddins2019}. It is termed parasitic since it is present even in the absence of any displacement during a dispersive or longitudinal qubit measurement.

The second component, $\mdrate$, is called the \textit{measurement-induced dephasing}, since it is only non-zero when the linear modes are displaced during a measurement. While the parasitic dephasing is independent of the measurement-induced dephasing, the reverse is not true, since the evolution of $\mean$ is dependent on the second-order moments in $\covar$.

Provided that $\hat{\rho}$ corresponds to a pure state, the dynamics of $\hat{\rho}_{eg}$, and hence the induced dephasing \cite{Gambetta2008}, can be extracted directly from the moments of the pointer states \cite{zurek_pointer_1981}, $\hat{\rho}_{ee}$ and $\hat{\rho}_{gg}$, which are true quantum states. If $\hat{\rho}$ is not pure, which occurs when damped harmonic modes are subject to squeezing or thermal fluctuations \cite{Isar1999}, then it is not possible to establish a direct connection between the moments of $\hat{\rho}_{eg}$ and those of the pointer states. The method presented here for calculating $\hat{\rho}_{eg}$ is therefore necessary to model the backaction from integrated qubit-amplifier systems.

\subsection{Measurement rate and efficiency}
\label{subsec:measrate_efficiency}
With the induced dephasing defined, the fraction of the qubit information that may be acquired is now characterized using the measurement efficiency
    \begin{equation}\label{eq:efficiency}
        \eta \defeq \frac{\mrate}{\drate}.
    \end{equation}
We note that this definition includes inefficiencies due to qubit backaction, similar to Ref.~\cite{Eddins2019}. In this definition, $\mrate$ is the measurement rate, and corresponds to the steady-state rate of change of the amplitude signal-to-noise ratio (SNR) squared:
    \begin{align}\label{eq:meas_rate_defn}
        \mrate 
        \defeq& \lim_{t \rightarrow \infty} \frac{\SNR^2(t)}{t} \nonumber \\
        =& \frac{\abs{\expt{\hat{\meas}[\omega_d]}_e-\expt{\hat{\meas}[\omega_d]}_g}^2 \!\! /4}{\bar{S}_{\meas \meas,e}[\omega_d]+\bar{S}_{\meas \meas,g}[\omega_d] + 2\bar{n}_{\add}},
    \end{align}
where $\hat{\meas}$ is the measurement operator, which is a function of all monitored quadratures. $\expt{\hat{\meas}[\omega]}_{\sigma}$ and $\bar{S}_{\meas \meas,\sigma}[\omega]$ are the measurement operator expectation value and noise power at the output of the readout network, respectively. Both quantities are conditioned on the state of the qubit, $\sigma = e,g$, and are evaluated at the frequency $\omega = \omega_d$ in the rotating frame. Added noise from downstream components in the readout chain is assumed to be uncorrelated with noise from the readout network, and is parameterized by the independent $\bar{n}_{\add}$.

While $\mrate$ is dependent on the choice of a specific $\hat{\meas}$, in practice, it is chosen to maximize the difference in displacement between the pointer states, equivalent to the signal component of $\mrate$ \cite{Bultink2018}. The monitored quadratures comprising $\hat{\meas}$ are modeled as output fields in the frequency domain, which may be calculated by solving the linear Fourier-transformed Heisenberg-Langevin equations and then applying quantum input-output theory \cite{gardiner_input_1985}. The result is a scattering equation which relates the output and input fields,
	\begin{equation}
	   \qvec_{\out}[\omega] = \mathbf{S}[\omega] \qvec_{\inn}[\omega] + \mathbf{T}[\omega].
	\end{equation}
$\qvec_{\inn}[\omega]$ and $\qvec_{\out}[\omega]$ are the vectors of quadratures for the input and output fields at the reference plane of the readout network, respectively. The readout network is now treated as a scattering center for these propagating input fields, represented by the scattering matrix $\mathbf{S}[\omega]$. The displacement of the output fields by sources other than the input drives, such as longitudinal couplings between the intracavity modes and the qubit, are represented by the matrix $\mathbf{T}[\omega]$. Further details are provided in Appendix~\ref{apdx:theory}.

Assuming the noise may be modeled as a stationary stochastic process, the output-noise power $\bar{S}_{\meas \meas,\sigma}[\omega]$ is equivalent to the Fourier transform of the autocorrelation function of $\hat{\meas}(t)$ by the Wiener–Khinchin theorem \cite{Kubo1991}. In the case of white noise, the output noise power matrix $\bar{\bm{S}}_{\out}[\omega]$ may then be calculated from the input noise power using the scattering matrix,
    \begin{equation}
        \bar{\bm{S}}_{\out}[\omega] = \frac{1}{2} \Big(\mathbf{S}[\omega] \bar{\bm{S}}_{\inn} \mathbf{S}^T[-\omega] + \mathbf{S}[-\omega] \bar{\bm{S}}_{\inn} \mathbf{S}^T[\omega] \Big),
    \end{equation}
where $\bar{\bm{S}}_{\inn}$ is the covariance matrix of the input fields.

Since the rate of information extracted from the qubit state cannot exceed the resulting backaction \cite{Clerk2010,Korotkov1999}, the measurement efficiency is limited to the range $0 \leq \eta \leq 1$, and hence the measurement rate is necessarily bounded from above by the backaction from the readout network according to Eq.~\eqref{eq:efficiency}, $\mrate \leq \drate$. Fundamentally, backaction from the measurement drive only gives rise to $\mdrate$, and so it alone sets the upper limit for the measurement rate, $\mrate \leq  \mdrate$. The following two requirements should therefore be met in order to maximize $\eta$:
    \begin{enumerate}[(1)]
        \item The parasitic dephasing rate $\pdrate$ must be minimized compared to the measurement-induced dephasing rate $\mdrate$, so that $\drate$ is dominated by the contribution from $\mdrate$.
        \item The measurement rate must be as close as possible to the measurement-induced dephasing rate, $\mrate / \mdrate \rightarrow 1$.
    \end{enumerate}
Careful engineering of the readout network is required in order to realize both conditions for a high measurement efficiency, which will be discussed in the next section.

\section{Building a readout network}
\label{sec:building_readout_network}
In this section, we consider a concrete implementation of a readout network for efficient qubit measurement, leveraging the tools and lessons learned from Sec.~\ref{sec:theory}. At its core, this readout network must ideally combine gain to overwhelm downstream system noise and nonreciprocity to protect the qubit from spurious backaction. To achieve nonreciprocity in such a network, interference between at least three parametrically coupled modes must be exploited \cite{ranzani_graph-based_2015, metelmann_nonreciprocal_2015}. To retain high efficiency and low backaction, these modes must have carefully-engineered couplings to the qubit and to the external environment. Lastly, to leverage the theory presented in Sec.~\ref{sec:theory}, the modes must have vanishing self-Kerr and only interact with other modes in the network via bilinear interactions. Below, we introduce such a network and elaborate on these constraints in more detail.

\subsection{Three-mode, phase-sensitive readout network}
\label{subsec:3mode_network}
\begin{figure}[t]
    \includegraphics[scale=1]{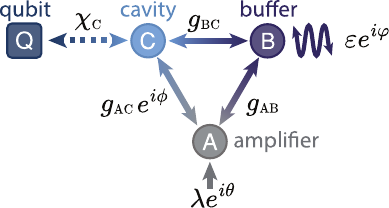}
    \caption{\textbf{A three-mode readout network for efficient qubit measurement}. The amplifier (A), buffer (B), and cavity (C) modes form an interferometer to directionally route and amplify readout signals containing information about the qubit (Q). The arrows with color gradients represent beam-splitter interactions between modes J and K with strength $g_{\J\K}$. The AC beam splitter is arbitrarily chosen to be the only beam splitter with nonzero phase, entirely constraining the interferometer phase $\phi$. The dashed arrow represents dispersive coupling of strength $\disp{\C}$, and all other dispersive shifts are neglected in this diagram. The solid gray single-sided arrow represents single mode squeezing at the A mode with strength $\lambda$ and phase $\theta$. The oscillating, double-sided arrow represents the input/output coupling to mode B, facilitating a drive of amplitude $\varepsilon$ and phase $\varphi$.}
\label{fig:desired_operation}
\end{figure}
We choose a minimal readout network that satisfies the constraints listed above, inspired by Refs.~\cite{Lecocq2020,Lecocq2021A} and shown in Fig.~\ref{fig:desired_operation}. The network consists of three linear modes: the \textit{cavity}  (C) mode that couples dispersively to the qubit, the \textit{amplifier} (A) mode wherein the qubit measurement signal is phase-sensitively amplified, and the \textit{buffer} (B) mode which acts as the input/output mode. These three modes are coupled via parametric frequency conversion to form an interferometer (IFO) that nonreciprocally routes readout signals and noise within the network. Only dispersive interactions between the qubit and the individual modes in the readout network are included, and any corrections to the dispersive terms are minimal and therefore omitted. The resulting interaction Hamiltonian in the rotating frame is then:
    \begin{equation}\label{eq:full_hamiltonian}
        \hat{H} = \underbrace{\hat{H}_{\ifo} + \hat{H}_{\mathrm{gain}} + \hat{H}_{\mathrm{drive}}}_{\textstyle \hat{H}_{\R}} + \hat{\sigma}_z \hat{H}_{\Q\R},
    \end{equation}
where
    \begin{align}
        \hat{H}_{\ifo} = \,& g_{\A\B} \hrm{a} \hat{b} + g_{\B\C} \hrm{b} \hat{c} + g_{\A\C} e^{i\phi} \hrm{a} \hat{c} + \hc, \label{eq:interferometer_hamiltonian} \\
        \hat{H}_{\mathrm{gain}} = \,& \frac{\lambda}{2} \left(e^{i \theta} \hat{a}^{\dagger \, 2} + e^{-i \theta} \hat{a}^2 \right),\\
        \hat{H}_{\mathrm{drive}} = \,& \varepsilon \left(e^{i \varphi} \hat{b} + e^{-i \varphi} \hrm{b}\right),\label{eq:drive_hamiltonian}\\
        \hat{H}_{\Q\R} = &\hspace{-2mm} \sum_{\K=\A,\B,\C} \hspace{-2mm} \disp{\K}\hrm{k}\hat{k}.
    \end{align}
Above, the canonical creation and annihilation operators $\hat{k}^{(\dagger)}$ for modes A, B, and C are denoted $\hat{a}^{(\dagger)}$, $\hat{b}^{(\dagger)}$, and $\hat{c}^{(\dagger)}$, respectively. $\phi \defeq \phi_{\A\B} + \phi_{\B\C} - \phi_{\A\C}$ is defined as the \textit{interferometer phase}, and $g_{\J\K} \in \mathbb{R}$ and $\phi_{\J\K}$ are the strength and phase of the pairwise beam-splitter coupling of arbitrary modes $\mathrm{J},\mathrm{K} \in \{\mathrm{A},\mathrm{B},\mathrm{C}\}$, $\mathrm{J} \neq \mathrm{K}$. $\lambda \in \mathbb{R}$ and $\theta$ define the strength and phase of the single-mode squeezing interaction at mode A. Optimally, the quadrature containing qubit information will be amplified to increase both $\expt{\hat{\meas}}_{\sigma}$ and $\bar{S}_{\meas \meas,\sigma}$ in Eq.~\eqref{eq:meas_rate_defn}, overwhelming the contributions of $\bar{n}_{\add}$ and improving the measurement rate. Dispersive couplings between the qubit and all three modes are included, however ideally the dispersive coupling between the qubit and the cavity mode should be the dominant interaction in $\hat{H}_{\Q\R}$ ($\disp{\C} \gg \disp{\B},\disp{\A}$).

These coherent processes compete with incoherent processes in the system, set by the total dissipation rate $\lw{\K}$ and thermal occupancy $\nth{\K}$ of each mode ($\mathrm{K}=\mathrm{A},\mathrm{B},\mathrm{C}$). The resulting general Lindbladian can be written as:
    \begin{align}\label{eq:L_jump}
        \lbld(\hat{\rho}) = & -i \comm{\hat{H}_{\R} + \hat{\sigma}_z \hat{H}_{\Q\R}}{\hat{\rho}} \nonumber \\[-0.5em]
        & 
        + \hspace{-3mm} \sum_{\K=\A,\B,\C} \hspace{-2.5mm} \! \lw{\K} \!\left((\nth{\K}+1) \diss[\hat{k}] + \nth{\K} \diss[\hrm{k}]\right) \! (\hat{\rho}),
    \end{align}
allowing the construction of the matrix $\dissmat$ through comparison with Eq.~\eqref{eq:cavity_lindblad}. Ideally, all thermal occupancies and internal losses are minimized, and only the B mode is strongly coupled to an input/output port with rate $\lw{\B}^\mathrm{ext}$ such that $\lw{\B}^{\ext}\approx\lw{\B} \gg \lw{\A},\lw{\C}$. $\varepsilon$ and $\varphi$ define the strength and phase of a resonant drive ($\omega = 0$ in the rotating frame) to this port, respectively. Additionally, to minimally degrade the measurement rate, signals present in the readout network must preferentially leave the system through the monitored B mode. This requires that all beam-splitter coupling rates also greatly exceed the total linewidths of modes A and C ($g_{\J\K}, \lw{\B} \gg \lw{\A},\lw{\C}$). In this regime, the readout network essentially behaves as a single-port device in which the nonreciprocal behavior is evident primarily in the internal dynamics. 

The desired internal dynamics can be understood in the following simplified picture: a signal injected into mode B travels first to mode C to pick up a qubit-state dependent phase shift, then to mode A to acquire phase-sensitive gain, and finally output through mode B. This configuration minimizes parasitic dephasing by suppressing the propagation of amplified noise from mode A to mode C. Achieving this circulation requires ideal interference within the readout network, placing further constraints on the beam-splitter interaction strengths and phases. 

These constraints can be found by maximizing the directional scattering asymmetry between modes A and C, quantified by the normalized degree of nonreciprocity \cite{orr_novel_2025}:
\begin{equation}\label{eq:normalized_degree_nonrecip}
    \nonrec{\K}{\J}[\omega] \defeq \frac{\norm{\smat_{\J\K}[\omega]}_F^2-\norm{\smat_{\K\J}[\omega]}_F^2}{\norm{\smat_{\J\K}[\omega]}_F^2+\norm{\smat_{\K\J}[\omega]}_F^2},
\end{equation}
where $\norm{\cdot}_F$ is the Frobenius norm, and $\smat_{\J\K}[\omega]$ is a $2 \times 2$ sub-matrix of the total scattering matrix $\smat[\omega]$ corresponding to the scattering from the input of some mode K to the output of another mode J. For example, $\smat_{\J\K}[\omega]$ may correspond to $(\hat{\jmath}_{\out} \,\,\, \hrm{\jmath}_{\out})^T \! = \smat_{\J\K}[\omega] (\hat{k}_{\inn} \,\,\, \hrm{k}_{\inn})^T$; however, note that the definition Eq.~\eqref{eq:normalized_degree_nonrecip} is independent of the choice of basis for $\smat[\omega]$. 

By construction $\nonrec{\K}{\J}[\omega] \in [-1,1]$, taking a null value for reciprocal scattering, a negative value for a dominant $\mathrm{J} \rightarrow \mathrm{K}$ scattering, and a positive value for dominant $\mathrm{K} \rightarrow \mathrm{J}$ scattering. Setting $\disp{\A}, \disp{\B} = 0$ for the sake of simplicity, the normalized degree of nonreciprocity between the modes A and C in the readout network at the resonance frequency $\nonrec{\C}{\A} = \nonrec{\C}{\A}[\omega = 0]$, is
\begin{equation}\label{eq:nonreciprocity_function_AC}
    \nonrec{\C}{\A} = -2 \sin(\phi) \left(\sqrt{\frac{\coop{\A\B} \coop{\B\C}}{\coop{\A\C}}} + \sqrt{\frac{\coop{\A\C}}{\coop{\A\B} \coop{\B\C}}}\right)^{\!\! -1},
\end{equation}
where $\coop{\J\K} \defeq 4 g_{\J\K}^2/\lw{\J}\lw{\K}$ is the cooperativity between modes J and K. The desired circulation requires that $\nonrec{\C}{\A} = 1$, which is met with the following conditions:
\begin{equation}\label{eq:nonreciprocity_conditions}
    \frac{\coop{\A\B} \coop{\B\C}}{\coop{\A\C}} = 1,
    \qquad
    \phi = -\frac{\pi}{2}.
\end{equation}
The constraint on the cooperativities effectively balances the scattering amplitudes along each arm of the interferometer, and the phase condition establishes maximal interference, with $\sgn(\phi)$ determining the prevailing direction of the nonreciprocal scattering. Such conditions can be understood as balancing coherent and dissipative processes to engineer nonreciprocity \cite{metelmann_nonreciprocal_2015}.
To summarize, this readout network requires three modes with low internal loss and tunable all-to-all parametric interactions. In this network, one mode should be primarily dispersively coupled to the qubit and another mode should be strongly coupled to its environment. Below we introduce an experimental implementation of such a readout network.

\subsection{Experimental implementation}
\label{subsec:network_expt_implementation}
\begin{figure*}[t]
    \includegraphics[scale=1]{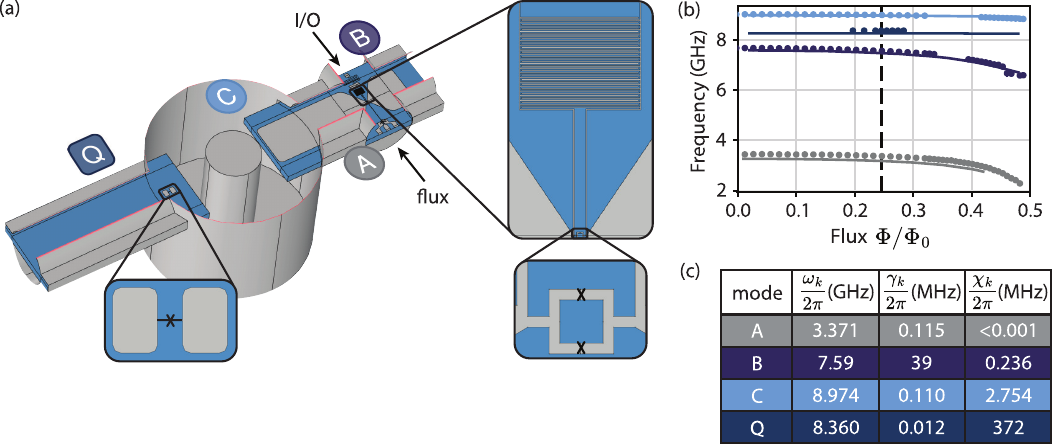}
    \caption{\textbf{Device overview.} \textbf{(a)} Sliced COMSOL model of the device. A transmon qubit is housed on a chip to the left, a $\lambda/4$ coaxial cavity in the center constitutes the cavity (C) mode, and a secondary chip on the right contains the amplifier (A) and buffer (B) modes. This chip also includes a dc SQUID and its inductively-coupled bias line, as well as a port to drive both the qubit and the B mode. \textbf{(b)} Measured and simulated frequencies of each mode and the qubit as a function of the DC flux bias through the SQUID. The black dashed line marks the operating flux bias of 0.246 $\Phi_0$. \textbf{(c)} Table of the mode frequencies, linewidths, and dispersive couplings with the qubit for each mode at the operating flux bias ({$\disp{\Q}$} denotes the self-Kerr or anharmonicity of the transmon).}
    \label{fig:dev_overview}
\end{figure*} 

The three-mode readout network is implemented with a hybrid 2D/3D device operating in a dilution refrigerator at $15~\mathrm{mK}$. As shown in Fig.~\ref{fig:dev_overview}(a), the device consists of a 3D coaxial $\lambda/4$ cavity \cite{reagor_quantum_2016} which houses the cavity (C) mode, as well as two chips. One chip contains a fixed-frequency transmon qubit (Q) dispersively coupled to the C mode. The other chip introduces a lumped resonance to act as the amplifier (A) mode and a distributed resonance to act as the buffer (B) mode, both of which have comparatively weak dispersive interactions with the qubit. Importantly, this latter chip includes a dc SQUID which participates in all three linear modes, enabling frequency tunability and parametric interactions. Lastly, the chip includes a fast-flux line to bias the dc SQUID and a port to drive both the B mode and the qubit. More details regarding the device design and EM simulations are included in Appendix~\ref{apdx:design_sim}.

In Fig.~\ref{fig:dev_overview}(b), we show the measured and simulated mode frequencies as a function of flux bias. As desired, the linear mode frequencies tune as a function of flux bias, and the transmon frequency remains approximately constant, retaining low sensitivity to flux noise. The linewidths of the linear modes and achievable parametric interaction strengths are also flux-dependent. We therefore choose a bias point that maximizes $\lw{\B}$ while both minimizing internal losses ($\lw{\A}$,$\lw{\C}$) and maintaining appreciable parametric interaction strengths. In the following, the flux bias is fixed to $0.246~\Phi_0$ and the extracted frequencies, linewidths and dispersive shifts at this operating point are tabulated in Fig.~\ref{fig:dev_overview}(c). Further details regarding the device characterization are available in Appendix~\ref{apdx:expt_calib}.

From the extracted device parameters, it is evident that $\lw{\B}\gg\lw{\A},\lw{\C}$ and $\disp{\C}\gg\disp{\B},\disp{\A}$, matching the requirements laid out in Sec.~\ref{subsec:3mode_network}. Following this basic characterization, we will now leverage the theory developed in Sec.~\ref{sec:theory} to study the performance of this readout network and its effect on the qubit coherence.
\begin{figure*}[t]
    \includegraphics[scale=1]{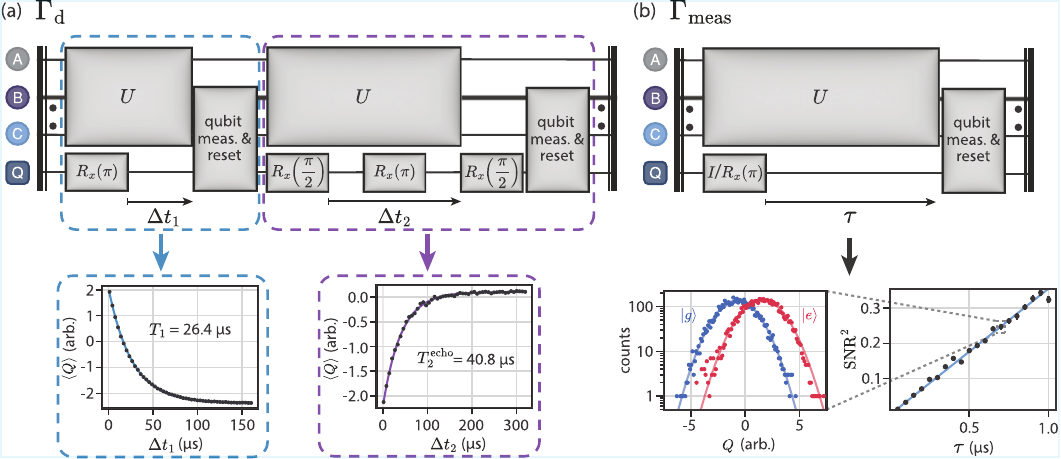}
    \caption{\textbf{Experimental sequences to quantify qubit measurement.} \textbf{(a)} $\drate$ is extracted with interleaved $T_1$ (blue dashed box) and $T_2^\echo$ (purple dashed box) measurements in the presence of arbitrary linear operations on the modes, denoted $U$. In reality, the operations in $U$ begin before the qubit state preparation to ensure the multimode system is in the steady state. The B mode rail is drawn thicker to represent the mode's comparatively large linewidth. Example $T_1$ and $T_2^\echo$ traces in the absence of any readout network drives or parametric interactions ($U=\idmat_3$) are shown below the sequence diagram. (b) To extract $\mrate$, the qubit is prepared in either the ground or excited state, a set of operations $U$ that contains a weak B mode drive is initiated, then the steady state signal is averaged over time $\tau$. Single shot histograms (plotted for the case of the interferometer with $\phi=0.55\pi$ and $\tau$ = 700 ns) are fit to extract $\SNR^2$ as a function of $\tau$, producing a linear relationship with slope $\mrate$.}
    \label{fig:mrate_mdeph}
\end{figure*}

\section{Experimental qubit measurement and dephasing}
\label{sec:qubit_meas}
In this section, we experimentally measure the rates $\mrate$ and $\drate$ in various readout network configurations, comparing them to theoretical predictions calculated with the methodology presented in Sec.~\ref{sec:theory}. Experimentally, the dephasing rate is extracted by interleaving relaxation ($T_1$) and Hahn echo coherence ($T_2^{\mathrm{echo}}$) measurements in the presence of steady-state readout network parametric interactions and drives $U$, see Fig.~\ref{fig:mrate_mdeph}(a). At the end of each sequence, the state of the qubit is read out and reset with the protocol detailed in Appendix~\ref{apdx:diagnostic_qubit_meas}. The Hahn echo sequence suppresses low-frequency dephasing, and assuming an absence of high-frequency contributions intrinsic to the qubit, dephasing due to the readout network can be approximated as $\drate \approx (T_2^{\mathrm{echo}})^{-1} - (2T_1)^{-1}$. The measurement rate $\mrate$ is extracted from single-shot qubit measurements for each qubit state $\sigma=e,g$ and various signal averaging times $\tau$, as shown in Fig.~\ref{fig:mrate_mdeph}.  The means $\mu_{\sigma} = \expt{\hat{\meas}[0]}_{\sigma}$ and variances $\Sigma_{\sigma} = (1/\tau)(\bar{S}_{\meas \meas,\sigma}[0]+\bar{n}_{\add})$ of the resulting distributions define the $\mathrm{SNR}^2$ in Eq.~\eqref{eq:meas_rate_defn}, which is linearly fit with respect to $\tau$ to extract $\mrate$.

In the following subsections, we use the qubit dephasing to independently calibrate the thermal occupancy of each mode with a single free parameter at a time. We then operate the readout network as a nonreciprocal three-mode interferometer, demonstrating quantitative understanding of the qubit dephasing and measurement rates.

\begin{figure}[t]
    \includegraphics[scale=1]{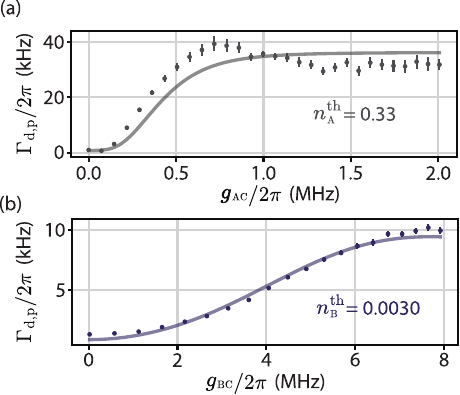}
    \caption{\textbf{Mean thermal occupancy extraction.} Experimental data and theoretical fits (with one free parameter each) of the qubit's parasitic dephasing rate $\pdrate$ as a function of \textbf{(a)} AC beam-splitter strength to extract $\nth{\A} = 0.33$ and \textbf{(b)} BC beam-splitter strength to extract $\nth{\B} = 0.0030$.}
    \label{fig:occupancies}
\end{figure}

\subsection{Thermal occupancy measurements}
\label{subsec:occupancies}
Thermal fluctuations in the readout network introduce excess qubit dephasing and reduce the qubit measurement rate. To quantify each mode's contributions to these fluctuations, the parasitic dephasing rate ($\pdrate$) of the qubit is extracted by measuring the qubit's dephasing rate ($\drate$) in some simple readout network configurations without measurement drives, such that $\drate(\varepsilon=0) = \pdrate$. 

In the absence of any parametric interactions, the readout network approximates a single mode (C) dispersively coupled to the qubit (due to the fact that $\disp{\C} \gg \disp{\A},\disp{\B}$). In this limit, the theory from Sec.~\ref{sec:theory} reproduces the known result for thermal state-induced dephasing expressed in Ref.~\cite{Clerk2007} (see Appendix~\ref{apdx:single_mode_thermal}). Assuming the magnitude of thermal fluctuations is small, the thermal occupancy of mode C, $\nth{\C}$, can be computed with the simple expression:
\begin{equation}\label{eq:nth_C}
    \nth{\C}\approx\frac{\lw{\C}^2+4\disp{\C}^2}{4\lw{\C}\disp{\C}^2}\pdrate = 0.0081.
\end{equation}
This occupancy is standard for state-of-the-art shielding, filtering, and thermalization of the cavity and its control lines.

By design, the qubit is significantly less sensitive to thermal fluctuations in modes A and B, owing to the weak dispersive interaction between each of these modes and the qubit. To enhance the qubit's sensitivity to these fluctuations, a single beam-splitter interaction between mode $\mathrm{K} = \mathrm{A},\mathrm{B}$ and mode C is introduced:
\begin{equation}
    \hat{H}_{\R} =
     g_{\K\C}\left(\hrm{k} \hat{c} + \hat{k} \hrm{c}\right),
\end{equation}
where $g_{\K\C}$ is the beam-splitter interaction strength. Conceptually, this beam-splitter interaction allows the participating modes to share their thermal fluctuations, dissipation rates, and dispersive shifts. As a consequence, the qubit's parasitic dephasing rate will nontrivially depend on $g_{\K\C}$. Figs.~\ref{fig:occupancies}(a) and (b) show the measured $\pdrate$ with respect to the coupling strengths $g_{\A\C}$ and $g_{\B\C}$, independently calibrated in Appendix~\ref{apdx:expt_calib}. Numerical fits of Eq.~\eqref{eq:pdrate} to the data yield the mode occupancies $\nth{\A}=0.33$ and $\nth{\B}=0.0030$. We attribute the larger thermal occupancy of mode A to its stronger coupling to the flux line, which is less filtered and attenuated than the signal line.

These studies of thermal-induced dephasing have facilitated the independent extraction of each mode's thermal occupancy, as well as the initial validation of the theoretical framework in Sec.~\ref{sec:theory} with data. Building on this agreement, we now study the readout network in a more complex mode of operation.

\subsection{Interferometer}
\label{subsec:interferometer}
We now demonstrate the nonreciprocal routing of noise and signals within the readout network by forming an interferometer using three beam-splitter interactions between modes A, B, and C. This results in the Hamiltonian $\hat{H}_{\R}=\hat{H}_{\ifo} + \hat{H}_{\mathrm{drive}}$, with each term being defined in Eq.~\eqref{eq:full_hamiltonian}. The beam-splitter coupling strengths are experimentally tuned to $(g_{\A\B}, g_{\B\C}, g_{\A\C})/2\pi = (14.5, 7.98, 6.2)$ MHz, approximating the optimal nonreciprocal scattering condition in Eq.~\eqref{eq:nonreciprocity_conditions}. The interferometer phase $\phi$ is experimentally calibrated up to an overall shift of $\pi$ by measuring the scattering parameter in reflection off of the B mode. Further details regarding the device tuneup are included in Appendix~\ref{apdx:mmcirc_tune}.

\begin{figure}[t]
    \includegraphics[scale=1]{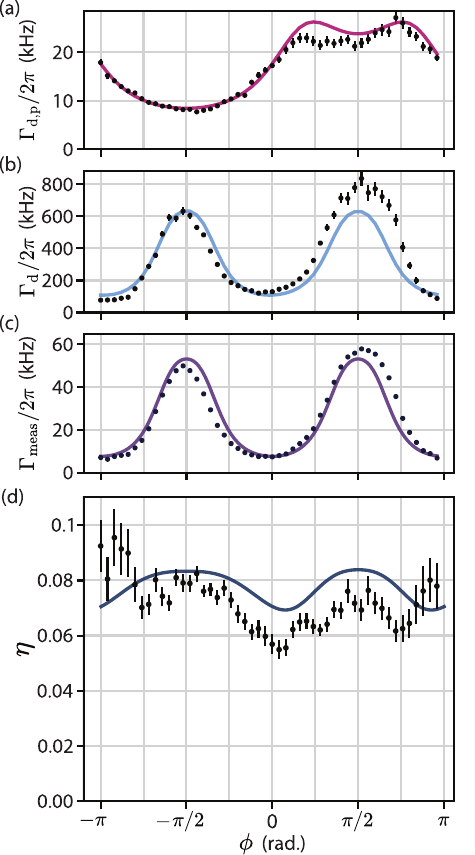}
    \caption{\textbf{Qubit measurement and backaction with a multimode interferometer.} Experimental data and theoretical predictions for the \textbf{(a)} parasitic dephasing rate, \textbf{(b)} total dephasing rate including parasitic and measurement contributions, \textbf{(c)} measurement rate, and \textbf{(d)} measurement efficiency of the qubit as a function of interferometer phase. The parametric coupling rates in this configuration are $(g_{\A\B}, g_{\B\C}, g_{\A\C})/2\pi = (14.5, 7.98, 6.2)$ MHz, the drive strength for the relevant quantities is fixed at $\varepsilon \approx 3.7$ MHz, and other device parameters are given in Fig.~\ref{fig:dev_overview}(c) and Appendix~\ref{apdx:expt_calib}.}
    \label{fig:mmcirc_rates}
\end{figure}

The nonreciprocal routing of thermal photons within the readout network is observed by measuring the parasitic dephasing rate $\pdrate$ as a function of the interferometer phase. In Fig.~\ref{fig:mmcirc_rates}(a), the data, in good agreement with the theoretical model, show a clear asymmetry of $\pdrate$ with respect to $\phi$. Due to the large imbalance in thermal occupancies ($\nth{\A} \gg \nth{\B},\nth{\C}$), the parasitic dephasing is greater when thermal photons in A are directed towards C: $\pdrate(\phi=-\pi/2) < \pdrate(\phi=+\pi/2)$. This asymmetry allows for the disambiguation of the $\pi$-offset in $\phi$, which is the only fit parameter in the model.

We now add a drive to the B mode, with strength $\varepsilon \approx 3.7$ MHz independently calibrated in Appendix~\ref{apdx:epsilonb_chib_calib}. This drive introduces a measurement-induced dephasing ($\mdrate$) contribution to $\drate$, allowing us to measure the qubit with rate $\mrate \leq \mdrate$. The measured total dephasing and measurement rates as a function of interferometer phase are shown in Fig.~\ref{fig:mmcirc_rates}(b) and (c) respectively. The theory predictions show considerable agreement with the measured data with no free parameters. At this drive strength, and for both circulation directions $\phi=\pm\pi/2$, the measurement-induced dephasing overwhelms the parasitic dephasing, $\mdrate \gg \pdrate$. Because $g_{\J\K}, \lw{\B} \gg \lw{\A},\lw{\C}$, measurement signals entering the network will measure the qubit and exit through the B mode regardless of the circulation direction, leading to approximately symmetric measurement and dephasing rates with respect to interferometer phase. 

The measurement efficiency $\eta$, defined in Eq.~\eqref{eq:efficiency} as the ratio of $\mrate$ and $\drate$, is shown in Fig.~\ref{fig:mmcirc_rates}(d) for this interferometer. The efficiency shows a weak dependence on $\phi$, further highlighting the fact that measurement signals preferentially leave mode B regardless of how they are routed within the network. The average efficiency of $\sim7\%$ is principally determined by the added noise of the subsequent measurement chain diagrammed in Appendix~\ref{apdx:experimental_diagram}. In Appendix~\ref{adpx:off_chip_noise_calib}, this added noise is independently calibrated as $\bar{n}_{\add} \approx 5.24$ quanta. 

\begin{figure*}[t]
    \includegraphics[width=1.\linewidth]{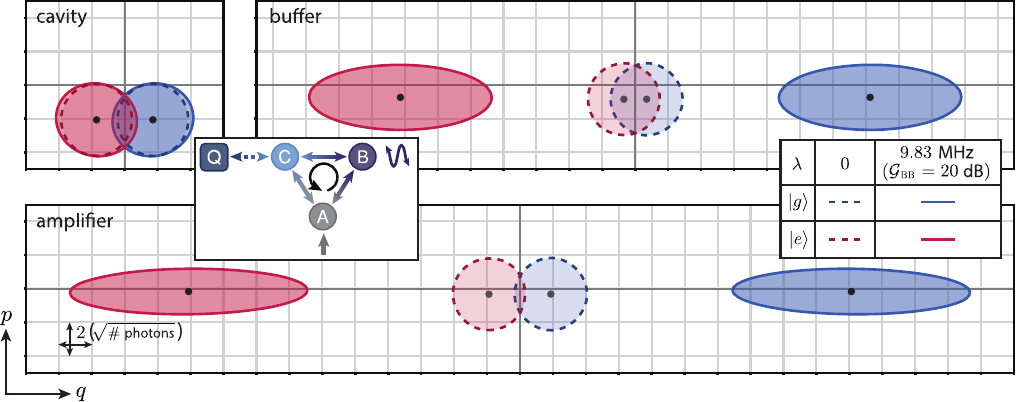}
    \caption{\textbf{Gaussian Wigner distributions for the three internal modes of the embedded amplifier.} Distributions plotted as 99\% confidence ellipses, conditioned on the state of the qubit, either the ground (blue) or excited (red) state, with black points representing the distribution centers. Dashed contours and gray center points correspond to the intracavity states absent any squeezing. The parameters are motivated by the three-mode readout network, modified to exactly satisfy the conditions for nonreciprocity between the A and C modes from Eq.~\eqref{eq:nonreciprocity_conditions}. Linewidths of $(\lw{\A}, \lw{\B}, \lw{\C})/2\pi = (0.115,39.0,0.110)$ MHz, and beam-splitter rates of $(g_{\A\B}, g_{\B\C}, g_{\A\C})/2\pi = (14.5, 7.98, 5.934)$ MHz are used. Only mode C is coupled to the qubit, $(\disp{\A}, \disp{\B}, \disp{\C})/2\pi = (0,0,2.754)$ MHz, and all baths are in the vacuum, $\nth{\K} = 0$. The drive amplitude is $\varepsilon / 2\pi = 20$ MHz, with phase $\varphi = 0$. The single-mode squeezing at mode A is tuned to realize a gain of $\mathcal{G}_{\B\B} = 20$ dB, equating to $\lambda/2\pi = 9.83$ MHz with a phase of $\theta = -\pi/2$. It may be noted that the displacement of the amplifier mode C state is identical to that of the interferometer, i.e. without any amplification, while the variance is close to that of a vacuum state.}
    \label{fig:wigner_ellipses}
\end{figure*}

This downstream added noise motivates the inclusion of gain at mode A, completing the readout network proposed in Sec.~\ref{sec:building_readout_network}. While we were able to observe the desired single-mode squeezing interaction at mode A, it was weak and overshadowed by a spurious gain process of unknown origin, which quickly brought the device into an unstable regime (see Appendix~\ref{apdx:spurious_gain}). With the use of nonlinear simulation techniques \cite{peng_x-parameter_2022} and Kerr-nulling mixing elements \cite{sivak_kerr-free_2019}, we suspect this process may be identified and avoided in future efforts. In the next section, we therefore utilize the experimentally-vetted theory to demonstrate the promise for high-efficiency qubit readout in future experiments, using a device with similar parameters.

\section{Embedded Amplifier Theory}
\label{sec:embedded_amp_theory}
The next step in understanding the full readout network is to add single mode squeezing at mode A of the interferometer discussed in the previous section. This system is now the phase-sensitive, nonreciprocal parametric amplifier modeled by the Hamiltonian in Eq.~\eqref{eq:full_hamiltonian}. To simplify this theoretical analysis, it is assumed that all baths are in the vacuum state, $\nth{\K} \approx 0$, and the dispersive couplings of modes A and B may be neglected, $\disp{\A}, \disp{\B} = 0$. The beam-splitter coupling rates are similar to those of the interferometer, only slightly altered to better match the nonreciprocity condition in Eq.~\eqref{eq:nonreciprocity_conditions}: $(g_{\A\B},g_{\B\C},g_{\A\C})/2\pi = (14.5, 7.98, 5.934)$ MHz. 

To keep the amplifier in the stable regime, we limit the squeezing interaction strength according to the critical stability criteria, discussed in Appendix~\ref{apdx:embedded_amplifier_stability}. Saturating this condition corresponds to $\lambda_{\mathrm{max}} / 2\pi = 10.84$~MHz, and results in gain in reflection off of mode B that grows without bound ($\mathcal{G}_{\B\B} \rightarrow \infty$). This upper limit on the squeezing results in the following inequality:
    \begin{equation}\label{eq:stability}
        \coop{\lambda} < 1 + \coop{\A\B},
    \end{equation}
where $\coop{\lambda} \defeq 2 \lambda/\lw{\A}$ is the squeezing cooperativity. This condition corresponds to the requirement that anti-damping from the squeezing cannot be greater than the enhanced mode A damping rate, $\lambda < \lw{\A,\mathrm{eff}}/2$ where $\lw{\A,\mathrm{eff}} = \lw{\A}(1+\coop{\A\B})$. In order to amplify the measurement field quadrature containing qubit information, the relative phase between the squeezing and the drive on mode B is set to $\theta - 2 \varphi = -\pi/2$. 

The simulated internal state of the embedded amplifier is shown in Fig.~\ref{fig:wigner_ellipses}. In the presence of a gain of $\mathcal{G}_{\B\B} = 20$~dB, the phase-sensitive amplification of the intracavity states is apparent in the A and B modes. In comparison, the nonreciprocity hampers amplification of the state of mode C. As a result, the variance of mode C only deviates slightly from that of a vacuum state, and the displacement of the intracavity state of mode C is identical to the case with no squeezing ($\lambda/2\pi = 0$ MHz). The consequence of this observed behavior on the dephasing rates, as well as the qubit measurement rate, will be discussed below, with supporting details provided in Appendix~\ref{apdx:embedded_amplifier}.

\subsection{Measurement and dephasing rates}
The discussion at the end of Sec.~\ref{sec:multimode_dephasing} provided a heuristic connection between the parasitic and measurement-induced dephasing rates and the variance and displacement of the internal states of any linear modes directly coupled to the qubit. As a consequence, the state of mode C observed in Fig.~\ref{fig:wigner_ellipses} indicates that the qubit backaction ideally has minimal dependence on amplifier gain. 

\begin{figure}
    \includegraphics{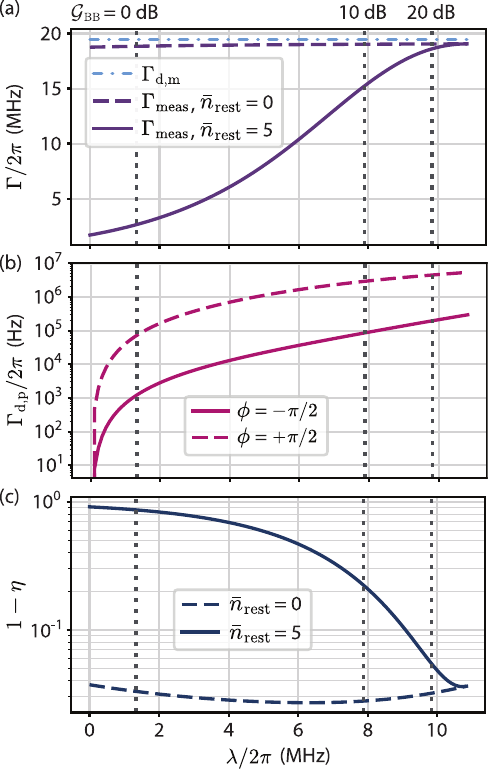}
    \caption{\textbf{Theoretical amplifier performance as a function of squeezing strength.} Plotted as a function of $\lambda$ is the \textbf{(a)} parasitic dephasing rate (pink) for both circulation directions $\phi = -\pi / 2$ (solid) and $\phi = \pi / 2$ (dashed), \textbf{(b)} measurement-induced dephasing (blue dotted-dashed) and measurement (purple) rates, and \textbf{(c)} inefficiency (dark blue), for fixed $\phi = -\pi / 2$ and  $\bar{n}_{\add}=0$ (dashed) or $\bar{n}_{\add}=5$ (solid). Vertical dashed gray lines show correspondence between squeezing strength and reflection gain for the values $\mathcal{G}_{\B\B} = 0, 10$, and  $20$ dB. Other parameters are identical to those in Fig.~\ref{fig:wigner_ellipses}, including the fixed drive strength $\varepsilon / 2\pi = 20$ MHz.}
    \label{fig:amplifier_rates_gain}
\end{figure}

In fact, provided the drive on mode B and all three beam-splitter interactions have zero detuning, the nonreciprocity of the network renders the measurement-induced dephasing entirely independent of any squeezing and thermal noise from mode A. As a result, $\mdrate$ is identical to the result obtained by turning off both the AB and AC beam splitters, effectively decoupling mode A from the system:
    \begin{equation}\label{eq:embedded_amp_mdrate_condition}
        \mdrate \! \left(\nonrec{\C}{\A} = 1\right)
        = \mdrate \Big(\coop{\A\C},\coop{\A\B} = 0\Big).
    \end{equation}
In contrast, the measurement rate is still dependent on the gain generated at mode A. However, for our system parameters, this dependence is minimal in the absence of added noise from the subsequent measurement chain, as illustrated in Fig.~\ref{fig:amplifier_rates_gain}(a). In this case, the measurement rate is reduced solely as a result of signal loss over the path C $\rightarrow$ A $\rightarrow$ B, which is minimized here due to the relatively large values of the beam-splitter cooperativities and the absence of thermal fluctuations. This contribution to the inefficiency becomes negligible as gain is increased. Conversely, the presence of substantial downstream added noise requires considerable amplification for $\mrate$ to approach its maximal value, which is again demonstrated in Fig.~\ref{fig:amplifier_rates_gain}(a).

The improvements in the measurement rate due to amplification necessarily come at the expense of increased parasitic dephasing. Choosing the ideal circulation direction to favor C $\rightarrow$ A scattering ($\phi = - \pi / 2$), in accordance with Eq.~\eqref{eq:nonreciprocity_conditions}, reduces the parasitic dephasing by one to two orders of magnitude in comparison to the opposing direction where A $\rightarrow$ C scattering is dominant ($\phi = + \pi / 2$), as shown in Fig.~\ref{fig:amplifier_rates_gain}(b). However, even for the favorable configuration, the finite bandwidth of the nonreciprocal interference provides only partial protection to mode C from the squeezed noise generated in mode A. As a result, even for the ideal configuration, there is residual $\pdrate$ present that scales with gain. 

\subsection{Measurement efficiency}
\label{subsec:embedded_amp_efficiency}
To quantify the tradeoff of increasing both the measurement rate and the parasitic dephasing with gain, we numerically simulate the measurement efficiency $\eta$ as a function of the squeezing interaction strength. Fig.~\ref{fig:amplifier_rates_gain}(c) presents the results for the inefficiency, $1-\eta$, for $\bar{n}_{\add} = 0$ and 5 photons of downstream added noise, along with a fixed drive strength of $\varepsilon/2\pi = 20$ MHz. At high gain, the system is rendered insensitive to downstream added noise, and the inefficiency converges to a value below $4\%$ for each value of $\bar{n}_{\add}$. Notably, for the case of $\bar{n}_{\add}=0$, the inefficiency reaches a minimum at $\lambda/2\pi = 6.29$ MHz ($\mathcal{G}_{\B\B} = 11.7$ dB). Beyond this squeezing strength, reductions in the inefficiency that come from increased robustness to internal losses are gradually overshadowed by the increased parasitic dephasing of the qubit.

\begin{figure}[t]
    \includegraphics[width=\linewidth]{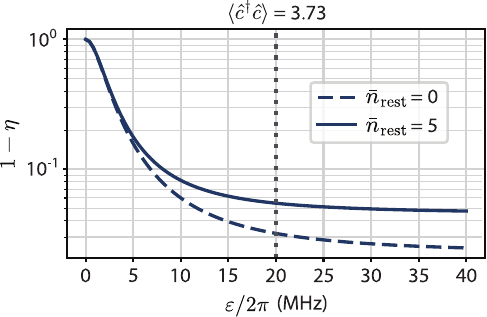}
    \caption{\textbf{Theoretical amplifier performance as a function of drive strength}. The inefficiency is plotted as a function of $\varepsilon$ for $\bar{n}_{\add}=0$ (dashed) or $\bar{n}_{\add}=5$ (solid). A vertical dashed gray line represents a reasonable working point of $\varepsilon / 2 \pi = 20$ MHz corresponding to an intracavity C mode occupancy of 3.73 photons. Other parameters are identical to those in Fig.~\ref{fig:wigner_ellipses}, including the fixed squeezing strength $\lambda / 2\pi = 9.83$ MHz ($\mathcal{G}_{\B\B} = 20$ dB).}
    \label{fig:embedded_amp_eff_drive}
\end{figure}

The degree to which the measurement-induced dephasing overwhelms the parasitic dephasing is determined in large part by the drive amplitude $\varepsilon$. This can be seen in the numerical simulation of the inefficiency as a function of drive strength shown in Fig.~\ref{fig:embedded_amp_eff_drive}. For a fixed gain of $\mathcal{G}_{\B\B} = 20$~dB ($\lambda = 9.83$~MHz), increasing $\varepsilon$ will decrease the inefficiency to an asymptotic value dependent on $\bar{n}_{\add}$, where the condition $\mdrate \gg \pdrate$ is met. In this regime, we may assume negligible parasitic dephasing, allowing us to obtain an approximate analytic expression for the measurement efficiency:
    \begin{align}\label{eq:embedded_amp_eff_A}
        \eta \approx & \, \frac{\coop{bc}}{\coop{bc} + 1} \\[-1.25em]
        & \times
        \frac{(-1 + \coop{ab} + \coop{\lambda})^2}{4 \coop{ab} \coop{\lambda} + (2 \bar{n}_\add+1) (1+\coop{ab}-\coop{\lambda})^2}. \nonumber
    \end{align}
An upper bound on the efficiency can be found by saturating the critical stability criterion from Eq.~\eqref{eq:stability}, resulting in
    \begin{equation}\label{eq:embedded_amp_eff_B}
        \lim_{\mathcal{G}_{\B\B} \rightarrow \infty} \eta \approx
        \frac{\coop{\B\C}}{\coop{\B\C} + 1}
        \times
        \frac{\coop{\A\B}}{\coop{\A\B} + 1}.
    \end{equation}
As expected, in the infinite gain limit, the efficiency becomes independent of $\bar{n}_{\add}$, and is principally limited by the beam-splitter cooperativities that can be achieved. Using the established device parameters and cooperativities, Eq.~\eqref{eq:embedded_amp_eff_B} provides an upper efficiency estimate of $\eta < 97.5 \%$.

This analysis has so far assumed that we are operating within the dispersive regime, however, increasing the drive strength will eventually lead to a breakdown of this model as spurious transmon state transitions become increasingly likely \cite{Blais2004,Koch2007,Sank2016, dumas_measurement-induced_2024, Khezri2023, nesterov_measurement-induced_2024, connolly_full_2025}. For the drive strength of $\varepsilon/2\pi = 20$ MHz used in Fig.~\ref{fig:amplifier_rates_gain}, the intracavity photon population for mode C is in the range of $\expt{\hrm{c} \hat{c}} \in [3.51,3.73]$ photons for $\lambda/2\pi \in [0,10.84]$ MHz, safely below the critical photon number regardless of the gain. For the amplifier under consideration, a range of drive strengths around $\varepsilon/2\pi = 20$~MHz are therefore weak enough to remain QND in most cases, yet strong enough such that $\mdrate \gg \pdrate$. 

\subsection{Limits to the efficiency}
In Sec.~\ref{subsec:measrate_efficiency}, we detailed two conditions for a high measurement efficiency. The second of these conditions demands that $\mrate/\mdrate \rightarrow 1$. 

The approximate expression for the efficiency in Eq.~\eqref{eq:embedded_amp_eff_B} would appear to indicate that this condition can be met simply by increasing the unconstrained beam-splitter cooperativities $\coop{\B\C}$ and $\coop{\A\B}$, allowing for an efficiency that comes arbitrarily close to unity. This improvement in the efficiency stems from the coherent processes overwhelming dissipation within the readout network, as discussed in Sec.~\ref{sec:building_readout_network}. However, apart from the experimental difficulties in achieving large parametric coupling rates, sufficiently large cooperativities also fundamentally curtail the ability of this system to meet the first condition from Sec.~\ref{subsec:measrate_efficiency}, namely $\mdrate \gg \pdrate$, and so will eventually reduce the efficiency. The reason for this behavior is given here, with supporting calculations provided in Appendix~\ref{apdx:embedded_amplifier_efficiency_limits}.

In the three-mode readout network, each beam-splitter interaction between mode B and modes A and C increases the effective linewidths of the latter modes, providing additional damping, $\lw{\K,\mathrm{eff}} = \lw{\K}(1+\coop{\B\K})$ where K=A,C. For the AB interaction, increasing the effective A mode damping improves the system stability according to Eq.~\eqref{eq:stability}, and hence a commensurate increase in squeezing strength is required to retain a constant gain $\mathcal{G}_{\B\B}$. This directly leads to an increase in parasitic dephasing, while the measurement-induced dephasing rate remains constant, as established in Eq.~\eqref{eq:embedded_amp_mdrate_condition}. Additionally, the required squeezing will eventually become so large that the system becomes unstable even for finite values of $\mathcal{G}_{\B\B}$. Suitably large $\coop{\A\B}$ therefore prevents the amplifier from fully combating $\bar{n}_{\add}$.
 
For the BC interaction, the enhanced C mode damping overwhelms its dispersive coupling to the qubit when $\lw{\C,\mathrm{eff}} \gg 2 \disp{\C}$. In this regime, analogous to the case of $\lw{} \gg 2\disp{}$ for a single-mode dispersive measurement, the separation between the pointer states becomes vanishingly small, reducing $\mrate$ and $\mdrate$ in the process. In comparison, the excess noise from the amplified intracavity state remains considerable. Therefore, for a fixed $\disp{\C}$, an arbitrarily large value of $\coop{\B\C}$ also overshadows the measurement process with parasitic dephasing.

The above complications limit the efficiency and render Eqs.~\eqref{eq:embedded_amp_eff_A} and~\eqref{eq:embedded_amp_eff_B} inapplicable for large cooperativities in the device considered here. However, despite these fundamental limitations, we have shown the theoretical promise of record-breaking efficiency for a reasonable extension of the experimentally-realized interferometer in Sec.~\ref{subsec:interferometer}.

\section{Conclusions and Outlook}
\label{sec:conclusion}
In this paper, we have presented a general framework to quantify qubit measurement and backaction metrics in a multimode readout system. While existing methods allowed for the computation of the measurement rate, establishing quantitative understanding of qubit backaction from the coupled readout system required an extension of existing theoretical methods. 

We then applied this theoretical framework to inform and understand the experimental implementation of a superconducting qubit dispersively coupled to a three-mode, parametric readout network. Readout network operations of increasing complexity were experimentally studied to extract system parameters and vet the theory, leading to a quantitative characterization of qubit measurement and backaction with a nonreciprocal interferometric readout. The inclusion of gain was theoretically explored, demonstrating the promise for efficient and low-backaction superconducting qubit readout in future experiments.

In addition to the experimental inclusion of gain, the work presented here suggests some natural improvements and extensions. For instance, favorable properties such as larger bandwidth for both gain and isolation can be achieved by incorporating additional modes and interactions \cite{roy_broadband_2015,kaufman_josephson_2023,abdo_nondegenerate_2025}. To this end, recent work has demonstrated the analytical and algorithmic-based synthesis of the linear response of coupled mode networks \cite{naaman_synthesis_2022, landgraf_automated_2025}, potentially accelerating the discovery of multimode systems with favorable qubit readout attributes. These methods, in conjunction with the theory presented here, could be integrated into a comprehensive workflow for readout network design. 

While extra modes and interactions can produce favorable readout network attributes, they also augment the system's complexity. The necessity of additional pumps particularly burdens the experimental operation of these devices through pump-induced frequency shifts and spurious mixing processes. Some of this complexity may be removed by employing resonant coupling, as opposed to beam-splitter couplings, for a subset of the inter-mode interactions \cite{kwende_josephson_2023}. In addition to aiding the realizability of more complex devices, this strategy could also simplify the readout network presented in this work. 

While this work focused on the readout of a single qubit, the theoretical framework presented is extensible to multi-qubit readout under some reasonable constraints. In addition to the requirements presented in Sec.~\ref{sec:theory}, the inter-qubit interactions remain $\hat{\sigma}_z$ in nature, and the measurement must occur well within the relaxation time of each qubit. By similar argument, the phase space method could be extended to the readout of multilevel systems (qudits), with analogous constraints imposed to those of a multi-qubit readout.

Lastly, while this work emphasizes readout in a continuous regime, the dynamical equations constructed in Sec.~\ref{sec:theory} equally apply to time-domain readout protocols. These protocols, an example of which is proposed in Ref.~\cite{bello_high-fidelity_2026}, are of interest as they allow for backaction that is principally determined by idle coupling between the readout cavity and the subsystem of the readout network with gain, or unintentional dispersive interactions. In a multimode parametric system, both of these can be suppressed to a high degree. A quantitative study of such a qubit readout performed with pulsed parametric interactions is an interesting subject for future exploration. 

\section{Acknowledgements}
\label{sec:thanks}
The authors thank Massachusetts Institute of Technology Lincoln Laboratory for providing the TWPA. A.M. and L.O. acknowledge funding by the Deutsche Forschungsgemeinschaft through the project 163436311-SFB 910 and the Emmy Noether program (Grant No. ME 4863/1-1).  F. L. and B. T. M. thank Jose Aumentado and John Teufel for their critical contributions to the design and troubleshooting of the experiment.

\appendix
\clearpage
\widetext
\section{General theory: supporting expressions}
\label{apdx:theory}
\subsection{The Wigner phase-space formulation}
In order to efficiently convert the dynamical equation for a Gaussian operator to a system of differential equations for its moments, we first perform a Wigner transformation to convert the dynamical equation to a PDE in phase space. To simplify this process, we use the fact that the product between operators is replaced by an associative and non-commutative ``star product'' of QPDs in phase space. For the Wigner representation this is the Moyal product, which may be defined as
    \begin{equation}\label{eq:moyal-product}
        f(\hat{\qvec}) g(\hat{\qvec})
        \xrightarrow[]{\text{Wigner}}
        \wig_{\! f}(\qvec) \star \wig_{\! g}(\qvec)
        = \wig_{\! f}(\qvec) \exp\left[\frac{i \hbar}{2} \lp{\qelem}_j \symelem_{jk} \rp{\qelem}_k \right] \wig_{\! g}(\qvec),
    \end{equation}
where $\wig_{\! f}(\qvec)$ and $\wig_{\! g}(\qvec)$ are the Wigner QPDs associated with $f(\hat{\qvec})$ and $g(\hat{\qvec})$, respectively. Here, $\pqelem_j \defeq \partial/\partial \qelem_j$ is shorthand for the partial derivative operator with respect to the phase-space coordinate $\qelem_j$, where the arrows above indicate that these derivatives are directional. Lastly, $\symelem_{jk} = -\symelem_{kj}$ denotes a matrix element of $\symform$. Examples of $\qvec$ and the associated $\symform$ may be found in Eqs.~\eqref{eq:quadrature_basis} and \eqref{eq:grouped_mode_basis} in the main text.

From this definition, the Wigner QPD for an operator $f(\hat{\qvec})$ is $\wig_{\! f}(\qvec) = f(\qvec \star)$. The linear operator $\hat{\qelem}_j$ therefore transforms to the phase-space symbol $\qelem_j$. Additionally defining the Wigner transform of an operator $\hat{\varrho}$ (not necessarily Hermitian) as $\wig(\qvec)$, the dynamical equation for $\hat{\varrho}$ may now be transformed to a phase-space PDE by treating the exponential in Eq.~\eqref{eq:moyal-product} as a power series. Taking $\hbar=1$, the coherent commutator terms in the dynamical equation then transform as follows
    \begin{equation}
        -i\comm{\hat{\qelem}_j}{\hat{\varrho}} 
        \rightarrow 
        - \pqelem_l \symelem_{lj} \wig (\qvec), 
        \qquad
        -i\comm{\hat{\qelem}_j \hat{\qelem}_k}{\hat{\varrho}}
        \rightarrow
        - \pqelem_l \left(\symelem_{lk} \qelem_j + \symelem_{lj} \qelem_k \right)  \wig (\qvec),
    \end{equation}
while anti-commutator terms transform as
    \begin{equation}
        \acomm{\hat{\qelem}_j}{\hat{\varrho}} 
        \rightarrow
        2 \qelem_j \wig (\qvec),
        \qquad
        \acomm{\hat{\qelem}_j \hat{\qelem}_k}{\hat{\varrho}} 
        \rightarrow
        \left(2 \qelem_j \qelem_k + i \symelem_{jk} + \frac{1}{2} \symelem_{lj} \symelem_{km} \pqelem_l \pqelem_m \right) \wig (\qvec).
    \end{equation}
The so-called jump terms take the form
    \begin{align}
        \hat{\qelem}_k \hat{\varrho} \hat{\qelem}_j
        \rightarrow
        \left(\qelem_j \qelem_k + \frac{i}{2} \symelem_{jk} + \frac{i}{2} \pqelem_l \left(\symelem_{lj} \qelem_k - \symelem_{lk} \qelem_j \right) - \frac{1}{4} \symelem_{lj} \symelem_{km} \pqelem_l \pqelem_m \right) \wig (\qvec)
    \end{align}
which, combined with the relevant anti-commutator, allows for dissipation terms to be transformed as
    \begin{align}
        \hat{\qelem}_k \hat{\varrho} \hat{\qelem}_j - \frac{1}{2} \acomm{\hat{\qelem}_j \hat{\qelem}_k}{\hat{\varrho}} 
        \rightarrow
        \frac{1}{2} \Big(i \pqelem_l \left(\symelem_{lj} \qelem_k - \symelem_{lk} \qelem_j \right) - \symelem_{lj} \symelem_{km} \pqelem_l \pqelem_m \Big) \wig (\qvec).
    \end{align}
With these relations, the dynamical equation for $\hat{\rho}_{eg}$ in Eq.~\eqref{eq:eff_master_offdiagonal} may be expressed as a differential equation for $W_{eg}(\qvec)$. For this, we start by writing the quadratic Hamiltonians $\hat{H}_{\R}$ and $\hat{H}_{\q\R}$ from Eqs.~\eqref{eq:cavity_lindblad} and~\eqref{eq:qubit_cavity_lindblad} in the form $\hat{H} =  \hat{\qelem}_j \HTwoElem_{jk} \hat{\qelem}_k/2 + \HOneElem_j \hat{\qelem}_j + \HNull$, where $\HTwoElem_{jk} = \HTwoElem_{kj}$. Eq.~\eqref{eq:eff_master_offdiagonal}, $\dot{\hat{\rho}}_{eg} = \lbld_{\R}(\hat{\rho}_{eg}) - i \acomm{\hat{H}_{\q\R}}{\hat{\rho}_{eg}}$, may now be expressed as:
    \begin{align}\label{eq:weg_fokker_planck}
        \frac{\partial}{\partial t} \wig_{eg}(\qvec)
        = \bigg[- 2 i \HNull_{\q\R} - \pqelem_j \symelem_{jl} (\HOneElem_{\R})_l - 2 i \qelem_j (\HOneElem_{\q\R})_j &- \frac{1}{2} \pqelem_j \symelem_{jl} \left(\frac{i}{2} (\HTwoElem_{\q\R})_{lm} + \real[\rate_{lm}] \right) \symelem_{mk} \pqelem_k \nonumber \\[-1.2em]
        &- i \qelem_j (\HTwoElem_{\q\R})_{jk} \qelem_k - \pqelem_j \symelem_{jl} \Big((\HTwoElem_{\R})_{lk} + \imag[\rate_{lk}]\Big) \qelem_k \bigg] \wig_{eg}(\qvec).
    \end{align}
The real and imaginary parts of $\rate_{jk}$ are inserted by identifying that $\rate_{jk} = \rate^*_{kj}$. Time evolution equations may also be obtained for the intracavity pointer state Wigner functions, denoted $W_{\sigma \sigma}(\qvec)$ for $\hat{\rho}_{\sigma \sigma}$ where $\sigma = e,g$. The reduced Lindblad master equation for these intracavity pointer states, $\dot{\hat{\rho}}_{\sigma \sigma} = \lbld_{\R}(\hat{\rho}_{\sigma \sigma}) - i \braket{\sigma| \hat{\sigma}_z| \sigma} \comm{\hat{H}_{\q\R}}{\hat{\rho}_{\sigma \sigma}}$, maps to the following PDE:
    \begin{equation}\label{eq:wee-gg_fokker_planck}
        \frac{\partial}{\partial t} \wig_{\sigma \sigma}(\qvec)
        = \bigg[- \pqelem_j \symelem_{jl} (\HOneElem_{\R} + s \HOneElem_{\q\R})_l - \frac{1}{2} \pqelem_j \symelem_{jl} \real[\rate_{lm}] \symelem_{mk} \pqelem_k
        - \pqelem_j \symelem_{jl} \Big((\HTwoElem_{\R} + s \HTwoElem_{\q\R})_{lk} + \imag[\rate_{lk}]\Big) \qelem_k \bigg] \wig_{\sigma \sigma}(\qvec)
    \end{equation}
where $s = \braket{\sigma| \hat{\sigma}_z | \sigma}$ yields $s=1$ for $\sigma = e$, and $s=-1$ for $\sigma=g$.

\subsection{Dynamical equations for moments of the Wigner function}
The Gaussian property of a Wigner QPD $\wig(\qvec)$ is preserved when $W(\qvec)$ evolves according to a PDE of the form
    \begin{equation}\label{eq:general_fokker_planck}
        \frac{\partial}{\partial t} \wig(\qvec) = \left[- g - \pqelem_j d_j -  \qelem_j f_j + \frac{1}{2} \pqelem_j C_{jk} \pqelem_k - \frac{1}{2} \qelem_j B_{jk}\qelem_k - \pqelem_j A_{jk} \qelem_k \right] \wig(\qvec).
    \end{equation}
Fourier transforming the above PDE to the reciprocal phase space coordinates yields another PDE for the characteristic function of $\wig(\qvec)$, $\charw(\fvec)$, as follows
    \begin{equation}
        \frac{\partial}{\partial t} \charw(\fvec) = \left[- g - i \felem_j d_j - i \pfelem_j f_j - \frac{1}{2} \felem_j C_{jk} \felem_k + \frac{1}{2} \pfelem_j B_{jk} \pfelem_k + \felem_j A_{jk} \pfelem_k\right] \charw(\fvec).
    \end{equation}
To derive the moment equations, the characteristic function for a Gaussian QPD is inserted into the above PDE. From Eq.~\eqref{eq:char_func_ansatz}, the characteristic function may be written as
    \begin{equation}
        \charw (\fvec)
        = \exp\left[-\frac{1}{2} \celem_{jk} \felem_j \felem_k - i \melem_k \felem_k - \nu \right]
    \end{equation}
where $\celem_{jk} = \celem_{kj}$ since the covariance matrix is symmetric. Evaluating the derivatives and collecting terms of identical order reveals the following differential equations for the individual moments,
    \begin{align}
        \frac{d}{dt} \celem_{jk} &= A_{jl} \celem_{lk} + \celem_{jl} A^T_{lk} - \celem_{jl} B_{lm} \celem_{mk} + C_{jk}, \nonumber \\[-0.5em]
        \frac{d}{dt} \melem_j &= A_{jk} \melem_k - \celem_{jk} B_{kl} \melem_l + d_j - \celem_{jk} f_k, \nonumber \\[-0.5em]
        \frac{d}{dt} \nu &= g + f_k \melem_k +  \melem_j B_{jk} \melem_k + \frac{1}{2} B_{jk} \celem_{kj},
    \end{align}
which are identical to those in Eq.~\eqref{eq:moment_odes}. The expressions in Eq.~\eqref{eq:array_definitions} may then be obtained by comparing terms in Eqs.~\eqref{eq:weg_fokker_planck} and~\eqref{eq:general_fokker_planck}. When using these equations to calculate the intracavity pointer states, Eq.~\eqref{eq:wee-gg_fokker_planck} shows that only the following arrays are non-zero:
    \begin{equation}\label{eq:pointer_states_moments_eqns}
        \bm{A} = \symform \left(\HTwo_{\R} + s \HTwo_{\q\R} + \imag[\dissmat] \right), \qquad
        \bm{C} = - \symform \real[\dissmat] \symform, \qquad
        \bm{d} = \symform \left(\HOne_{\R} + s \HOne_{\q\R}\right).
    \end{equation}

\subsection{Heisenberg-Langevin equations}
\label{apdx:HLEs}
While the Heisenberg-Langevin equations (HLEs) are used to model the output fields of the readout network required to calculate the measurement rate, they may also be used to model the intracavity pointer states. These dynamics must therefore be consistent with those obtained from the Lindblad master equation, and hence the dynamical equations for the means and covariances obtained from the HLEs must be identical. In order to check that both formalisms yield the same intracavity dynamics, we start by deriving the HLEs following the approach of Ref.~\cite{gardiner_input_1985}. We consider a system of $N$ linear modes coupled to $M$ independent baths, which are treated as noninteracting harmonic oscillators. The total Hamiltonian is written as $\hat{H} = \hat{H}_{\mathrm{s}} + \hat{H}_{\mathrm{b}} + \hat{H}_{\mathrm{sb}}$, where:
    \begin{equation}
        \hat{H}_{\mathrm{s}} = \frac{1}{2} \hat{\qvec}^T \HTwo_{\mathrm{s}} \hat{\qvec} + \hat{\qvec}^T \HOne_{\mathbf{s}}, \qquad
        \hat{H}_{\mathrm{b}} = \frac{1}{2} \int_{-\infty}^{\infty} \omega \hat{\qvec}_{\mathrm{b}}^T(\omega) \hat{\qvec}_{\mathrm{b}}(\omega) \, d\omega, \qquad
        \hat{H}_{\mathrm{sb}} = \frac{1}{\sqrt{2\pi}} \int_{-\infty}^{\infty} \hat{\qvec}^T \HTwo_{\mathrm{sb}} \hat{\qvec}_{\mathrm{b}}(\omega) \, d\omega,
    \end{equation}
Here, $(\HTwo_{\mathrm{s}})^T = \HTwo_{\mathrm{s}} \in \mathbb{R}^{2N \times 2N}$, $\HOne_{\mathbf{s}} \in \mathbb{R}^{2N}$, and $\HTwo_{\mathrm{sb}} \in \mathbb{R}^{2N \times 2M}$. The Markov approximation has already been applied, which renders the system-bath coupling independent of frequency. In the above, $\hat{\qvec}$ is the vector of $2N$ system quadrature operators, while $\hat{\qvec}_{\mathrm{b}}(\omega)$ is the vector of $2M$ bath quadrature operators at the frequency $\omega$, which obey $\comm{\hat{\qelem}_{\mathrm{b},j}(\omega)}{\hat{\qelem}_{\mathrm{b},k}(\omega')} = i \symelem_{jk} \delta(\omega - \omega')$. The Heisenberg equation for $\hat{\qvec}_{\mathrm{b}}(\omega)$ is formally solved and substituted into the dynamical equation for $\hat{\qvec}$ to obtain the HLEs:
    \begin{equation}\label{eq:heisenberg-langevin}
        \frac{d}{dt} \hat{\qvec}(t) = \bm{A} \hat{\qvec}(t) + \symform_{2N} \HOne_{\mathbf{s}} + \symform_{2N} \HTwo_{\mathrm{sb}} \hat{\qvec}_{\inn}(t)
        \quad \text{where} \quad
        \bm{A} = \symform_{2N} \! \left(\HTwo_{\mathrm{s}} + \frac{1}{2} \HTwo_{\mathrm{sb}} \symform_{2M} \HTwo_{\mathrm{sb}}^T \right).
    \end{equation}
Here, $\symform_{2N}$ and $\symform_{2M}$ are the matrix representation of the symplectic forms for the system and bath operators, respectively. The vector $\hat{\qvec}_{\inn}(t)$ contains the $2M$ input-field quadrature operators, defined as
    \begin{equation}
        \hat{\qelem}_{\inn,j}(t)
        \defeq \frac{1}{\sqrt{2\pi}} \! \int_{-\infty}^{\infty} \!\! \left(e^{-i \omega (t-t_0) \symform}\right)_{\!\! jk} \!\! \hat{\qelem}_{\mathrm{b},k}(\omega;t_0) \, d\omega 
        = \frac{1}{\sqrt{2\pi}} \! \int_{-\infty}^{\infty} \!\! \Big(\delta_{jk} \cos\boldsymbol{(}\omega(t-t_0)\boldsymbol{)} + \symelem_{jk} \sin\boldsymbol{(}\omega(t-t_0)\boldsymbol{)} \! \Big) \hat{\qelem}_{\mathrm{b},k}(\omega;t_0) \, d\omega,
    \end{equation}
where $\hat{\qelem}_{\mathrm{b},j}(\omega;t_0)$ is the value of $\hat{\qelem}_{\mathrm{b},j}(\omega)$ at an initial time $t_0 < t$. The input fields satisfy the canonical commutation relations, $\comm{\hat{\qelem}_{\inn,j}(t)}{\hat{\qelem}_{\inn,k}(t')} = i \symelem_{jk} \delta(t - t')$, and are assumed to be Gaussian noise processes, with first and second moments given by
    \begin{equation}
        \expt{\hat{\qelem}_{\inn,j}(t)} = \melem_{\inn,j},
        \qquad
        \frac{1}{2} \expt{\hat{\qelem}_{\inn,j}(t) \hat{\qelem}_{\inn,k}(t') + \hat{\qelem}_{\inn,k}(t') \hat{\qelem}_{\inn,j}(t)} 
        - \expt{\hat{\qelem}_{\inn,j}(t)} \expt{\hat{\qelem}_{\inn,k}(t')} 
        = (\bar{S}_{\inn})_{jk} \delta(t-t').
    \end{equation}
With these correlators, and the HLEs in Eq.~\eqref{eq:heisenberg-langevin}, we can now derive dynamical equations for the intracavity moments. For the means, $\melem_j \defeq \expt{\hat{\qelem}_j}$, this is simply the expectation value of the HLE for $\hat{\qelem}_j$. For the covariances, $\celem_{jk} \defeq \frac{1}{2} \expt{\hat{\qelem}_j \hat{\qelem}_k + \hat{\qelem}_k \hat{\qelem}_j} - \expt{\hat{\qelem}_j} \expt{\hat{\qelem}_k}$, we must use the fact the HLEs are consistent with calculus \cite{gardiner_input_1985}, $\frac{d}{dt}(\hat{\qelem}_j \hat{\qelem}_k) = (\frac{d}{dt} \hat{\qelem}_j) \hat{\qelem}_k + \hat{\qelem}_j (\frac{d}{dt} \hat{\qelem}_k)$. When evaluating terms of the form $\frac{1}{2} \expt{\hat{\qelem}_j (\frac{d}{dt} \hat{\qelem}_k) + (\frac{d}{dt} \hat{\qelem}_k) \hat{\qelem}_j} - \expt{\hat{\qelem}_j} \expt{\frac{d}{dt} \hat{\qelem}_k}$, $\frac{d}{dt} \hat{\qelem}_k$ is replaced by the HLE, while $\hat{\qelem}_j(t)$ is replaced by the formal solution to its HLE. We assume that the initial-time value of $\hat{\qelem}_j(t)$ is uncorrelated with the input-field, $\frac{1}{2} \expt{\hat{\qelem}_j(t_0) \hat{\qelem}_{k,\inn}(t) + \hat{\qelem}_{k,\inn}(t) \hat{\qelem}_j(t_0)} - \expt{\hat{\qelem}_j(t_0)} \expt{\hat{\qelem}_{k,\inn}(t)} = 0$, while the same-time correlation between the system and environment is calculated to be $\frac{1}{2} \expt{\hat{\qelem}_j(t) \hat{\qelem}_{k,\inn}(t) + \hat{\qelem}_{k,\inn}(t) \hat{\qelem}_j(t)} - \expt{\hat{\qelem}_j(t)} \expt{\hat{\qelem}_{k,\inn}(t)} = \frac{1}{2} (\symelem_{2N} \HTwoElem_{\mathrm{sb}})_{jl} (\bar{S}_{\inn})_{lk}$. Combining terms then yields the following expressions for the intracavity moments:
    \begin{equation}\label{eq:HL_moments_eqns}
        \frac{d}{dt} \mean = \bm{A} \mean + \symform_{2N} (\HOne_{\mathbf{s}} + \HTwo_{\mathrm{sb}} \mean_{\inn}),
        \qquad
        \frac{d}{dt} \covar = \bm{A} \covar + \covar \bm{A}^T - \symform_{2N} \HTwo_{\mathrm{sb}} \bar{\bm{S}}_{\inn} \HTwo_{\mathrm{sb}}^T \symform_{2N}.
    \end{equation}
In order for these dynamical equations to match those of the intracavity pointer states generated from the Lindblad master equation, represented by the arrays in Eq.~\eqref{eq:pointer_states_moments_eqns}, it is required that $\HTwo_{\R} + s \HTwo_{\Q\R} = \HTwo_{\mathrm{s}}$ and $\HOne_{\R} + s \HOne_{\Q\R} = \HOne_{\mathbf{s}} + \HTwo_{\mathrm{sb}} \mean_{\inn}$. Additionally, the matrix of decay rates is related to the input-noise correlators and system-bath coupling rates via $\dissmat = \HTwo_{\mathrm{sb}} (\bar{\bm{S}}_{\inn} + i \symform_{2M}/2) \HTwo_{\mathrm{sb}}^T$. The input fields obey the Robertson-Schr\"{o}dinger uncertainty relation if $\bar{\bm{S}}_{\inn} + i \symform_{2M}/2 \geq 0$, and so it follows that $\dissmat \geq 0$. Since $\dissmat$ is Hermitian and positive semidefinite, the dynamics of the equivalent Lindblad master equation will be completely positive and trace-preserving. This relation also demonstrates that $\HTwo_{\mathrm{sb}}$ is not uniquely determined by $\dissmat$ and $\bar{\bm{S}}_{\inn}$, and hence a Lindblad master equation does not correspond to a unique set of HLEs.

\subsection{Input-output theory and scattering}
\label{apdx:input-output_scattering}
Following standard input-output theory, the HLEs in Eq.~\eqref{eq:heisenberg-langevin} may also be written in time-reversed form using output-field quadrature operators, $\hat{\qelem}_{\out,j}(t) \defeq \int_{-\infty}^{\infty} \left(\exp[-i \omega (t-t_1) \symform]\right)_{jk} \hat{\qelem}_{\mathrm{b},k}(\omega;t_1) \, d\omega/\sqrt{2\pi}$, where $\hat{\qelem}_{\mathrm{b},k}(\omega;t_1)$ is the value of $\hat{\qelem}_{\mathrm{b},k}(\omega)$ at some future time $t_1>t$. By equating both the time-forward and reversed HLEs, the input-output relation is obtained:
    \begin{equation}\label{eq:input_out_relation}
        \hat{\qvec}_{\out}(t) - \hat{\qvec}_{\inn}(t) = \symform_{2M} \HTwo_{\mathrm{sb}}^T \hat{\qvec}(t).
    \end{equation}
To derive a scattering relation between the input and output fields, the HLEs in Eq.~\eqref{eq:heisenberg-langevin} are first Fourier-transformed, yielding a set of linear equations in frequency space. Solving these linear equations and then substituting them into the Fourier-transformed input-output relation Eq.~\eqref{eq:input_out_relation} produces the following scattering equation:
    \begin{align}\label{eq:scattering_relation}
        \hat{\qvec}_{\out}[\omega] = \smat[\omega] \hat{\qvec}_{\inn}[\omega] + \mathbf{T}[\omega]
        \quad \text{where} \quad
        \smat[\omega] &= - \symform_{2M} \HTwo_{\mathrm{sb}}^T \left(\bm{A} + i \omega \idmat_{2N}\right)^{-1} \! \symform_{2N} \HTwo_{\mathrm{sb}} + \idmat_{2N} \\[-0.5em]
        \mathbf{T}[\omega] &= - \symform_{2M} \HTwo_{\mathrm{sb}}^T \left(\bm{A} + i \omega \idmat_{2N}\right)^{-1} \! \symform_{2N} \HOne_{\mathrm{s}}. \nonumber
    \end{align}
We have defined the Fourier-transformed operators as $\hat{r}_j[\omega] \defeq \int^{\infty}_{-\infty} e^{i \omega t} \hat{r}_j(t) dt/\sqrt{2\pi}$. 

In this work, each system mode is coupled to its own bath, so that the number of system modes and baths is the same, $N=M$. Our choice of quadrature basis for the system and bath corresponds to the convention defined in Eq.~\eqref{eq:quadrature_basis}, and therefore we use $\hat{\qvec} = (\hat{q}_1,\hat{p}_1,\ldots,\hat{q}_N,\hat{p}_N)$ and $\hat{\qvec}_{\mathrm{b}}(\omega) = (\hat{q}_{\mathrm{b},1}(\omega),\hat{p}_{\mathrm{b},1}(\omega),\ldots,\hat{q}_{\mathrm{b},N}(\omega),\hat{p}_{\mathrm{b},N}(\omega))$. The bath and system modes are assumed to exchange a conserved number of excitations, i.e. the system-bath couplings take the form of beam-splitter interactions with arbitrary phases:
    \begin{equation}
        \hat{H}_{\mathrm{sb}} = \frac{1}{\sqrt{2\pi}} \sum_{\K \in N} \int_{-\infty}^{\infty} \sqrt{\lw{\K}} \Big(\! \cos(\phi_{\K}) \left[\hat{q}_{\K}\hat{q}_{\mathrm{b},\K}(\omega) + \hat{p}_{\K} \hat{p}_{\mathrm{b},\K}(\omega)\right] - \sin(\phi_{\K}) \left[\hat{q}_{\K} \hat{p}_{\mathrm{b},\K}(\omega) - \hat{p}_{\K} \hat{q}_{\mathrm{b},\K}(\omega)\right] \! \Big) \, d\omega.
    \end{equation}
The matrix representing the couplings in $\hat{H}_{\mathrm{sb}}$ is therefore $\HTwo_{\mathrm{sb}} = \bigoplus_{\K \in N} \sqrt{\lw{\K}} \left(\begin{smallmatrix} \cos(\phi_{\K}) & -\sin(\phi_{\K}) \\ \sin(\phi_{\K}) & \cos(\phi_{\K}) \end{smallmatrix}\right)$. Without loss of generality, all phases are set to $\phi_{\K} = -\pi/2$ so that $\symform_{2N} \HTwo_{\mathrm{sb}} = - \sqrt{\boldsymbol{\upgamma}}$ where $ \boldsymbol{\upgamma} = \bigoplus_{\K \in N} \lw{\K} \idmat_2$. The Heisenberg-Langevin equations in Eq.~\eqref{eq:heisenberg-langevin} are now
    \begin{equation}
        \frac{d}{dt} \hat{\qvec}(t) = \bm{A} \hat{\qvec}(t) + \symform_{2N} \HOne_{\mathbf{s}} - \sqrt{\boldsymbol{\upgamma}} \hat{\qvec}_{\inn}(t)
        \quad \text{where} \quad
        \bm{A} = \symform_{2N} \HTwo_{\mathrm{s}} - \frac{1}{2} \boldsymbol{\upgamma}.
    \end{equation}
Consequently, the input-output relation in Eq.~\eqref{eq:input_out_relation} are $\hat{\qvec}_{\out}(t) - \hat{\qvec}_{\inn}(t) = \sqrt{\boldsymbol{\upgamma}} \hat{\qvec}(t)$. Finally, the matrices in the scattering equation Eq.~\eqref{eq:scattering_relation} simplify to
    \begin{equation}\label{eq:scattering_thermal}
        \smat[\omega] = \sqrt{\boldsymbol{\upgamma}} \left(\bm{A} + i \omega \idmat_{2N}\right)^{-1} \!\!\!\! \sqrt{\boldsymbol{\upgamma}} + \idmat_{2N},
        \qquad
        \mathbf{T}[\omega] = \sqrt{\boldsymbol{\upgamma}} \left(\bm{A} + i \omega \idmat_{2N}\right)^{-1} \! \symform_{2N} \HOne_{\mathrm{s}},
    \end{equation}
Provided that each input field is in a thermal state, then the input covariance matrix is $\bar{\bm{S}}_{\inn} = \bigoplus_{\K \in N} (\nth{\K}+1/2) \idmat_2$. As a result, the dissipation matrix in the corresponding Lindbladian takes the form $\dissmat = \sqrt{\boldsymbol{\upgamma}} (\bar{\bm{S}}_{\inn} + i \symform_{2N}/2) \sqrt{\boldsymbol{\upgamma}} = \bigoplus_{\K \in N} \lw{\K} ((\nth{\K}+1/2) \idmat_2 + i \symform_2/2)$.

\section{Reproducing expressions for dephasing in simple systems}
\label{apdx:theory_application}
\subsection{A single mode in a coherent state}
\label{apdx:single_mode_coherent}
Here we calculate the qubit dephasing rate in the case of a single mode (labeled the C mode) in a coherent state with qubit state-dependent amplitudes. Using the formalism developed in Sec.~\ref{sec:theory}, we show that the resulting dephasing rate is equivalent to Eq.~(5.20) in Ref.~\cite{Gambetta2006}. We begin with the following Lindbladian, written in the rotating frame:
    \begin{align}
        \lbld(\hat{\rho})
        = -i \comm{\hat{H}_{\R} + \hat{\sigma}_z \hat{H}_{\q\R}}{\hat{\rho}} 
        &+ \lw{\C} \diss[\hat{c}] (\hat{\rho})\nonumber\\
        = -i \comm{\hat{H}_{\R} + \hat{\sigma}_z \hat{H}_{\q\R}}{\hat{\rho}} 
        &+ \frac{\lw{\C}}{2} \left(\hat{q}_{\C} \hat{\rho} \hat{q}_{\C} - \frac{1}{2} \acomm{\hat{q}_{\C}^2}{\hat{\rho}} 
        + \hat{p}_{\C} \hat{\rho} \hat{p}_{\C} - \frac{1}{2} \acomm{\hat{p}_{\C}^2}{\hat{\rho}} \right)\nonumber\\
        &+ \frac{i\lw{\C}}{2} \left(\hat{p}_{\C} \hat{\rho} \hat{q}_{\C} - \frac{1}{2} \acomm{\hat{q}_{\C} \hat{p}_{\C}}{\hat{\rho}} 
        - \hat{q}_{\C} \hat{\rho} \hat{p}_{\C} + \frac{1}{2} \acomm{\hat{p}_{\C} \hat{q}_{\C}}{\hat{\rho}}\right),
    \end{align}
where the Hamiltonian terms consist of a drive with complex strength $\varepsilon$ and detuning $\Delta$, as well as the dispersive interaction between the cavity mode and qubit:
    \begin{align}
        \hat{H}_{\R} &=  -\Delta\hrm{c}\hat{c} + \varepsilon \hrm{c} + \varepsilon^*\hat{c} =  -\frac{\Delta}{2}\left(\hat{q}_{\C}^2+\hat{p}_{\C}^2-1\right) +  \sqrt{2}\left(\real[\varepsilon]\hat{q}_{\C} + \imag[\varepsilon]\hat{p}_{\C}\right), \label{eq:H_C} \\
        \hat{H}_{\q\R} &= \disp{\C} \hrm{c} \hat{c} = \frac{\disp{\C}}{2}\left(\hat{q}_{\C}^2+\hat{p}_{\C}^2-1\right). \label{eq:H_QC}
    \end{align}
Using Eqs.~\eqref{eq:cavity_lindblad} and \eqref{eq:qubit_cavity_lindblad}, we construct the following coefficient matrices in the basis $\hat{\qvec}=(\hat{q}_{\C},\hat{p}_{\C})$:
    \begin{equation}\label{eq:1mode_coefficients}
        \HTwo_{\R} = -\Delta\idmat, \qquad
        \HOne_{\R} = \sqrt{2}\begin{pmatrix} \real[\varepsilon] & \imag[\varepsilon]\end{pmatrix}^T, \qquad
        \HTwo_{\q\R} = \disp{\C} \idmat, \qquad \HNull_{\q\R} = - \frac{\disp{\C}}{2}, \qquad
        \dissmat = \frac{\lw{\C}}{2}\left(\idmat + i \jmat\right),
    \end{equation}
where the $2 \times 2$ matrices $\idmat$ and $\jmat$ are as follows:
    \begin{equation}\label{eq:i_j_mat_definitions}
        \idmat \defeq \begin{pmatrix} 1 & 0 \\ 0 & 1 \end{pmatrix}, \qquad \jmat \defeq \begin{pmatrix} 0 & 1 \\ -1 & 0 \end{pmatrix}.
    \end{equation}
In addition, $\symform = \jmat$ for this single mode. The matrices necessary to solve for the moments of the system may then be constructed using Eq.~\eqref{eq:array_definitions}:
    \begin{equation}
        \bm{A} = -\frac{\lw{\C}}{2}\idmat - \Delta \jmat, \qquad
        \bm{B} = 2i\disp{\C}\idmat, \qquad
        \bm{C} = \frac{1}{2}\left(\lw{\C}+i\disp{\C}\right)\idmat, \qquad
        \bm{d} = \sqrt{2}\begin{pmatrix} \imag[\varepsilon] & -\real[\varepsilon]\end{pmatrix}^T.
    \end{equation}
Recalling that $\covar = \covar^T$, these matrices generate the following differential equations for the first and second moments:
    \begin{align}
        \frac{d}{dt} \melem_{q_{\C}} &= -\left(2 i \chi \celem_{q_{\C}q_{\C}} + \frac{\lw{\C}}{2} \right) \melem_{q_{\C}} -\left(2 i \chi \celem_{q_{\C}p_{\C}} + \Delta \right)\melem_{p_{\C}} +\sqrt{2} \imag[\varepsilon], \nonumber \\
        \frac{d}{dt} \melem_{p_{\C}} &= -\left(2 i \chi \celem_{p_{\C}p_{\C}} + \frac{\lw{\C}}{2} \right) \melem_{q_{\C}} -\left(2 i \chi \celem_{q_{\C}p_{\C}} - \Delta \right)\melem_{q_{\C}} -\sqrt{2} \real[\varepsilon], \nonumber \\
        \frac{d}{dt} \celem_{q_{\C}q_{\C}} &= -2 \Delta \celem_{q_{\C}p_{\C}} -\lw{\C} \celem_{q_{\C}q_{\C}} - 2 i \disp{\C} (\celem_{q_{\C}q_{\C}}^2 + \celem_{q_{\C}p_{\C}}^2) + \frac{1}{2}(\lw{\C} + i \disp{\C}), \nonumber \\
        \frac{d}{dt} \celem_{p_{\C}p_{\C}} &= 2 \Delta \celem_{q_{\C}p_{\C}} -\lw{\C} \celem_{p_{\C}p_{\C}} - 2 i \disp{\C} (\celem_{p_{\C}p_{\C}}^2 + \celem_{q_{\C}p_{\C}}^2) + \frac{1}{2}(\lw{\C} + i \disp{\C}), \nonumber \\
        \frac{d}{dt} \celem_{q_{\C}p_{\C}} &= -\lw{\C} \celem_{q_{\C}p_{\C}} - 2 i \disp{\C} \celem_{q_{\C}p_{\C}} (\celem_{q_{\C}q_{\C}} + \celem_{p_{\C}p_{\C}}).
    \end{align}
We first solve for the covariance in the steady state, yielding multiple solutions for each matrix element. The only valid solutions satisfy $\real[\celem_{q_{\C}q_{\C}}], \real[\celem_{p_{\C}p_{\C}}] > 0$ as this retains the integrability of $\wig(\qvec)$. This results in the expected covariance matrix for a coherent state, $\covar = \idmat/2$, and accordingly, from Eq.~\eqref{eq:pdrate}, no parasitic dephasing ($\pdrate=0$). Plugging the covariance matrix into the equations for the steady state equations for the first moments, we find:
    \begin{equation}
        \mu_{q_{\C}} = \sqrt{2} \frac{\left(i \disp{\C} + \lw{\C}/2\right)\imag[\varepsilon] + \Delta \real[\varepsilon]}{\left(i\disp{\C}+\lw{\C}/2\right)^2 + \Delta^2},
        \qquad
        \mu_{p_{\C}} = -\sqrt{2} \frac{\left(i \disp{\C} + \lw{\C}/2\right)\real[\varepsilon] + \Delta \imag[\varepsilon]}{\left(i\disp{\C}+\lw{\C}/2\right)^2 + \Delta^2}.
    \end{equation}
Computing the measurement-induced dephasing with Eq.~\eqref{eq:mdrate} then produces:
    \begin{equation}\label{eq:coherent_state_mdrate}
        \mdrate = \frac{2|\varepsilon|^2\disp{\C}^2\lw{\C}}{ \left[\lw{\C}^2/4 + ( \Delta - \disp{\C} )^2 \right] \left[ \lw{\C}^2/4 + ( \Delta + \disp{\C} )^2 \right]} = \frac{ \left( n_{\C,e} + n_{\C,g} \right) \disp{\C}^2 \lw{\C} }{ \lw{\C}^2/4 +   \disp{\C}^2 + \Delta^2 },
        \quad \text{where} \quad
        n_{\C,e/g} = \frac{ |\varepsilon|^2 }{ \lw{\C}^2/4 + \left( \Delta \pm \disp{\C} \right)^2}
    \end{equation}
is the qubit state-dependent intracavity occupancy of mode C. This is in agreement with Eqs.~(5.20) and (5.21) of Ref.~\cite{Gambetta2006}.

\subsection{A single mode in a thermal state}
\label{apdx:single_mode_thermal}
Here we analyze the case of a single mode (labeled the C mode) occupied by a thermal state characterized by the mean photon number $\nth{\C}$. In doing so, we reproduce Eq.~(44) in Ref.~\cite{Clerk2007}. We begin with the following Lindbladian, written in the rotating frame:
    \begin{align}\label{eq:L_jump_C_thermal}
        \lbld(\hat{\rho})
        = -i \comm{\hat{\sigma}_z \hat{H}_{\q\R}}{\hat{\rho}} &+ \lw{\C} (\nth{\C}+1) \diss[\hat{c}] (\hat{\rho}) + \lw{\C} \nth{\C} \diss[\hrm{c}] (\hat{\rho}) \! \nonumber\\
        = -i \comm{\hat{\sigma}_z \hat{H}_{\q\R}}{\hat{\rho}} &+ \lw{\C} \left(\nth{\C}+\frac{1}{2}\right) \left(\hat{q}_{\C} \hat{\rho} \hat{q}_{\C} - \frac{1}{2} \acomm{\hat{q}_{\C}^2}{\hat{\rho}} 
        + \hat{p}_{\C} \hat{\rho} \hat{p}_{\C} - \frac{1}{2} \acomm{\hat{p}_{\C}^2}{\hat{\rho}} \right)\nonumber\\
        &+ \frac{i\lw{\C}}{2} \left(\hat{p}_{\C} \hat{\rho} \hat{q}_{\C} - \frac{1}{2} \acomm{\hat{q}_{\C} \hat{p}_{\C}}{\hat{\rho}} 
        -\hat{q}_{\C} \hat{\rho} \hat{p}_{\C} + \frac{1}{2} \acomm{\hat{p}_{\C} \hat{q}_{\C}}{\hat{\rho}}\right),
    \end{align}
where the $\hat{H}_{\q\R}$ is given in Eq.~\eqref{eq:H_QC}. Using Eqs.~\eqref{eq:cavity_lindblad} and \eqref{eq:qubit_cavity_lindblad}, we can construct the following coefficient matrices in the basis $\hat{\qvec}=(\hat{q}_{\C},\hat{p}_{\C})$:
    \begin{equation}
        \HTwo_{\q\R} = \disp{\C} \idmat, \qquad \HNull_{\q\R} = - \frac{\disp{\C}}{2}, \qquad
        \dissmat = \lw{\C}\left(\nth{\C}+\frac{1}{2}\right) \idmat + i \frac{\lw{\C}}{2} \jmat,
    \end{equation}
where the $2 \times 2$ matrices $\idmat$ and $\jmat$ are defined in Eqs.~\eqref{eq:i_j_mat_definitions}, and $\symform = \jmat$. We then use Eq.~\eqref{eq:array_definitions} to construct matrices necessary to solve for the second moment of the system:
    \begin{equation}
        \bm{A} = -\frac{\lw{\C}}{2}\idmat, \qquad
        \bm{B} = 2i\disp{\C}\idmat, \qquad
        \bm{C} = \left[\lw{\C}\left(\nth{\C}+\frac{1}{2}\right)+i \frac{\disp{\C}}{2}\right]\idmat.
    \end{equation}
These matrices generate a set of differential equations for the covariance matrix entries. As in the previous subsection, the only physically valid solution results in off-diagonal components which are zero. The diagonal entries therefore have identical dynamical equations of the form
    \begin{equation}\label{eq:cov_diag_eom}
        \frac{d}{dt} \celem_{q_{\C} q_{\C}} = -2i\disp{\C}\celem_{q_{\C} q_{\C}}^2-\lw{\C}\celem_{q_{\C} q_{\C}}+\frac{i\disp{\C}}{2}+\lw{\C}\left(\nth{\C}+\frac{1}{2}\right),
    \end{equation}
which can be solved in the steady state and utilized with Eq.~\eqref{eq:pdrate} to yield
    \begin{equation}\label{eq:nth_c_full}
        \pdrate = \frac{\lw{\C}}{2}\real\left[\sqrt{\left(1+\frac{2i\disp{\C}}{\lw{\C}}\right)^2+\frac{8i\disp{\C}\nth{\C}}{\lw{\C}}}-1\right],
    \end{equation}
in agreement with Eq.~(44) of Ref.~\cite{Clerk2007}. Expanding $\pdrate$ to first order around $\nth{\C}=0$ produces the expression in Eq.~\eqref{eq:nth_C}.

\section{The embedded amplifier}
\label{apdx:embedded_amplifier}
\subsection{Analytic solution for the dephasing}
The coefficient arrays for the embedded amplifier described in Sec.~\ref{sec:embedded_amp_theory} are obtained using the Hamiltonian in Eq.~\eqref{eq:full_hamiltonian}. The arrays $\bm{A}$ and $\bm{d}$ take the forms
    \begin{align}
        \bm{A} &=
        \begin{pmatrix} 
            -(\lw{\A}/2) \idmat + \lambda[-\cos(\theta) \xmat + \sin(\theta) \zmat] & g_{\A\B} \jmat & g_{\A\C} [\cos(\phi) \jmat + \sin(\phi) \idmat] \\
            g_{\A\B} \jmat & -(\lw{\B}/2) \idmat & g_{\B\C} \jmat \\
            g_{\A\C} [\cos(\phi) \jmat - \sin(\phi) \idmat] & g_{\B\C} \jmat & -(\lw{\C}/2) \idmat
        \end{pmatrix},
        \nonumber \\
        \bm{d} &= \sqrt{2} \varepsilon \begin{pmatrix} 0 & 0 & \sin(\varphi) & -\cos(\varphi) & 0 & 0 \end{pmatrix}^T,
    \end{align}
where the $2 \times 2$ matrices $\idmat$ and $\jmat$ are defined in Eq.~\eqref{eq:i_j_mat_definitions}, and $\xmat$ and $\zmat$ are defined as
    \begin{equation}
        \xmat \defeq \begin{pmatrix} 0 & 1 \\ 1 & 0 \end{pmatrix}, \qquad 
        \zmat \defeq \begin{pmatrix} 1 & 0 \\ 0 & -1 \end{pmatrix}.
    \end{equation}
Thermal noise from the environment is included for each mode. Additionally, to simplify our analysis we assume only mode C to have a non-negligible coupling to the qubit. The other arrays required to solve the moments, $\bm{B}$ and $\bm{C}$, are then:
    \begin{equation}
        \bm{B} =
            \begin{pmatrix} 
            \nullmat & \nullmat & \nullmat \\ 
            \nullmat & \nullmat & \nullmat \\ 
            \nullmat & \nullmat & 2 i \disp{\C} \idmat 
            \end{pmatrix},
        \qquad
        \bm{C} = \frac{1}{2} 
            \begin{pmatrix} 
            \lw{\A} (2 \nth{\A} + 1) \idmat & \nullmat & \nullmat \\ 
            \nullmat & \lw{\B} (2 \nth{\B} + 1) \idmat & \nullmat \\ 
            \nullmat & \nullmat & [i \disp{\C} + \lw{\C} (2 \nth{\C} + 1)] \idmat 
            \end{pmatrix}.
    \end{equation}
The equation which characterizes the steady-state of the covariance matrix is known as a continuous algebraic Riccati equation (CARE) \cite{Bittanti1991,Lancaster1995}, which may be solved by linearizing the differential equation for $\covar$: 
    \begin{equation}
        \frac{d}{dt}\covar = \bm{A} \covar + \covar \bm{A}^T - \covar \bm{B} \covar + \bm{C}
        \,\, \rightarrow \,\,
        \frac{d}{dt} \begin{pmatrix} \bm{U} \\ \bm{V} \end{pmatrix} 
        = \mathbf{H} \begin{pmatrix} \bm{U} \\ \bm{V} \end{pmatrix}
        \text{ where } \,
        \mathbf{H} = \begin{pmatrix} \bm{A}^T & -\bm{B} \\ -\bm{C} & -\bm{A} \end{pmatrix}.
    \end{equation}
$\bm{U}$ and $\bm{V}$ are both $6 \times 6$ matrices, and provided that $\bm{U}$ is invertible, then the solution for the covariance matrix is $\covar = \bm{V} \bm{U}^{-1}$. In order for $\bm{V} \bm{U}^{-1}$ to be the steady-state solution for the CARE, $(\bm{U} \,\, \bm{V})^T$ must span the space of stable solutions for $\mathbf{H}$. For simplicity, the eigenvectors corresponding to the stable eigenvalues of $\mathbf{H}$ were used as the columns of $(\bm{U} \,\, \bm{V})^T$. Tractable analytic solutions are only possible when the loop-phase is $\phi = \pm\pi/2$. In this case, the eigenvalues are independent of the phase of the amplifier $\theta$. This is evident from the characteristic polynomial of $\mathbf{H}$, which may be expressed as
    \begin{equation}
        0 = \Big(s^6 + v_2(+\lambda) s^4 + v_1(+\lambda) s^2 + v_0\boldsymbol{(}+\lambda,\sgn(\phi)\boldsymbol{)}\Big) \Big(s^6 + v_2(-\lambda) s^4 + v_1(-\lambda) s^2 + v_0\boldsymbol{(}-\lambda,\sgn(\phi)\boldsymbol{)}\Big).
    \end{equation}
The twelve eigenvalues of $\mathbf{H}$ have the form $\pm \sqrt{\polyroot_k(\pm \lambda)} \,,\, k=1,2,3$, where $\polyroot_k(\pm\lambda)$ are the roots of the cubic polynomial $0 = s^3 + v_2(\pm \lambda) s^2 + v_1(\pm \lambda) s + v_0\boldsymbol{(}\pm \lambda,\sgn(\phi)\boldsymbol{)}$. Six of these eigenvalues have a negative real part, and so their associated eigenvectors form the stable subspace of $\mathbf{H}$, and hence the columns of $(\bm{U} \,\, \bm{V})^T$. From here, $\covar = \bm{V} \bm{U}^{-1}$ may be solved, followed by $\mean$. Using Eq.~\eqref{eq:pdrate}, the parasitic dephasing is then:
    \begin{equation}\label{eq:embedded_amp_pdrate}
        \pdrate = - \frac{1}{2} (\lw{\A} + \lw{\B} + \lw{\C}) + \frac{1}{2} \! \sum_{k=1,2,3} \hspace{-1mm} \mathrm{Re} \! \left[\sqrt{\polyroot_k(+\lambda)} + \sqrt{\polyroot_k(-\lambda)}\right].
    \end{equation}
The measurement induced dephasing is also obtained using Eq.~\eqref{eq:mdrate}:
    \begin{align}\label{eq:embedded_amp_mdrate}
        \mdrate =  \varepsilon^2 \disp{\C}^2
        \Bigg(&\frac{u_1\boldsymbol{(}+\lambda,\sgn(\phi)\boldsymbol{)} \big[\sgn(\phi) g_{\A\B} g_{\A\C} - g_{\B\C} (\lw{\A}/2 + \lambda)\big]^2}{\abs{\polyroot_1(+\lambda) \, \polyroot_2(+\lambda) \, \polyroot_3(+\lambda)}^2} \big(1 - \sin(\theta - 2 \varphi)\big) \nonumber \\[-0.5em]
        &+ \frac{u_1\boldsymbol{(}-\lambda,\sgn(\phi)\boldsymbol{)} \big[\sgn(\phi) g_{\A\B} g_{\A\C} - g_{\B\C} (\lw{\A}/2 - \lambda)\big]^2}{\abs{\polyroot_1(-\lambda) \, \polyroot_2(-\lambda) \, \polyroot_3(-\lambda)}^2} \big(1 + \sin(\theta - 2 \varphi)\big)\Bigg)
    \end{align}
where $\abs{\polyroot_1(\lambda) \, \polyroot_2(\lambda) \,\polyroot_3(\lambda)}^2
        = \abs{v_0\boldsymbol{(}\lambda,\sgn(\phi)\boldsymbol{)}}^2 = \big[u_0(\lambda) - \disp{\C}^2 u_2(\lambda)\big]^2 + \big[\disp{\C} u_1\boldsymbol{(}\lambda,\sgn(\phi)\boldsymbol{)}\big]^2$, and
    \begin{align}
        & u_0(\lambda) = \left[g_{\A\B}^2 \lw{\C}/2 + g_{\B\C}^2 (\lw{\A}/2 + \lambda) + g_{\A\C}^2 \lw{\B}/2 + (\lw{\A}/2 + \lambda) (\lw{\B}/2) (\lw{\C}/2) \right]^2 \nonumber \\[-0.5em]
        & u_1\boldsymbol{(}\lambda,\sgn(\phi)\boldsymbol{)} = \Big(\lw{\A} (2 \nth{\A} + 1) \left[g_{\A\B} g_{\B\C} + \sgn(\phi) g_{\A\C} \lw{\B}/2\right]^2 \nonumber \\[-0.5em]
        & \hspace{27mm} + \lw{\B} (2 \nth{\B} + 1) \left[\sgn(\phi) g_{\A\B} g_{\A\C} - g_{\B\C} (\lw{\A}/2 + \lambda)\right]^2 \nonumber \\[-0.5em]
        & \hspace{27mm} + \lw{\C} (2 \nth{\C} + 1) \left[g_{\A\B}^2 + (\lw{\A}/2 + \lambda) (\lw{\B}/2) \right]^2 \Big) \nonumber \\[-0.5em]
        & u_2(\lambda) = \left[g_{\A\B}^2 + (\lw{\A}/2 + \lambda) (\lw{\B}/2) \right]^2.
    \end{align}
    
\subsection{Amplifier stability and gain}
\label{apdx:embedded_amplifier_stability}
In order for the embedded amplifier to be stable, the eigenvalues of the dynamical matrix for the pointer states must have a negative real part:
    \begin{equation}\label{eq:embedded_amp_dynmat}
    \bm{A} =
        \begin{pmatrix} 
            -(\lw{\A}/2) \idmat + \lambda[-\cos(\theta) \xmat + \sin(\theta) \zmat] & g_{\A\B} \jmat & g_{\A\C} [\cos(\phi) \jmat + \sin(\phi) \idmat] \\
            g_{\A\B} \jmat & -(\lw{\B}/2) \idmat & g_{\B\C} \jmat \\
            g_{\A\C} [\cos(\phi) \jmat - \sin(\phi) \idmat] & g_{\B\C} \jmat & -(\lw{\C}/2) \idmat + s \disp{\C} \jmat 
        \end{pmatrix},
    \end{equation}
where $s = \braket{\sigma| \hat{\sigma}_z|\sigma}$ for $\sigma = e,g$. The inclusion of qubit detuning results in complicated inequalities when using the Routh-Hurwitz criterion for stability. As a result, we instead focus on the critical criteria for stability \cite{Fuller1968}, where one or both of $\det[\bm{A}]$ and the 5\textsuperscript{th} Hurwitz determinant go exactly to zero on the border between the regions of stability and instability, and are otherwise non-zero. Applying the nonreciprocity condition from Eq.~\eqref{eq:nonreciprocity_conditions}, we find that
$\det[\bm{A}] \propto (1 + \coop{\A\B} - \coop{\lambda}) (1 + \coop{\A\B} + \coop{\lambda}) [(1 +\coop{\B\C})^2 + \mathcal{X}^2] (\lw{\A} \lw{\B} \lw{\C})^2$, which goes to zero exactly when $\coop{\lambda} = 1 + \coop{\A\B}$. Hence, Eq.~\eqref{eq:stability} is a sufficient but not necessary criterion to mark the transition from stability to instability for the embedded amplifier.

We can compute the scattering matrix in the quadrature basis using Eq.~\eqref{eq:scattering_thermal}, where the matrix $\bm{A}$ is defined in Eq.~\eqref{eq:embedded_amp_dynmat} and $\boldsymbol{\upgamma} = \diag(\lw{\A},\lw{\A},\lw{\B},\lw{\B},\lw{\C},\lw{\C})$. We focus solely on scattering on resonance where the nonreciprocity condition for C $\rightarrow$ A has been applied. The scattering matrix may then be written in block form as follows:
    \begin{equation}
        \smat[0] = 
        \begin{pmatrix}
        \smat_{\A\A} & \smat_{\A\B} & \smat_{\A\C} \\
        \smat_{\B\A} & \smat_{\B\B} & \smat_{\B\C} \\
        \nullmat & \smat_{\C\B} & \smat_{\C\C}
        \end{pmatrix}
        = - \smat_{\mathrm{sqz}} \smat_{\mathrm{bs}}.
    \end{equation}
To simplify the expression for the total scattering of the embedded amplifier, $\smat[0]$ may be decomposed into a product of two scattering matrices: $\smat_{\mathrm{sqz}}$, which represents the scattering for modes A and B coupled via a beam splitter with amplification at mode A (where coupling to mode C and the qubit is omitted, $g_{\A\C},g_{\B\C} = 0$ and $\disp{\C} = 0$), and $\smat_{\mathrm{bs}}$, which corresponds to the scattering between the beam-splitter coupled modes B and C with the qubit detuning present (here, coupling to mode A and the amplification are omitted, $g_{\A\C},g_{\A\B} = 0$ and $\lambda = 0$). These scattering matrices may be written as:
    \begin{align}
        \smat_{\mathrm{sqz}} &= -\frac{1}{\Lambda}
            \begin{pmatrix}
                \substack{\textstyle{(1-\coop{\A\B}^2+\coop{\lambda}^2) \idmat} \\[0.1em] \textstyle{+ \, 2 \coop{\lambda} [-\cos(\theta) \xmat + \sin(\theta) \zmat]}}
                & \substack{\textstyle{2 \sqrt{\coop{\A\B}} (1+\coop{\A\B}) \jmat} \\[0.1em] \textstyle{+ \, 2 \coop{\lambda} \sqrt{\coop{\A\B}} [\cos(\theta) \zmat + \sin(\theta) \xmat]}}
                & \nullmat \\[1.2em]
                \substack{\textstyle{2 \sqrt{\coop{\A\B}} (1+\coop{\A\B}) \jmat} \\[0.1em] \textstyle{- \, 2 \coop{\lambda} \sqrt{\coop{\A\B}} [\cos(\theta) \zmat + \sin(\theta) \xmat]}} 
                & \substack{\textstyle{(1-\coop{\A\B}^2-\coop{\lambda}^2) \idmat} \\[0.1em] \textstyle{+ \, 2 \coop{\lambda} \coop{\A\B} [-\cos(\theta) \xmat + \sin(\theta) \zmat]}}
                & \nullmat \\[1.2em]
                \nullmat & \nullmat & \Lambda \idmat
            \end{pmatrix} \,\,,\,\,
        \Lambda = (1+\coop{\A\B})^2 - \coop{\lambda}^2,  
    \nonumber \\
        \smat_{\mathrm{bs}} &= -\frac{1}{\mathrm{X}}
            \begin{pmatrix}
                \mathrm{X} \idmat & \nullmat & \nullmat \\
                \nullmat
                & (1-\coop{\B\C}^2 + \mathcal{X}^2) \idmat - 2 s \mathcal{X} \coop{\B\C} \jmat
                & - 2 \textstyle{\sqrt{\coop{\B\C}}} [s \mathcal{X} \idmat - (1+\coop{\B\C}) \jmat] \\[0.2em]
                \nullmat 
                & - 2 \textstyle{\sqrt{\coop{\B\C}}} [s \mathcal{X} \idmat - (1+\coop{\B\C}) \jmat]
                & (1-\coop{\B\C}^2 - \mathcal{X}^2) \idmat + 2 s \mathcal{X} \jmat
            \end{pmatrix} \,\,,\,\,
        \mathrm{X} = (1+\coop{\B\C})^2 + \mathcal{X}^2,    
    \end{align}
where we define $\mathcal{X} \defeq 2 \disp{\C}/\lw{\C}$. Applying the basis transformation $\mathbf{P} \smat[0] \mathbf{P}^\dagger$ where $\mathbf{P} = \idmat_3 \otimes \frac{1}{\sqrt{2}} \left(\begin{smallmatrix} 1 & i \\ 1 & -i \end{smallmatrix} \right)$ takes the scattering matrix to the creation and annihilation operator basis, $(\hat{a},\hrm{a},\hat{b},\hrm{b},\hat{c},\hrm{c})$. We can now extract the power gain in reflection off of mode B (in decibels) on resonance, defined as $\mathcal{G}_{\B\B} (\mathrm{dB}) \defeq 10 \log_{10} \abs{\smelem_{\B\B}[0]}^2$, where the scattering matrix element is $\hat{b}_{\out} = \smelem_{\B\B}[0] \hat{b}_{\inn}$. Given the same parameters, the gain (expressed as the original power ratio) is:
    \begin{equation}\label{eq:gain}
        \mathcal{G}_{\B\B} = \frac{\left[(1 - \coop{\A\B})(1 + \coop{\A\B}) - \coop{\lambda}^2\right]^2 \left[(1 - \coop{\B\C})^2 + \mathcal{X}^2\right]}{\left(1 + \coop{\A\B} - \coop{\lambda}\right)^2 \left(1 + \coop{\A\B} + \coop{\lambda}\right)^2 \left[(1 + \coop{\B\C})^2 + \mathcal{X}^2\right]},
    \end{equation}
The above expression goes to infinity as the critical stability condition in Eq.~\eqref{eq:stability} is saturated, yielding the maximum measurement efficiency in Eq.~\eqref{eq:embedded_amp_eff_B}. However, in certain parameter regimes, instability occurs for values of $\coop{\lambda}$ well below the critical condition in Eq.~\eqref{eq:stability} as a result of the 5\textsuperscript{th} Hurwitz determinant going to zero before $\det[\bm{A}]$, such as in Fig.~\ref{fig:dephasing_coop}. In this case, infinite gain cannot be realized.

\subsection{Optimum measurement efficiency and its limits}
\label{apdx:embedded_amplifier_efficiency_limits}
\begin{figure}[htbp]
    \includegraphics{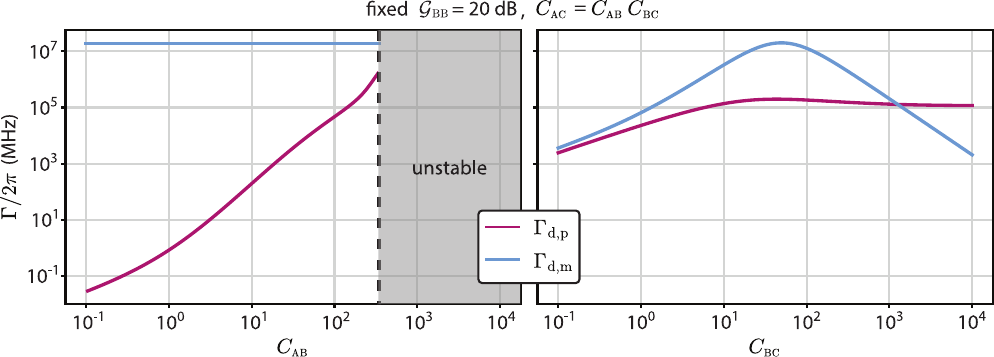}
    \caption{\textbf{Qubit dephasing rates as a function of the unconstrained cooperativities $\coop{\A\B}$ and $\coop{\B\C}$}. The remaining cooperativities $\coop{\A\C}$ and $\coop{\lambda}$ are constrained in each simulation to optimize nonreciprocal scattering between A and C (see Eq.~\eqref{eq:nonreciprocity_conditions}) and retain $\mathcal{G}_{\B\B} = 20$ dB, respectively. All other parameters are detailed in Sec.~\ref{sec:embedded_amp_theory}. For $\coop{\A\B} \gtrsim 340$, the amplifier becomes unstable at the values of $\coop{\lambda}$ required to retain $\mathcal{G}_{\B\B} = 20$ dB.}
    \label{fig:dephasing_coop}
\end{figure}

From the main text, we are interested in the efficiency when $\nonrec{\C}{\A} = 1$, requiring $\coop{\A\C} = \coop{\A\B} \coop{\B\C}$ and $\phi = -\pi/2$. Further assuming that $\nth{\B},\nth{\C}=0$, the measurement induced dephasing in Eq.~\eqref{eq:embedded_amp_mdrate} reduces to
    \begin{equation}\label{eq:embd_amp_mdrate}
        \mdrate = \frac{8 (\varepsilon^2/\lw{\B}) \mathcal{X}^2 \coop{\B\C}(1+\coop{\B\C})}{\left[(1+\coop{\B\C})^2+\mathcal{X}^2\right]^2}.
    \end{equation}
For the measurement rate we can determine that the optimal amplifier phase is $\theta = -\pi/2 + 2 \varphi$. For these parameters the ideal measurement operator for a measurement of the output of mode B is $\hat{\meas}[0] = \cos(\varphi) \hat{q}_{\B,\out}[0] + \sin(\varphi) \hat{p}_{\B,\out}[0]$. The measurement rate is then:
    \begin{equation}\label{eq:embd_amp_mrate}
        \mrate = \frac{8 (\varepsilon^2/\lw{\B}) \mathcal{X}^2 \coop{\B\C}^2}{\left[(1+\coop{\B\C})^2+\mathcal{X}^2\right]^2} \times
        \frac{(-1 + \coop{\A\B} + \coop{\lambda})^2}{4 \coop{\A\B} (\coop{\lambda}+2\nth{\A}) + (2 \bar{n}_\add+1) (1+\coop{\A\B}-\coop{\lambda})^2}.
    \end{equation}   
As discussed in Sec.~\ref{subsec:embedded_amp_efficiency}, the approximate expression for the measurement efficiency given by $\eta \approx \mrate / \mdrate$ in Eq.~\eqref{eq:embedded_amp_eff_A} is not valid for sufficiently large beam-splitter cooperativities. This is because one of the conditions for high efficiency from Sec.~\ref{subsec:measrate_efficiency}, $\pdrate \ll \mdrate$, does not hold. Here we provide evidence to support this argument. In Fig.~\ref{fig:dephasing_coop}, the parasitic and measurement-induced dephasing rates (for a fixed drive strength of $\varepsilon/2\pi = 20$ MHz) are plotted as functions of $\coop{\A\B}$ and $\coop{\B\C}$.

As the cooperativity of the AB beam-splitter interaction $\coop{\A\B}$ increases, $\coop{\A\C}$ must be increased correspondingly to retain the optimal nonreciprocal scattering condition in Eq.~\eqref{eq:nonreciprocity_conditions}. In addition, because $\mathcal{G}_{\B\B}$ monotonically decreases as $\coop{\A\B}$ increases, $\coop{\lambda}$ must also be increased to keep $\mathcal{G}_{\B\B}$ constant (at 20dB in Fig.~\ref{fig:dephasing_coop}), until it is no longer possible to retain the gain in a stable amplifier configuration (which occurs at $\coop{\A\B} \gtrsim 340$ for this system). This constraint results in a parasitic dephasing that grows monotonically with $\coop{\A\B}$, while the measurement induced dephasing remains constant, as shown in Eq.~\eqref{eq:embd_amp_mdrate}. Eventually, $\pdrate$ becomes a significant part of the total induced dephasing, reducing the efficiency.

Considering now the BC beam-splitter interaction, the limit on $\eta$ in this case is set by a comparison of the dissipation rate and dispsersive shift of the dressed C mode, given by $\lw{\C,\mathrm{eff}} \defeq \lw{\C}(1+\coop{\B\C})$ and $\disp{\C,\mathrm{eff}} \approx \disp{\C}$, respectively. For qubit readout with a single mode, $\mdrate$ and $\mrate$ are maximized when the dispersive shift matches the mode's dissipation rate, $\disp{\C} = \lw{\C}/2$. For the embedded amplifier, the equivalent condition is $\disp{\C} = \lw{\C,\mathrm{eff}}/2$, requiring $\coop{\B\C} = 2 \disp{\C} / \lw{\C} - 1 \approx 49$. For constant values of $\disp{\C}$, larger values of $\coop{\B\C}$ will therefore move the embedded amplifier measurement into the small-pull limit, $\disp{\C} \ll \lw{\C,\mathrm{eff}}/2$.
In this regime, the photon lifetime in mode C is short compared to the time required to attain substantial qubit information, eventually reducing $\mdrate$, and hence $\mrate$, to zero. While the parasitic dephasing peaks near the same cooperativity, it asymptotes to a nonzero value for large $\coop{\B\C}$. This is because $\pdrate$ is mainly dependent on $\mathcal{G}_{\B\B}$, a quantity which becomes effectively independent of $\coop{\B\C}$ when the beam-splitter rate is large. Values of $\coop{\B\C} \gg 2\disp{\C} / \lw{\C} - 1$ therefore result in a larger contribution of $\pdrate$ to the total dephasing.

\section{Experimental device calibrations}
\label{apdx:expt_calib}
\subsection{Preliminary characterization vs. SQUID bias}
\label{sec:prelim_char}
\begin{figure}[htbp]
    \includegraphics[scale=0.99]{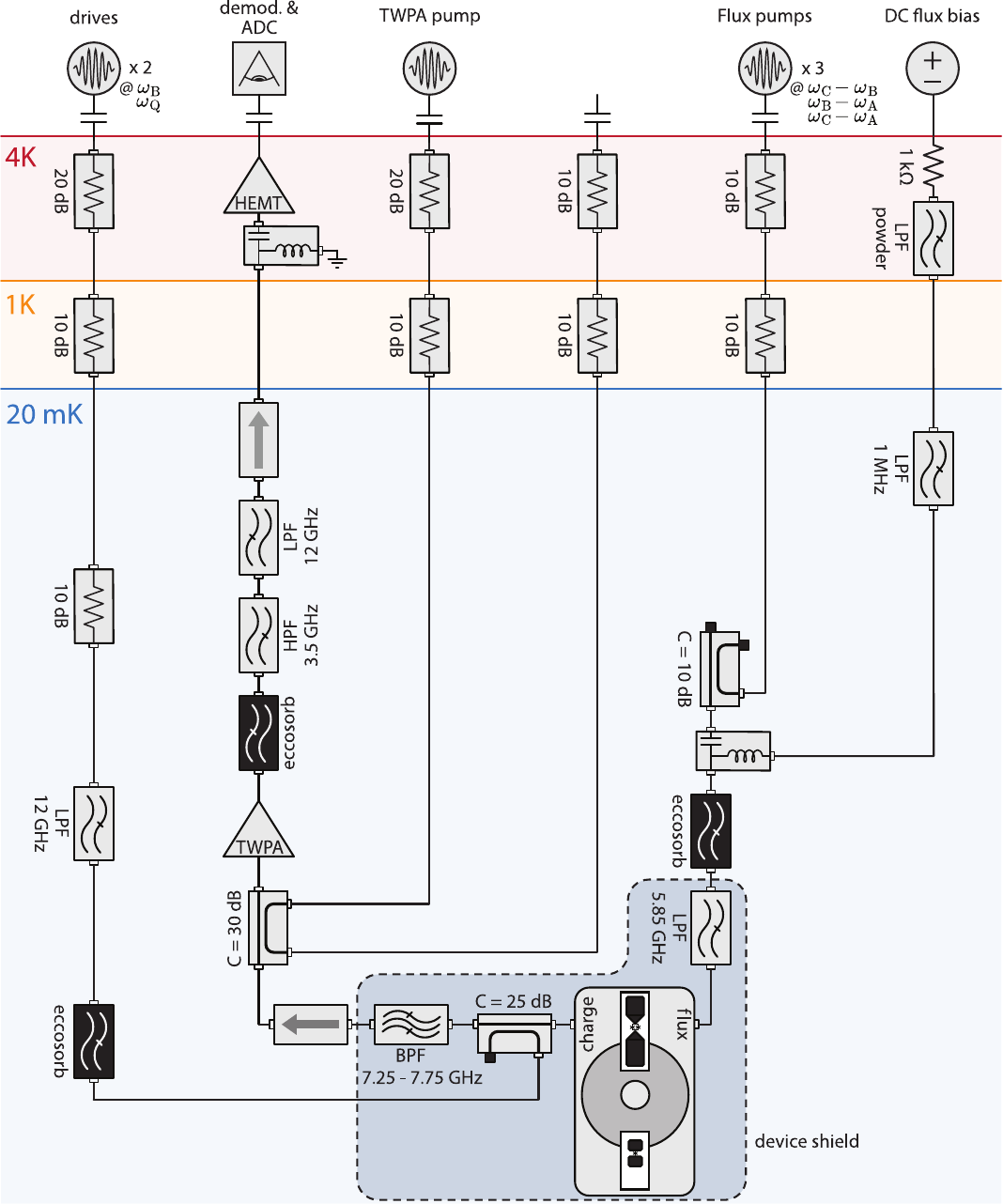}
    \caption{\textbf{Cryogenic wiring diagram} utilized to deliver to the device drives at the buffer and qubit frequencies, as well as the DC flux bias and three parametric pumps at the AB, BC, and AC beam-splitter frequencies. Outgoing signals near $\freq{\B}$ are captured with low added noise through the use of a TWPA.}
    \label{fig:fridge_diagram}
\end{figure}

Here we discuss the preliminary measurements of the linear mode frequencies and linewidths as a function of flux bias. We note that for these preliminary measurements, we replace the band-pass Purcell filter and directional coupler of the experimental setup depicted in Fig.~\ref{fig:fridge_diagram} with a single-stage circulator. We then measure the scattering parameter in reflection off of mode B to extract the B mode frequency and linewidth. Because the A and C mode frequencies lie outside the band of the circulators in the measurement setup, we extract their frequencies and linewidths by including an AB or BC beam-splitter interaction and fitting the scattering off of B using coupled-mode theory \cite{ranzani_graph-based_2015, Lecocq2017}, which is equivalent to the scattering theory presented in Appendix~\ref{apdx:input-output_scattering} for these device configurations. These produce the measured linear mode frequencies and linewidths in Fig.~\ref{fig:linewidths_v_bias}, which we fit with EM simulations presented in Appendix~\ref{apdx:design_sim}. We note that both the A and C modes being outside the bandwidth of the measurement setup may account in part for the discrepancy in their measured and simulated linewidths. Changing to the standard measurement configuration shown in Fig.~\ref{fig:fridge_diagram}, we extract the transmon's $g\leftrightarrow e$ transition frequency with a Ramsey sequence, and fit it with the same EM simulations, both of which are shown in Fig.~\ref{fig:linewidths_v_bias}(a). Repeating these measurements as a function of dc SQUID flux bias informs the chosen operation point of $\Phi = 0.246 \Phi_0$. In the next section, we describe the method we employed to extract device parameters with a higher degree of accuracy at this chosen flux bias. 

\begin{figure*}[ht]
    \includegraphics[scale=1]{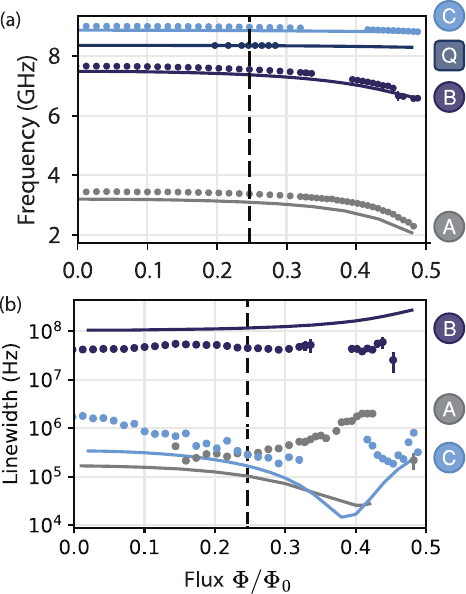}
    \caption{\textbf{Frequencies and linewidths vs. SQUID bias.} Measured and EM-simulated values for \textbf{(a)} the frequency of modes A, B, C and the qubit, and \textbf{(b)} the linewidth of modes A, B, and C, as a function of the flux bias through the dc SQUID. The dashed vertical line denotes the operation point of $\Phi = 0.246 \Phi_0$.}
    \label{fig:linewidths_v_bias}
\end{figure*} 

\subsection{Catch-and-release protocol}
\label{apdx:c_r_eom}
As stated in the previous section, many static (linewidths, bare frequencies, dispersive shifts) and dynamic (beam-splitter coupling strengths, etc.) characteristics of a readout network can be calibrated with continuous-wave S or X parameter measurements paired with fits from a steady-state linear system model \cite{ranzani_graph-based_2015, Lecocq2017}. However our chosen filtering and signal routing scheme (see Appendix~\ref{apdx:experimental_diagram}) allows only for measurements in reflection off of mode B, and also results in a large degree of Fano interference in this scattering parameter.

To avoid this issue, we employ pulsed drives and parametric interactions in a``catch-and-release" protocol, akin to those that have been demonstrated in other platforms \cite{novikova_optimal_2007, palomaki_coherent_2013, sirois_coherent-state_2015, peronnin_sequential_2020}. These protocols not only remove continuous interference effects, but can also precisely calibrate device parameters. In this section, we analyze a simplified model of the protocol to demonstrate how it can extract the frequency and linewidth of mode $\mathrm{K} = \mathrm{A,C}$. 

Here we consider a system consisting of a drive of strength $\varepsilon$ and frequency $\freq{d}$ near mode B, and a beam-splitter pump with strength $g$ and frequency $\freq{p}$ near $|\freq{\K} - \freq{\B}|$ to couple mode K with mode B. Considering the evolution of the mean fields in the rotating frame $k \defeq \langle \hat{k} \rangle$, $b \defeq \langle \hat{b} \rangle$: 
\begin{align}\label{eq:two_mode_bs_eom}
    \frac{d}{dt} b &= \left( i \dtn{\B} - \frac{\lw{\B}}{2} \right) b - i g k + \varepsilon, \nonumber \\
    \frac{d}{dt} k &= \left( i \dtn{\K} - \frac{\lw{\K}}{2} \right) k - i g b,
\end{align}
where
\begin{equation}\label{eq:detuning_KB}
    \dtn{\B} \defeq \freq{d} - \freq{\B} + s \disp{\B}. \qquad
    \dtn{\K} \defeq \freq{d} + \sgn( \freq{\K} - \freq{\B}) \freq{p} - \freq{\K} + s \disp{\K}.
\end{equation}
In the above, $s = \braket{\sigma|\hat{\sigma}_z|\sigma}$ takes the values $s=1$ for $\sigma = e$, and $s=-1$ for $\sigma=g$. Adiabatically eliminating mode B, we arrive at a single equation for K 
\begin{equation}\label{eq:adiabatic_elim_BK}
    \frac{d}{dt} k = \left( i \dtn{\K,\mathrm{eff}} - \frac{\lw{\K,\mathrm{eff}}}{2} \right) k + \varepsilon_{\mathrm{eff}},
\end{equation}
where
\begin{equation}\label{eq:adiabaitc_elim_BK_defs}
    \dtn{\K,\mathrm{eff}} \defeq \dtn{\K} - \eta_{\K\B} \dtn{\B}, \qquad \lw{\K,\mathrm{eff}} \defeq \lw{\K} + \eta_{\K\B} \lw{\B}, \qquad \varepsilon_{\mathrm{eff}} \defeq \varepsilon \sqrt{\eta_{\K\B}}, \qquad \eta_{\K\B} \defeq \frac{ g^2 }{ \lw{\B}^2/4 + \dtn{\B}^2 }.
\end{equation}
The catch-and-release model is as follows. The beam-splitter interaction and B drive are off for $t<0$, then both are concurrently turned on as a square pulse from time $t=0$ to $t=t_0$. After these first pulses end there is a passive delay, followed by another square beam-splitter pulse beginning at $t=t_1$. The full procedure is visualized in Fig.~\ref{fig:catch_release}. The resulting solution of Eq.~\eqref{eq:adiabatic_elim_BK} yields a piecewise function:
\begin{equation}
    k(t)=\frac{\varepsilon_{\mathrm{eff}}}{i\dtn{\K,\mathrm{eff}}+\lw{\K,\mathrm{eff}}/2} \begin{cases} 
      1 - e^{ ( i \dtn{\K,\mathrm{eff}} - \lw{\K,\mathrm{eff}}/2 ) t } & 0<t<t_0 \\
      e^{ ( i \dtn{\K} - \lw{\K}/2 )(t-t_0)}\left( 1 - e^{ ( i \dtn{\K,\mathrm{eff}} - \lw{\K,\mathrm{eff}}/2 ) t_0 } \right) & t_0<t<t_1 \\
       e^{ ( i \dtn{\K,\mathrm{eff}} - \lw{\K,\mathrm{eff}}/2 )(t-t_1)}e^{ ( i \dtn{\K} - \lw{\K}/2 )(t_1-t_0)}\left( 1 - e^{ ( i \dtn{\K,\mathrm{eff}} - \lw{\K,\mathrm{eff}}/2 ) t_0 } \right) & t>t_1
    \end{cases}
\end{equation}
These equations describe an internal state $k(t)$ that approaches its Lorentzian steady-state value for $0<t<t_0$, decays at a rate given by the bare mode linewidth $\lw{\K}$ for $t_0<t<t_1$, and then decays at rate $\lw{\K,\mathrm{eff}}$ for $t>t_1$. During this last step, energy is preferentially released out of the B mode to the monitored feedline ($\lw{\B, \mathrm{ext}} \approx \lw{\B}$ and $\eta_{\K\B} \lw{\B} \gg \lw{\K}$). Using input-output theory (see Appendix~\ref{apdx:input-output_scattering}), ignoring the prefactors independent of the evolution time $t_{\mathrm{ev}} \defeq t_1-t_0$ and interrogation time $\tau\defeq t-t_1$, this output field is given by
\begin{equation}\label{eq:b_out}
    b_{\mathrm{out}}(t_{\mathrm{ev}}, \tau) = \sqrt{\eta_{\K\B} \lw{\B}} k(t) \propto e^{ ( i \dtn{\K,\mathrm{eff}} - \lw{\K,\mathrm{eff}}/2 )\tau}e^{ ( i \dtn{\K} - \lw{\K}/2 )t_{\mathrm{ev}}}.
\end{equation}
We linearly amplify this field, and conduct homodyne detection by first sending the signal to the RF port of an IQ mixer. For the sake of simplicity, for this analysis we set the LO frequency to the original B drive frequency $\omega_d$. In the frame of the downconverted signal, and ignoring fast-oscillating sidebands, the voltages produced at the I and Q ports of the mixer can be expressed as \cite{Blais2021}:
\begin{equation}\label{eq:v_complex}
    V(t_{\mathrm{ev}}, \tau) \defeq V_I(t_{\mathrm{ev}}, \tau)+iV_Q(t_{\mathrm{ev}}, \tau) \propto b_{\mathrm{out}}(t_{\mathrm{ev}}, \tau).
\end{equation}
To produce a set of IQ coordinates for each value of $t_{\mathrm{ev}}$, we perform a weighted integration of these voltages with respect to $\tau$. 
\begin{equation}
    z_{\mathrm{IQ}}(t_{\mathrm{ev}}) \defeq I(t_{\mathrm{ev}}) + iQ(t_{\mathrm{ev}}) =  \int_0^{\tau^\prime} w^*(\tau)V(t_{\mathrm{ev}},\tau)d\tau.
\end{equation}
Assuming the signal is on top of white noise, the weighting function $w(\tau)$ that maximizes SNR contains the same amplitude envelope and oscillation frequency as these output voltages with respect to the interrogation time \cite{Bultink2018}:
\begin{equation}
    w(\tau) = e^{ ( i \dtn{\K,\mathrm{eff}} - \lw{\K,\mathrm{eff}}/2 ) \tau }.
\end{equation}
Averaging the weight-biased signal over a long duration $\tau^\prime\gg2\pi/\lw{\K,\mathrm{eff}}$ so that the upper integration bound can be extended $\tau^\prime \rightarrow \infty$ produces
\begin{align}\label{eq:iq_coordinate}
    z_{\mathrm{IQ}}(t_{\mathrm{ev}}) \propto e^{ ( i \dtn{\K} + \lw{\K}/2 )t_{\mathrm{ev}}}.
\end{align}
We note that this proportional relationship holds regardless of the weight function used in the integration. Due to the adiabatic elimination of mode B and the exclusion of the upper sidebands in this simple model, the experimentally-observed voltages at the I and Q ports of the mixer as a function of $\tau$ will in general differ from the simple exponential form given in Eq.~\eqref{eq:b_out}. This merely changes the form of the optimal weight function employed in the time-averaging, and does not affect the dynamics of the system during $t_{\mathrm{ev}}$. A treatment of a similar experiment without the adiabatic elimination of the buffer mode is performed in Ref.~\cite{peronnin_sequential_2020}. Moving forward, it is instructive to write the amplitude and phase of the demodulated shot-averaged signal $z_{\mathrm{IQ}}(t_{\mathrm{ev}})$:
\begin{equation}\label{eq:cr_fits}
    |z_{\mathrm{IQ}}(t_{\mathrm{ev}})| \propto e^{- \lw{\K} t_{\mathrm{ev}} / 2}, \qquad \arg(z_{\mathrm{IQ}}(t_{\mathrm{ev}})) = \dtn{\K} t_{\mathrm{ev}}.
\end{equation}
Therefore, the bare K mode linewidth $\lw{\K}$ and detuning $\dtn{\K}$ can be directly extracted by fitting the resulting exponential amplitude decay and linear phase accumulation of the demodulated and shot-averaged IQ coordinates with respect to $t_{\mathrm{ev}}$. This is apparent in the application of the catch-and-release protocol to modes $\mathrm{K} = \mathrm{A},\mathrm{C}$ in Fig.~\ref{fig:catch_release}. To extract the dispersive shift, this protocol can be performed for each qubit state. We can then use Eq.~\eqref{eq:detuning_KB} to compute the bare mode frequency $\freq{\K}$ and dispersive shift $\disp{\K}$:
\begin{align}
    \freq{\K} &= \freq{d} + \sgn(\freq{\K} - \freq{\B}) \freq{p} - \frac{1}{2} \left( \dtn{\K}(s=1) + \dtn{\K}(s=-1) \right), \label{eq:cr_fit_freq} \\
    \disp{\K} &= \frac{1}{2} \left( \dtn{\K}(s=1) - \dtn{\K}(s=-1) \right). \label{eq:cr_fit_disp}
\end{align}
It is worth emphasizing that these expressions only hold for $g_{\B\C} \ll \lw{\B}$.

\subsection{Catch and release: A and C mode characterization}
\label{apdx:cr_a_c_char}
Here we utilize the catch-and-release protocol described above to find the frequencies, dispersive shifts, and linewidths of the amplifier and cavity modes ($\freq{\K}$, $\disp{\K}$, and $\lw{\K}$, where $\mathrm{K} = \mathrm{A},\mathrm{C}$). We note that in general, the frequency shift due to the dispersive interaction also changes the environment that the mode sees, and thus the linewidth. The linewidths quoted in the main body (i.e. Fig.~\ref{fig:dev_overview}(c)) are the averaged values with respect to qubit state $(\lw{\K}(s=1) + \lw{\K}(s=-1)) / 2$.

In the experimental implementation, the duration of the initial beam-splitter and B drive is $t_0=60$ ns including 10 ns $\cos^2$ rise and fall envelopes. We keep the drive weak ($|\varepsilon| = 4.6$ kHz) to avoid contributions to the dynamics from the A or C mode's unknown self-Kerr nonlinearity, an effect which can appear as an inflated amplitude decay rate in the demodulated and shot-averaged IQ data. We sweep the idling time up to $t_{\mathrm{ev}} = 10$ $\mu$s and probe the released signal for $\tau=300$ ns. We construct the weighted averaging functions from shot-averaged ($5\times10^5$ shots for the A mode, $10^6$ shots for the C mode) time-traces of the downconverted signal with the qubit in the ground state. We then repeat this procedure for both states of the qubit, whose measurement and reset protocol is detailed in Appendix~\ref{apdx:diagnostic_qubit_meas}. Fig.~\ref{fig:catch_release} shows the simplified experimental sequence and resulting data for modes A and C, demonstrating the exponential decay in amplitude and linear accumulation of phase with respect to $t_{\mathrm{ev}}$. We fit the data to Eq.~\eqref{eq:iq_coordinate} and Eq.~\eqref{eq:cr_fits}, then utilize Eqs.~\eqref{eq:cr_fit_freq} and ~\eqref{eq:cr_fit_disp} to obtain the results tabulated in Table \ref{tbl:mode_params}.

\begin{table}[htpb!]
\begin{tabular}{ |c||c|c| }
 \hline
 mode & A & C\\
 \hline
$\freq{k}/2\pi$ & 3.37097441 GHz $\pm$ 410 Hz & 8.97391248 GHz $\pm$ 440 Hz\\
\hline
$\disp{k}/2\pi$ & 520 $\pm$ 410 Hz & 2.75411 $\pm$ 0.00044 MHz\\
\hline
$\lw{k}(s=-1)/2\pi$ & 115.14 $\pm$ 1.16 kHz & 103.79 $\pm$ 1.20 kHz\\
\hline
$\lw{k}(s=+1)/2\pi$ & 114.71 $\pm$ 1.14 kHz & 115.72 $\pm$ 1.29 kHz\\
\hline
$\lw{k}/2\pi$ & 114.92 $\pm$ 0.81 kHz & 109.76 $\pm$ 0.88 kHz\\
\hline
\end{tabular}

\caption{Parameters of the amplifier (A) and cavity (C) modes extracted through catch-and-release measurements for each qubit state, as well as their qubit state-averaged values. Uncertainties quoted are $1\sigma$ confidence intervals.}
\label{tbl:mode_params}
\end{table}
The vanishing dispersive shift of mode A is apparent in the near indistinguishability of the signal's phase accumulation with respect to the qubit's state in the AB beam-splitter measurement. By contrast, the BC beam-splitter data show a phase accumulation that is highly distinguishable with respect to the qubit state, due to the considerable dispersive shift of mode C. The amplitude decay and thus linewidth for both modes only slightly depends on qubit state, with mode C experiencing a larger variation due to its larger dispersive shift paired with a general frequency-dependence of the real part of the environmental impedance seen by the cavity. 

As a brief aside, it is worth noting that the BC catch-and-release protocol can serve as a qubit measurement \cite{peronnin_sequential_2020}. Assuming $\lw{\C}=0$ and the ability to instantaneously inject signal into mode C, maximal SNR would be achieved for a delay time of $t_\mathrm{ev,opt} = \pi/2\disp{\C} \approx 90$ ns. Experimentally, due to the finite $\lw{\C}$ and effective C mode drive duration, we observe optimal SNR at a delay time of $ t_\mathrm{ev,opt}\approx40$ ns.

\begin{figure}[htpb!]
    \includegraphics[scale=1]{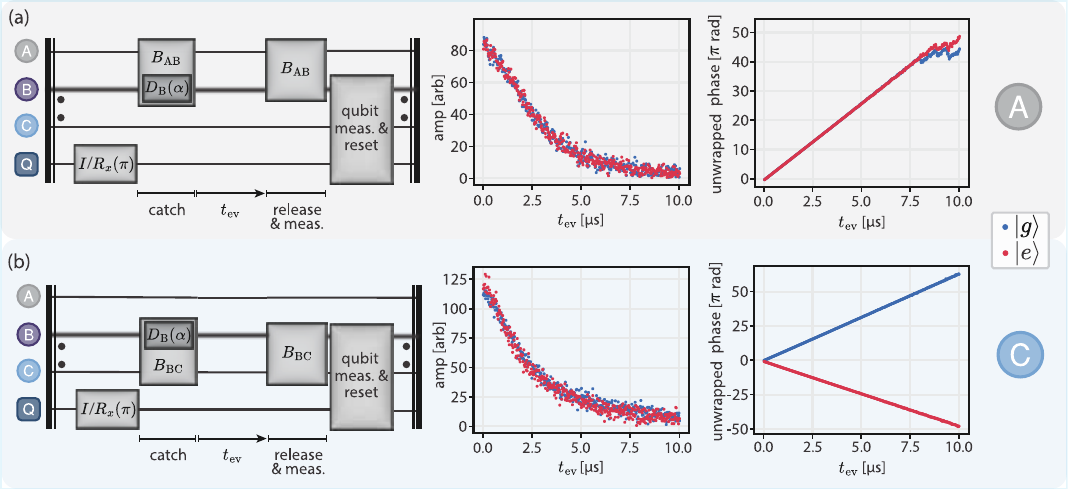}
    \caption{\textbf{A and C mode characterization}. Simplified experimental sequence of and data for the catch-and-release protocols with \textbf{(a)} mode A and \textbf{(b)} mode C. The protocol includes qubit state preparation, a simultaneous B mode drive and beam-splitter between modes B and $\mathrm{K} = \mathrm{A},\mathrm{C}$, a time delay $t_{\mathrm{ev}}$ of variable duration, and a BK beam splitter to release the measurement signal. Demodulating and shot-averaging this signal, and repeating for various values of $t_{\mathrm{ev}}$ and each qubit state, produces the plotted data.}
    \label{fig:catch_release}
\end{figure}

\subsection{BC beam-splitter calibration (weak coupling regime)}
\label{apdx:bc_calib}
\begin{figure}[htbp]
    \includegraphics[scale=1]{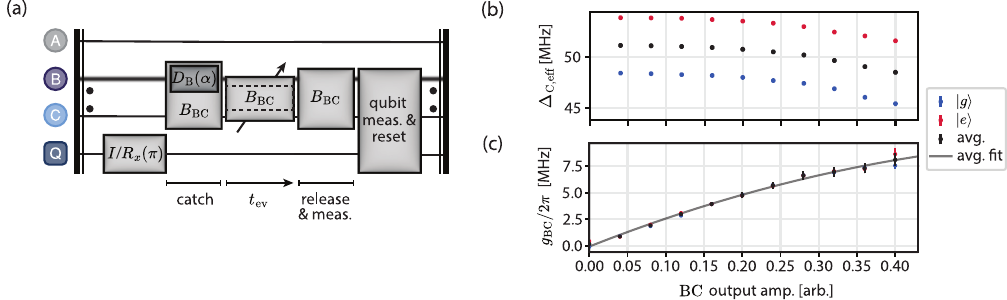}
    \caption{\textbf{BC beam-splitter calibration}. \textbf{(a)} Diagram of the general experimental procedure, consisting of a tunable BC beam-splitter pump during the C mode evolution time ($t_{\mathrm{ev}}$) in a catch-and-release protocol. \textbf{(b)} $\dtn{\C,\mathrm{eff}}$ (defined in the main text), a quantity used to calibrate the pump-induced frequency shift, and \textbf{(c)} $g_{\B\C}$ are both plotted as a function of the BC beam-splitter pump amplitude (in arbitrary units) and qubit state. We note that the measurement in (b) retains a static pump frequency, whereas the measurement in (c) varies the pump frequency as a function of amplitude to ideally keep the BC beam-splitter on-resonance. We fit the $g_{\B\C}$ data averaged over the qubit state with a quadratic function with respect to BC output amplitude for interpolation in future measurements.}
    \label{fig:bc_calib}
\end{figure}

As shown in Eq.~\eqref{eq:adiabaitc_elim_BK_defs}, a beam-splitter interaction between modes B and C alters the effective dissipation rate of mode C. In addition, higher order nonlinearities of the dc SQUID result in a pump power-dependent frequency shift (rectification) of each mode from $\freq{\B}$ and $\freq{\C}$ to $\freq{\B}^{\prime}$ and $\freq{\C}^{\prime}$ respectively. The pump frequency and amplitude are thus interdependent, inspiring a calibration of both of these effects to reliably implement a BC beam-splitter interaction with a desired strength and detuning, as in the B mode thermal occupancy calibration of Sec.~\ref{subsec:occupancies}. To estimate the resonant beam-splitter pump frequency $\freq{\B\C}^{\mathrm{opt}} \defeq | \freq{\C}^\prime - \freq{\B}^\prime |$ and strength $g_{\B\C}$ as a function of pump amplitude, we include a BC beam-splitter pump during the evolution time ($t_{\mathrm{ev}}$) of a catch-and-release measurement between modes B and C, as depicted in Fig.~\ref{fig:bc_calib}(a). This alters the phase accumulation and increases the amplitude decay rate of the demodulated signal expressed in Eq.~\eqref{eq:cr_fits}:
\begin{equation}\label{eq:cr_fits_bc}
    |z_{\mathrm{IQ}}(t_{\mathrm{ev}})| \propto e^{- \lw{\C,\mathrm{eff}} t_{\mathrm{ev}} / 2}, \qquad \arg(z_{\mathrm{IQ}}(t_{\mathrm{ev}})) = \dtn{\C,\mathrm{eff}} t_{\mathrm{ev}}.
\end{equation}
First, we experimentally track $\dtn{\C,\mathrm{eff}}$. For this first measurement, we keep the frequency of the BC beam-splitter pump constant as we sweep its amplitude. This produces a value of $\dtn{\C,\mathrm{eff}}(s)$ for each BC pump amplitude and qubit state ($s=-1,1$), shown in Fig.~\ref{fig:bc_calib}(b). Using Eqs.~\eqref{eq:detuning_KB} and \eqref{eq:adiabaitc_elim_BK_defs}, and generalizing mode C $\rightarrow$ K, the qubit state-dependent pump frequency producing a resonant beam-splitter interaction between modes B and K ($\freq{\B\K}^{\mathrm{opt}}(s)$) is given by
\begin{equation}\label{eq:opt_bs_freq_bc}
    \freq{\B\K}^{\mathrm{opt}}(s) \defeq | \freq{\K}^\prime - \freq{\B}^\prime | \approx \freq{p} + \sgn( \freq{\K} - \freq{\B} ) \left[ ( \freq{d} - \freq{\B}^\prime ) ( 1 - 4 g_{\B\K}^2 / \lw{\B}^2 ) - \dtn{\K,\mathrm{eff}(s)} \right],
\end{equation}
where the second term can be thought of as an effective pump detuning. This allows one to remove the unknown term $\freq{\C}^\prime$, however the other unknown terms of $\freq{\B}^{\prime}$ and $g_{\B\C}$ remain. Here we will make a strong approximation and neglect the terms dependent on $\freq{\B}^{\prime}$ and $g_{\B\C}$. Indeed, by comparing the  center of the feature in Fig.~\ref{fig:qbsig} with respect to drive frequency (a proxy for $\freq{\B}'$), to the bare B mode frequency, we expect a maximal pump-induced B-mode rectification of approximately $\freq{\B} - \freq{\B}' \lesssim 13$ MHz, which is still substantially less than $\lw{\B}$. We therefore expect that the BC beam-splitter detunings resulting from this imperfect calibration do not substantially affect calibration of $g_{\B\C}$ discussed in the following paragraph, nor the dephasing dynamics in Fig.~\ref{fig:occupancies}(b). With the assumption of $\disp{\B}\approx0$, a more accurate calibration would include an initial independent measurement of $\freq{\B}^\prime$ using a detuned BC beam-splitter interaction ($\dtn{\C} \gg \lw{\B}$) to rectify the system with negligible beam-splitter dynamics. Subsequently, the B mode can be driven on-resonance for the catch-and-release measurements using the same BC beam-splitter pump strength employed to calibrate $\freq{\B}'$ (assuming negligible or calibrated-out flux-line ripple), effectively removing the term in Eq.~\eqref{eq:opt_bs_freq_bc} dependent on $g_{\B\C}$.

After we perform the calibration of $\freq{\B\K}^{\mathrm{opt}}(s)$, we employ these values in another catch-and-release protocol, altering the BC beam-splitter pump frequency as a function of its amplitude, to calibrate $g_{\B\C}(s)$. Assuming the beam-splitter interaction is on-resonance for each amplitude and qubit state, we extract the beam-splitter coupling rate as
\begin{equation}
    g_{\B\C}(s) = \sqrt{ \left( \lw{\C,\mathrm{eff}}(s) - \lw{\C} \right) \left( \lw{\B} / 4 + \disp{\B}^2 / \lw{\B} \right) }.
\end{equation}
The resulting values are plotted in Fig.~\ref{fig:bc_calib}(c). It is once again worth noting that while the BC beam-splitter pump frequencies employed for the extraction of $g_{\B\C}(s)$ in Fig.~\ref{fig:bc_calib}(c) were qubit-state dependent, the pump frequencies chosen for the B mode occupancy measurement were averaged with respect to the qubit state: $\freq{\B\C,\mathrm{avg}}^\mathrm{opt} = (1/2)(\freq{\B\C}^\mathrm{opt}(s=-1) + \freq{\B\C}^\mathrm{opt}(s=1))$.

\subsection{Catch and release: AC beam-splitter calibration (strong coupling regime)}
\label{apdx:ac_calib}
\begin{figure}[htbp]
    \includegraphics[scale=1]{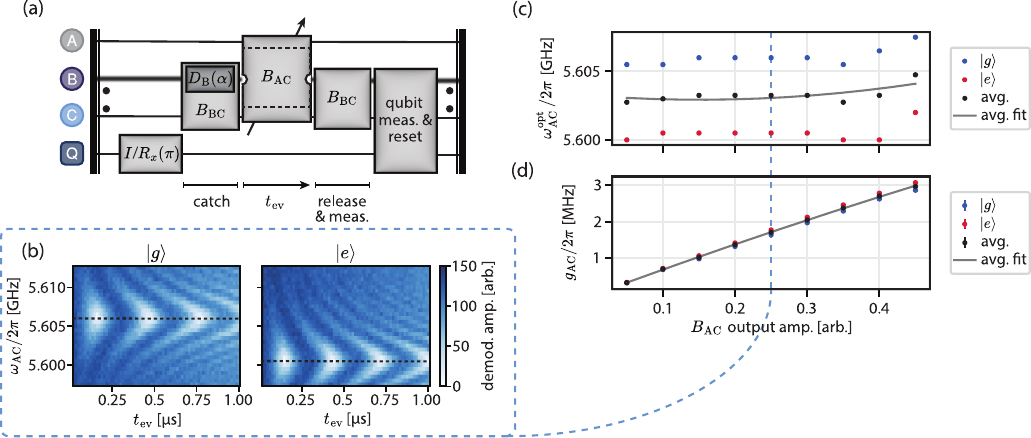}
    \caption{\textbf{AC beam-splitter calibration}. \textbf{(a)} Diagram of the general experimental procedure, consisting of an amplitude and frequency-varied AC beam-splitter pump during the C mode evolution time ($t_{\mathrm{ev}}$) in a catch-and-release protocol. \textbf{(b)} Magnitude of the signal released from the C mode over a 2D scan of AC beam-splitter frequency $\freq{\A\C}/2\pi$ (y axis) and evolution time (x axis) at a beam-splitter amplitude of 0.25 [arb.] ($g_\mathrm{AC}\approx1.7$ MHz) for both qubit states. \textbf{(c)} Pump frequency required to achieve a resonant AC beam-splitter interaction, $\freq{\A\C}^\mathrm{opt}(s)$, and \textbf{(d)} AC beam-splitter interaction strength, $g_\mathrm{AC}$, plotted as a function of output pump strength in arbitrary amplitude units. We average each quantity with respect to the qubit state, and fit the average data to a quadratic function for interpolation in subsequent measurements.}
    \label{fig:ac_calib}
\end{figure}

Similarly, to analyze the frequency shifts and parametric coupling rate of modes A and C in the presence of an AC beam-splitter pump, we include this pump  during the evolution time $t_{\mathrm{ev}}$ of the catch-and-release sequence, shown in Fig.~\ref{fig:ac_calib}(a). To understand the signal output at the end of this measurement, we begin with the equations of motion for the mean fields $a \defeq \expt{\hat{a}}$, $c \defeq \expt{\hat{c}}$, including some beam-splitter detuning $\dtn{\A\C} \defeq \freq{p} - \freq{\C} 
+ \freq{\A}$:
\begin{align}
    \frac{d}{dt} a &= \left( i \dtn{\A\C} - \frac{\lw{\A}}{2} \right) a - ig_{\A\C}c \nonumber\\
    \frac{d}{dt} c &= \left( i s \disp{\C} - \frac{\lw{\C}}{2} \right) c - ig_{\A\C}a
\end{align}
In the beginning of the catch-and-release protocol, we perform an effective drive on the C mode. Correspondingly assuming $a(t_{\mathrm{ev}}=0) = 0$ and $c(t_{\mathrm{ev}}=0) \neq 0$, the equations of motion for C can be solved as
\begin{equation}
    c(t) \propto e^{-\lw{}t_{\mathrm{ev}}/2} \left[ \cos \left( \frac{\Omega t_{\mathrm{ev}}}{2} \right) - \frac{ i ( \dtn{\A\C} - s \disp{\C} ) }{ \sqrt{ \Omega } } \sin \left( \frac{\Omega t_{\mathrm{ev}}}{2} \right) \right], \qquad \Omega \defeq \sqrt{ 4 g_{\A\C}^2 + ( \dtn{\A\C} - s \disp{\C} )^2}.
\end{equation}
Following the logic of Appendix~\ref{apdx:c_r_eom}, the magnitude of the released signal is 
\begin{equation}\label{eq:cr_ac}
    |z_{\mathrm{IQ}}(t_{\mathrm{ev}})| \propto e^{- \lw{} t_{\mathrm{ev}} / 2} \sqrt{ \cos^2 \left( \frac{\Omega t_{\mathrm{ev}}}{2} \right) + \frac{ ( \dtn{\A\C} - s \disp{\C} )^2 }{ \Omega } \sin^2 \left( \frac{\Omega t_{\mathrm{ev}}}{2} \right) }.
\end{equation}
As a function of $t_{\mathrm{ev}}$, we therefore expect magnitude oscillations whose frequency grows as $g_{\A\C}$ increases, and whose amplitude grows and frequency diminishes as $\dtn{\A\C} \rightarrow s \disp{\C}$. Experimentally performing this catch-and-release sequence for various values of $t_{\mathrm{ev}}$ and AC beam-splitter pump frequencies produces the colorplot in Fig.~\ref{fig:ac_calib}(b) (for a fixed output pump amplitude of 0.25 [arb.] or $g_\mathrm{AC}\approx 1.7$ MHz, in this case). The pump frequency and evolution time-dependence of the amplitude in Eq.~\eqref{eq:cr_ac} is captured in the chevron-like features of Fig.~\ref{fig:ac_calib}(b). Taking a linecut at frequency $\freq{\A\C}^{\mathrm{opt}}$ for which $\dtn{\A\C} = s \disp{\C} $, we recognize that
\begin{equation}\label{eq:cr_ac_linecut}
    |z_{\mathrm{IQ}}(t_{\mathrm{ev}})| \propto e^{- \lw{} t_{\mathrm{ev}} / 2} | \cos ( g_{\A\C} t_{\mathrm{ev}} ) |.
\end{equation}
Fitting this linecut data to Eq.~\eqref{eq:cr_ac_linecut} allows us to extract $g_{\A\C}$ for a given output AC pump amplitude. Repeating this procedure allows us to extract $\freq{\A\C}^\mathrm{opt}$ and $g_{\A\C}$ as a function of AC pump amplitude, producing the data in Fig.~\ref{fig:ac_calib}(c). To interpolate these points in subsequent measurements, we fit this data with a quadratic function for both parameters.

\subsection{Drive strength and $\disp{\B}$ calibration}
\label{apdx:epsilonb_chib_calib}
\begin{figure}[htbp]
    \includegraphics[scale=1]{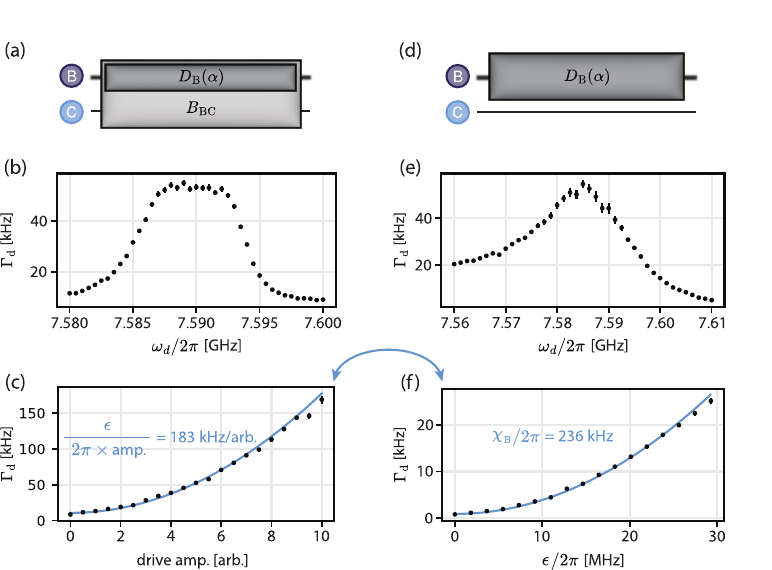}
    \caption{\textbf{(a)-(c) Drive strength $\varepsilon$ and (d)-(f) spurious B dispersive shift $\disp{\B}$ calibrations.} \textbf{(a)} Diagram of the concurrent B mode drive and fixed BC beam-splitter ($g_{\B\C}=7.075 ~\mathrm{MHz}$) sent during the qubit's evolution in $T_1$ and $T_2^\mathrm{echo}$ measurements. \textbf{(b)} Measured qubit dephasing as a function of drive frequency $\freq{d}$ at fixed amplitude (5.0 [arb.]) to determine the drive frequency corresponding to a zero-detuning effective drive on the dressed C mode. \textbf{(c)} Measured qubit dephasing rate $\drate$ as a function of drive strength  with a fixed drive frequency $\freq{d}/2\pi=7.59$ GHz, fit to find the coefficient between $\varepsilon$ and arbitrary output amplitude. \textbf{(d)-(f)} is a repetition of (a)-(c), without the BC beam-splitter during the qubit's evolution in $T_1$ and $T_2^\mathrm{echo}$ measurements. In the sweep of $\freq{d}$ shown in (e), we choose a drive output amplitude of 250.0 [arb.], and in the sweep over $\varepsilon$ shown in (f) we set the drive frequency to $\freq{d}/2\pi=7.585$ GHz, and fit the resulting data for $\disp{\B}$. As indicated by the double-sided arrow, the fits of the data in (c) and (f) are interdependent and must be sequentially performed for continuously-updated values of $\varepsilon$ and $\disp{\B}$.}
    \label{fig:epsilonb_chib_calib}
\end{figure}

To minimize the number of free parameters in the comparison between theory and experiment for the measurement $\mrate$ and dephasing $\drate$ rates in the main body, we separately calibrated the drive strength at the reference plane of the device ($\varepsilon$), as well as the spurious dispersive shift to mode B ($\disp{\B}$). We extract these parameters with iterative fits of codependent measurements using the full theory presented in Sec.~\ref{sec:theory}.

To calibrate $\varepsilon$, we monitor $\drate$ in the presence of a weak B mode drive (5.0 in arbitrary output amplitude units) and BC beam-splitter ($g_{\B\C}=7.075$ MHz), as shown in Fig.~\ref{fig:epsilonb_chib_calib}(a). To ensure the effective C mode drive is approximately on-resonance, we operate in the regime $\lw{\C,\mathrm{eff}}\approx 2\disp{\C}$, and sweep the frequency of the weak B drive $\freq{d}$ to find where the qubit's dephasing rate is maximal (assuming a resonant BC beam-splitter interaction). The data, shown in Fig.~\ref{fig:epsilonb_chib_calib}(b), suggest an optimal drive frequency of approximately 7.59 GHz. We then hold the drive constant at this frequency, and monitor $\drate$ as we sweep the amplitude over a range of weak values (to prevent the appearance of self Kerr-induced dynamics). As shown in Fig.~\ref{fig:epsilonb_chib_calib}(c), $\drate$ increases proportionally to the square of drive amplitude, per the expected behavior. Initially assuming $\disp{\B}=0$, we then numerically fit the data to the theory with one fit parameter, allowing us to extract a preliminary coefficient relating $\varepsilon$ and output drive amplitude in arbitrary units.

However, in reality $\disp{\B} \neq 0$, a fact which contributes to further quadratic growth of the qubit's dephasing rate as a function of drive amplitude. To disambiguate these two effects, we repeat the same pair of measurements without the BC beam-splitter. Due to the fact that $\lw{\B}\gg 2\disp{\B}$, much larger drive strengths are required to noticeably dephase the qubit. The sweep of $\freq{d}$ with a drive amplitude of 250.0 [arb.] is shown in Fig.~\ref{fig:epsilonb_chib_calib}(e), resulting in a chosen drive frequency of $\freq{d}/2\pi = 7.585$ GHz for the sweep of drive amplitude in Fig.~\ref{fig:epsilonb_chib_calib}(f). The $\varepsilon$ coefficient from the previous fit renormalizes the x axis, allowing for a single-free parameter fit for $\disp{\B}$.

Because of the interdependence of the fits for $\varepsilon$ and $\disp{\B}$, we iteratively fit the data from each measurement over multiple rounds. We include the preliminary value of $\disp{\B}$ as a static parameter in a second fit for the $\varepsilon$ coefficient, producing a more accurate value which we then include as a static parameter in the fit for $\disp{\B}$. We repeat this process until the change in the $\varepsilon$ coefficient fit parameter from one iteration to the next is within the confidence interval given by 1$\sigma$, which in our case takes three rounds of successive fitting. From this iterative procedure we extract $\varepsilon / (2\pi\times\mathrm{amp.}) = 183.0\pm0.6$ kHz/arb., and $\disp{\B}=236$ kHz.

\subsection{Calibrating noise of the subsequent measurement chain, $\bar{n}_{\add}$}
\label{adpx:off_chip_noise_calib}
To compare the experimentally-measured and theoretically-computed $\mrate$ for the case of the interferometer in Sec.~\ref{subsec:interferometer} with no additional fit parameters, we must independently characterize the added noise of the measurement chain downstream from the readout network, $\bar{n}_\mathrm{add}$. We extract this value by probing $\mrate$ (see Fig.~\ref{fig:mrate_mdeph}(b)) in the presence of a BC beam-splitter of strength $g_{\B\C} = 7.075$ MHz and a B mode drive of strength $\varepsilon = 2.77$ MHz. For this configuration, $\mrate = 32.732$ kHz. Utilizing the theory from Sec.~\ref{sec:theory} with this value and the previously-calibrated linewidths, thermal occupancies, and cross Kerrs tabulated in Fig.~\ref{fig:dev_overview}(c) and Table \ref{tbl:mode_params}, we performed a numerical inversion of $\Gamma_\mathrm{m}(\bar{n}_\mathrm{add})$ to extract $\bar{n}_{\add}=5.24$ photons.

\subsection{Mismatched circulator tuning}
\label{apdx:mmcirc_tune}
The strategies discussed in Appendices~\ref{apdx:bc_calib} and \ref{apdx:ac_calib} can be used to calibrate the coupling strength and frequency rectification from a single beam-splitter pump. However, when adding multiple pumps of various frequencies, the rectification effects combine in a manner dependent on all of the higher-order nonlinearities of each mode -- a calibration of which is outside the scope of this work. Because of this, we use a separate heuristic to account for the rectification-induced shifts in beam-splitter pump frequencies and strengths when operating the interferometer in Sec.~\ref{subsec:interferometer}.

As mentioned in Appendix~\ref{sec:prelim_char}, the Fano interference present in our system renders the steady-state reflection coefficient off of mode B ($S_{\B\B}$) difficult to fit using coupled-mode theory. Luckily, the fields that bypass the reflection off of the device and interfere with our measurement signal are independent of the qubit state. With this in mind, a reasonable heuristic for diagnosing multiple pump-induced shifts is an effective qubit measurement signal $S_{\B\B,\mathrm{diff}} \equiv S_{\B\B,\ket{e}} - S_{\B\B,\ket{g}}$. By sweeping both the interferometer phase $\phi$, as well as the drive frequency sent to the B mode $\omega_d$, the magnitude of the qubit signal $\left|S_{\B\B,\mathrm{diff}}\right|$ provides visual features whose characteristics vary as a function of pump frequency and strength.

To theoretically analyze the qubit measurement signal, we compute $S_{\B\B,\ket{\sigma}}$, where $\sigma=e,g$, for the interferometer with the linear-response theory presented in Appendix~\ref{apdx:input-output_scattering}, which is equivalent to coupled mode theory in this system \cite{ranzani_graph-based_2015, Lecocq2017, Lecocq2020}.  We then took the magnitude of the difference between the two qubit state-dependent reflection coefficients $|S_{\B\B,\mathrm{diff,theory}}| \defeq A\left|S_{\B\B,\ket{e},\mathrm{theory}} - S_{\B\B,\ket{g},\mathrm{theory}}\right|$ where $A$ is an arbitrary prefactor to account for the total gain level of our measurement path. We then tune the following quantities in the theory until there is a substantial match with the data:
\begin{itemize}
    \item amplitude prefactor on qubit signal,
    \item experimental offset in interferometric phase (up to $\pi$),
    \item assumed rectified frequencies of A, B, and C modes (this parameterizes AB, BC, and AC pump frequencies),
    \item parametric coupling strengths $g_{\A\B}$, $g_{\B\C}$, and $g_{\A\C}$.
\end{itemize}

All in all, there are 8 free parameters in this fit. In theory, it is possible to constrain $A$ from our independent measurements of $\varepsilon$ and $\bar{n}_{\add}$ in Appendices~\ref{apdx:epsilonb_chib_calib} and \ref{adpx:off_chip_noise_calib} respectively. However, we found that this parameter, along with the experimental offset in interferometer phase, were orthogonal from the rest of the parameters and didn't contribute to the tuning complexity. In the tuning process, we found more success performing this fit ``by hand" compared to defining the cost function as $\left||S_{\B\B,\mathrm{diff}}|-|S_{\B\B,\mathrm{diff,theory}}|\right|$ and performing a simple numerical minimization. This seems to be due to the large number of local minima in the cost function with respect to the free parameters; however, more sophisticated methods of fitting are worth exploring. 

\begin{figure}[htpb!]
    \includegraphics[scale=1]{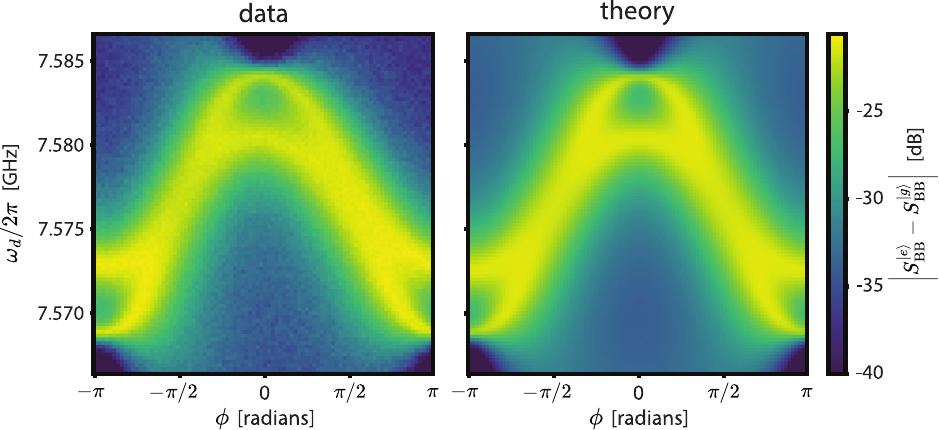}
    \caption{\textbf{Mismatched circulator qubit signal}. Measured data and theory of the qubit signal, $\left|S_{\B\B,\ket{e}} - S_{\B\B,\ket{g}}\right|$ plotted against the probe tone frequency $\omega_d/2\pi$ (y axis) and interferometer phase $\phi$ (x axis). An arbitrary coefficient scales qubit signal theory to account for the net gain of the signal path. We employ this heuristic of qubit measurement signal to find beam-splitter detunings and deviations from ideal coupling rates caused by rectification.}
    \label{fig:qbsig}
\end{figure}

Once we fit the data, we use the results to steer the experimental pump parameters toward those desired for ideal device operation. Because of the aforementioned rectification effects, the ideal pump amplitudes and frequencies are interdependent, necessitating small changes in pump amplitude and frequency for each fit iteration. The final data/theory comparison resulting from this fitting and steering procedure is shown in Fig.~\ref{fig:qbsig}. The theory contains zero pump detunings, and strengths of $(g_{\A\B},g_{\B\C},g_{\A\C})/2\pi = (7.98, 14.5, 6.2)$ MHz.

\section{Diagnostic qubit readout}
\label{apdx:diagnostic_qubit_meas}
Regardless of the readout network's operation, a standard configuration to read out the qubit's state is necessary to perform the $T_1$ and $T_2^\mathrm{echo}$ measurements. In principle, this readout could be performed with a reflection measurement off of mode C. However, the Purcell filtering scheme we have chosen (see Appendix~\ref{apdx:experimental_diagram}) limits us to high SNR measurements only in reflection off of mode B. However, a dispersive measurement of the qubit using mode B directly would also be low fidelity as $\lw{\B}\gg\disp{\B}$. 

We therefore include a single BC beam-splitter interaction to effectively increase the external coupling rate of the C mode. This can be seen by taking Eq.~\eqref{eq:adiabatic_elim_BK} for the case of K $\rightarrow$ C:
\begin{equation}\label{eq:adiabatic_elim_drive}
    \frac{d}{dt} c = \left( i s \disp{\C,\mathrm{eff}} - \frac{\lw{\C,\mathrm{eff}}}{2} \right) c + \varepsilon
\end{equation}
where
\begin{equation}
    \lw{\C,\mathrm{eff}} := \frac{ g^2 }{ \lw{\B}^2 / 4 + \disp{\B}^2 } \lw{\B} + \lw{\C}, \qquad \disp{\C,\mathrm{eff}} := \disp{\C} - \frac{ g^2 }{ \lw{\B}^2 / 4 + \disp{\B}^2 } \disp{\B}.
\end{equation}
Assuming $\disp{\B} \ll \disp{\C}, \lw{\B}$ and negligible parasitic dephasing, the total qubit dephasing rate $\drate \approx \mdrate$ can be estimated from Eq.~\eqref{eq:coherent_state_mdrate}:
\begin{equation}\label{eq:eff_mdrate_c}
    \mdrate \approx \frac{4 n_{\C, \mathrm{eff}} \disp{\C}^2 \lw{\B} (4 g^2 + \lw{\B} \lw{\C})}{(4 g^2 + \lw{\B} \lw{\C})^2 + 4 \lw{\B}^2 \disp{\C}^2}
\end{equation}
where the effective intracavity photon number is given by
\begin{equation}
    n_{\C, \mathrm{eff}} = \frac{4|\varepsilon|^2}{(4 g^2 / \lw{\B} + \lw{\C})^2 + \disp{\C}^2}.
\end{equation}
Assuming that we hold $n_{\C, \mathrm{eff}}$ constant by sweeping $|\varepsilon|$ alongside $g$, maximizing Eq.~\eqref{eq:eff_mdrate_c} with respect to $g$ produces
\begin{equation}
    g_{\mathrm{max}(\mdrate)} = \frac{1}{2} \sqrt{ \lw{\B} ( 2\disp{\C} - \lw{\C} ) } \quad \Rightarrow \quad \lw{\C,\mathrm{eff}} = 2\disp{\C}.
\end{equation}
Because $\mrate \propto \mdrate$ for this case, maximal fidelity occurs at maximum $\mdrate$. In our device, this condition is met for $g=7.25$ MHz. For diagnostic qubit measurement, we therefore set the beam-splitter coupling to this value, aided by the BC strength calibration discussed in Appendix~\ref{apdx:bc_calib}.

\begin{figure}[htpb!]
    \includegraphics[scale=1]{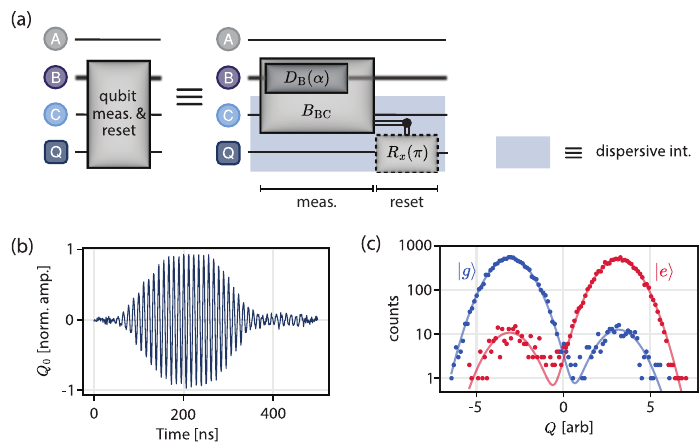}
    \caption{\textbf{Diagnostic qubit measurement and reset.} \textbf{(a)} Diagram of the measurement protocol including a displacement drive on mode B, $D_{\B}(\alpha)$, and beam-splitter between the modes B and C modes, $B_{\B\C}$, followed by a $\pi$ pulse conditioned on the qubit being measured in the excited state. The shaded blue box represents the dominant dispersive interaction that is always on between mode C and the qubit; this is omitted in other experimental sequence diagrams for the sake of simplicity. \textbf{(b)} Weighted averaging function including the IF frequency for the qubit's signal, taken from a difference between the demodulated time traces when the qubit is in the ground and excited states, averaged over $10^4$ shots for each qubit state. \textbf{(c)} Single shot histogram distributions over the quadrature containing qubit information $Q$ in arbitrary units of amplitude, for $10^4$ interleaved measurements of each qubit state. These distributions exhibit a readout and reset fidelity of $\mathcal{F}_r = 0.952$ for a repetition cycle of 5.36 $\mu$s $< T_1$ (26 $\mu$s).}
    \label{fig:meas_reset}
\end{figure}

In the experimental sequence for diagnostic qubit readout, we pulse this beam-splitter interaction with a duration of 550 ns, beginning concurrently with a B mode drive which lasts 500 ns. Each pulse has 50 ns $\cos^2$ rise/fall envelopes. We digitally downconvert the resulting signal to an IF frequency of 90 MHz and subsequently digitize it. To obtain a time-weighted averaging function that maximizes qubit readout SNR, we repeat this measurement $10^4$ times for each state of the qubit with a passive-reset delay between each shot. We then take the difference of the resulting shot-averaged time traces with respect to the qubit's state, producing the weighted averaging function (including oscillations at the IF frequency) shown in Fig.~\ref{fig:meas_reset}(b). To quantify the readout and active reset fidelity $\mathcal{F}_r$, we prepare and measure the qubit's ground or excited state, then actively reset the qubit to ground at the end of the sequence. We repeat this procedure every 5.36 $\mu s$ (which is a fraction of $T_1 \approx 26 \mu s$) over $10^4$ shots for each qubit state, prepared in an interleaved manner. We then threshold the resulting single shot values to extract a single-shot readout and reset fidelity $\mathcal{F}_r = 1 - \mathrm{P}(g|e) - \mathrm{P}(e|g) = 0.952$. This fidelity includes assignment, state preparation, and QND (i.e. relaxation and measurement-drive excitation) errors within the two-state manifold. A simplified diagram of the measurement and active reset sequence, and the single shot histograms are shown in Fig.~\ref{fig:meas_reset}(a) and (c) respectively.

\section{Spurious Gain} 
\label{apdx:spurious_gain}
\begin{figure}[htpb!]
    \includegraphics[scale=1]{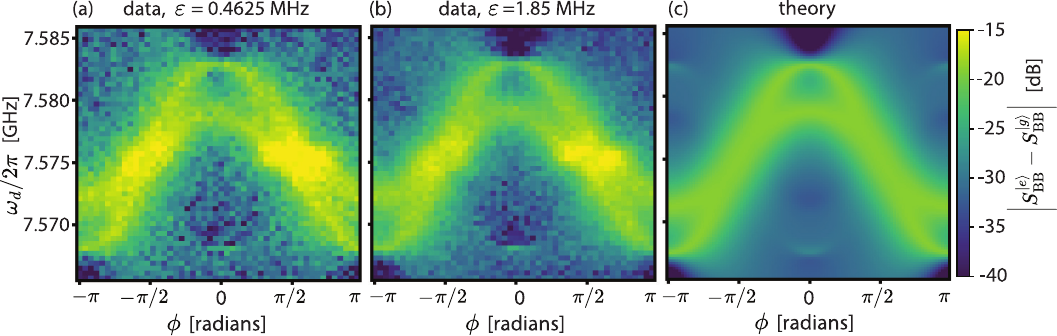}
    \caption{\textbf{Spurious gain in embedded amplifier}. Measured qubit signal ($\left|S_{\B\B,\ket{e}} - S_{\B\B,\ket{g}}\right|$) data with a drive amplitude of \textbf{(a)} $\varepsilon = 0.4625$ MHz and \textbf{(b)} $\varepsilon = 1.85$ MHz, and \textbf{(c)} theory plotted as a function of the probe tone frequency $\omega_d/2\pi$ (y-axis) and interferometric phase $\phi$ (x-axis).}
    \label{fig:spurious_gain}
\end{figure}

While attempting to introduce a single-mode squeezing interaction in mode A, we noticed a spurious gain process that overwhelmed the desired interaction. To demonstrate this, in Fig.~\ref{fig:spurious_gain}, we plot the measured and simulated qubit measurement signal $S_{\B\B,\mathrm{diff}}$ as a function of B mode drive frequency and interferometer phase (as in Appendix~\ref{apdx:mmcirc_tune}), for two different drive powers ($\varepsilon = 0.4625, 1.85 $ MHz). This beam-splitter pump configuration is the same as the interferometer $(g_{\A\B},g_{\B\C},g_{\A\C})/2\pi = (7.98, 14.5, 6.2)$ MHz, with an additional pump at $2 \freq{\A}$. In the theory, we set the squeezing strength to $\lambda / 2\pi = 2$ MHz (corresponding to $\sim0.85$ dB of gain at $\phi = \pm \pi / 2$) to qualitatively match the data and remain approximately consistent with the measured gain from the idler of mode A's single mode squeezing process (see next paragraph). Despite this, the $\sim 5$ dB gain features around $\phi \approx \pm \pi / 2$ and $\freq{d} \approx \freq{\B}'$ were not reproducible in the theory, even over a large range of beam-splitter detunings and strengths. For measurements with larger gain-pump amplitudes, this spurious process dominates the dynamics of the device, quickly bringing it into an unstable regime.

Comparing Fig.~\ref{fig:spurious_gain}(a) and (b), these spurious gain features do not change when we change the drive amplitude by a factor of four ($\approx12$ dB). It is therefore unlikely to be a signature of self-Kerr of the modes, or cross-Kerr between any of the modes, due to the presence of the SQUID. In an attempt to identify the idler of this gain process, we analyzed the output spectrum of the device and identified numerous peaks. If we included a slightly off-resonance B mode drive, we could identify a small peak corresponding to the idler of the weak single-mode squeezing at mode A. From the difference in signal and idler peak heights, we estimated the gain from this process to be about 1 dB before the onset of instability from the other spurious process. We identified all other present peaks as spurious tones generated by the room temperature setup, which did not appreciably affect this spurious process when suppressed via filtering or active cancellation. Note that our spectrum analysis was limited to the bandwidth of the Purcell filter: 7.25 - 7.75 GHz, and it is very possible that the idler of the spurious process is present outside that bandwidth.

\section{Device design and simulation details}
\label{apdx:design_sim}
In this section we include details regarding the finite-element simulation of the device described in Sec.~\ref{subsec:network_expt_implementation}.

\subsection{Eigenmode simulation results}
\begin{figure}[htbp]
    \includegraphics[scale=1]{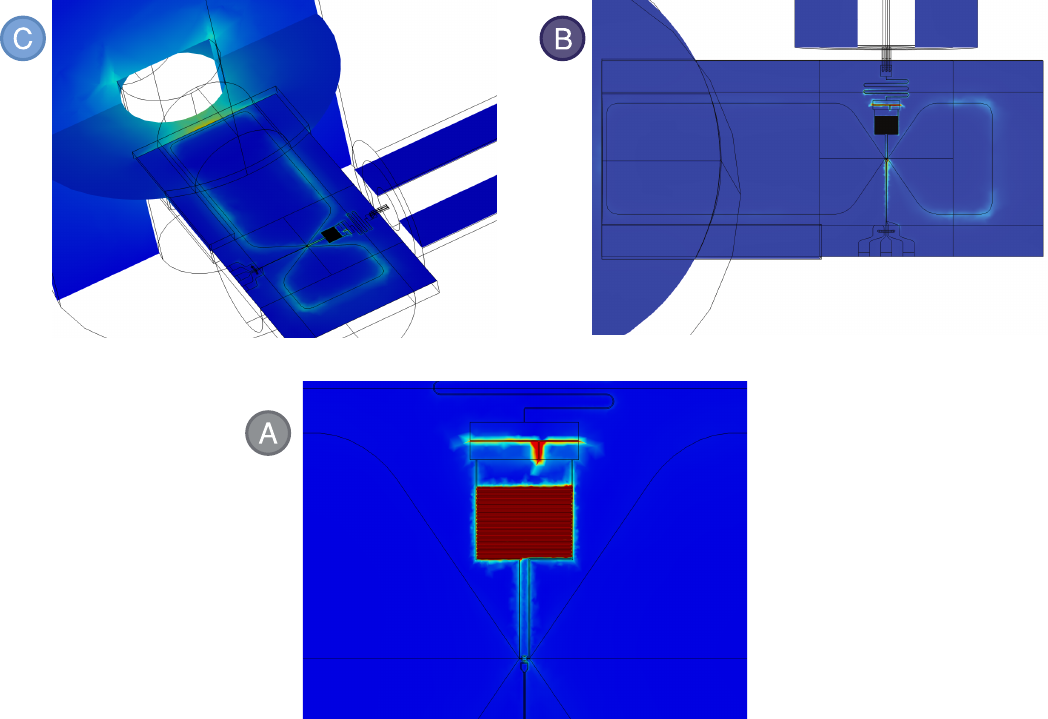}
    \caption{\textbf{Simulated electric field mode profiles} for the cavity (C), buffer (B), and amplifier (A) modes.}
    \label{fig:mode_profiles}
\end{figure}

To simulate the frequencies, linewidths, and the qubit junction's participation in all of the modes in the device, we used the COMSOL Eigenfrequency module. To compute the qubit's anharmonicity and cross-Kerr with each mode, we followed the energy participation ratio method detailed in Ref.~\cite{minev_energy-participation_2021}. These result in the electric field profiles plotted in Fig.~\ref{fig:mode_profiles}, and the simulated device parameters summarized in Table~\ref{tbl:EM_outputs} for the operation flux of $\Phi = 0.246 \Phi_0$.
\begin{table}
\begin{tabular}{ |c||c|c|c| }
    \hline
    mode & $\freq{\K} / 2\pi$ & $\lw{\K} / 2\pi$ & $\disp{\K} / 2\pi$ \\
    \hline
    A & 3.1843 GHz & 103.2 kHz& 1.5 mHz\\
    \hline
    B & 7.4676 GHz & 116.8 MHz & 14.8 kHz\\
    \hline
    C & 8.9558 GHz & 166.7 kHz & 8.98 MHz\\
    \hline
    Q & 8.2463 GHz (g$\leftrightarrow$e) & 18.1 kHz & 319.3 MHz\\
    \hline
\end{tabular}
\caption{\textbf{EM simulation outputs at $\Phi = 0.246 \Phi_0$}. Note, modes A, C, and Q have port boundary conditions in the simulation that differs from the experiment due to external filtering of the I/O and flux lines at the respective frequencies.}
\label{tbl:EM_outputs}
\end{table}
It is worth noting that experimentally, the frequencies of modes A, C, and Q lie outside the bandwidth of the circulators and filtered drive and flux lines. These modes will therefore see a different boundary condition than the 50 Ohm environment enforced at both ports in the simulation, leading to potentially noticeable discrepancies in the simulated and measured parameters for these modes. In general, the frequencies of the simulated modes slightly underestimate the measured device frequencies. The simulation overestimates the linewidth of mode B, and estimates the measured A and C mode linewidths relatively well both with and without the Purcell filter (see Fig.~\ref{fig:dev_overview}(c) and Fig.~\ref{fig:linewidths_v_bias} respectively). The simulation underestimates the qubit's self Kerr (anharmonicity) and cross Kerr with mode B, but overestimates the cross Kerr with mode C. Some of these discrepancies could be due the small detuning between the qubit and the C mode in the COMSOL model, as the qubit frequency without the loading of the junction nonlinearity is 8.566 GHz. The simulation may therefore overestimate the degree of hybridization of the qubit and C mode, diluting the nonlinearity of the qubit, and increasing its dispersive coupling with mode C.

\subsection{I/O and qubit drive port}
\begin{figure}[htbp]
    \includegraphics[scale=0.4]{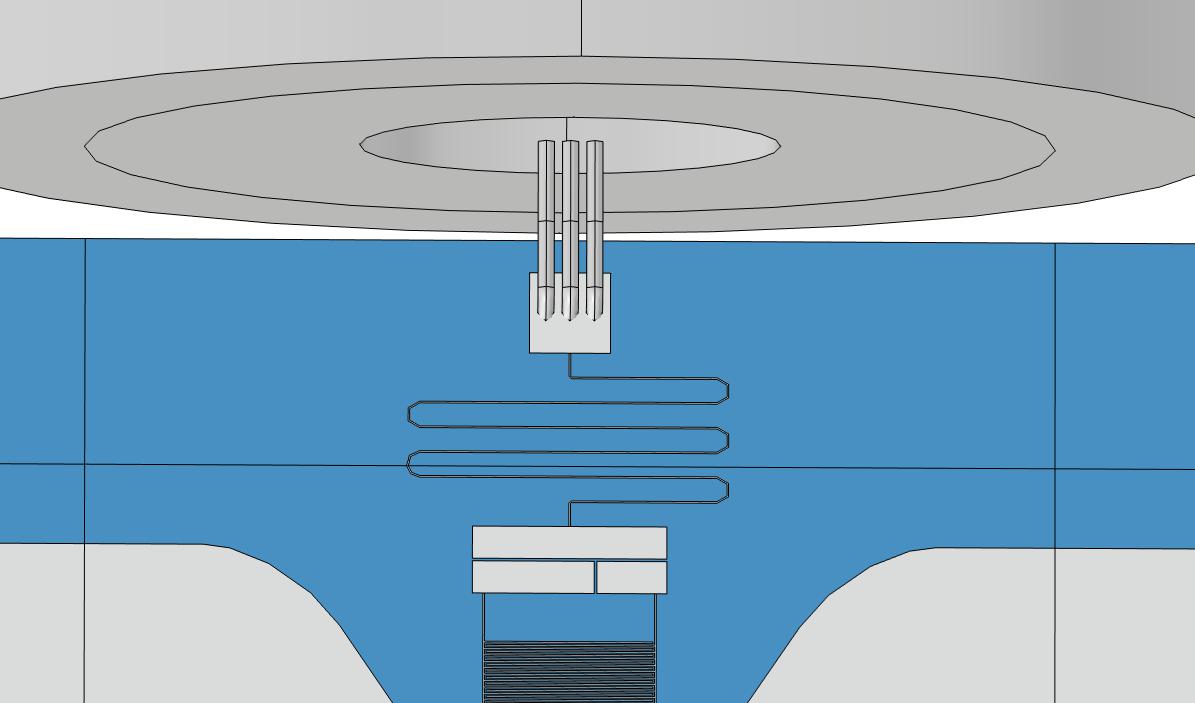}
    \caption{\textbf{I/O and qubit drive port}. COMSOL model of charge-coupled port to act as the I/O port and qubit drive. Also pictured are the meander inductor, and two parallel capacitors connected to the terminals of a separate IDC, as described in the text.}
    \label{fig:charge_port}
\end{figure}

The model for the charge-driven I/O and qubit-drive port is shown in Fig.~\ref{fig:charge_port}. We model this port as a 50 $\Omega$ coaxial line with a center conductor radius 0.65 mm, PTFE ($\epsilon_r = 2.1$) dielectric sleeve with an outer radius of 2.05 mm, a length of 2 cm, and a TEM boundary condition, which may be excited to simulate the device's driven response. The center pin is connected to an on-chip pad with three modeled wirebonds of radius 25 $\mu$m. This pad connects to a meander inductor, which provides the inductance for a single-pole, series LC filter with a simulated center frequency of 5.98 GHz and linewidth of 1.47 GHz, setting some degree of Purcell protection for the qubit and the A and C modes. The capacitance of this single-pole filter is provided by two capacitors in parallel below the meander inductor, which were also designed to destructively interfere the differential voltages on the interdigitated capacitor (IDC) leads for the A mode excitation. This minimizes mode A's coupling to the I/O port, without substantially sacrificing the external coupling rate of mode B, due to the fact that the two modes excite different field profiles within the IDC.

\subsection{Flux biasing}
\begin{figure}[htbp]
    \includegraphics{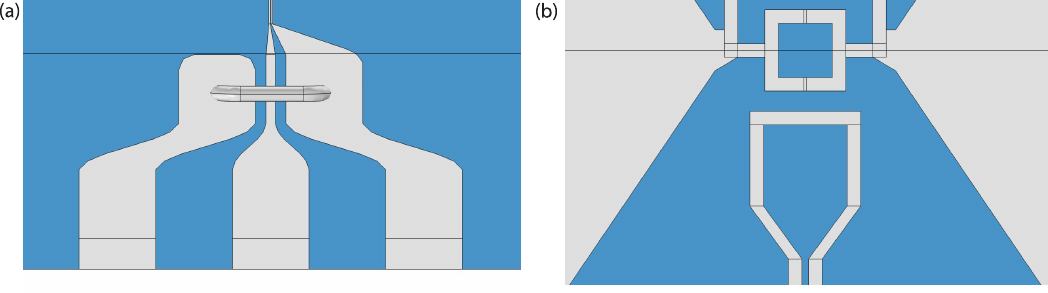}
    \caption{\textbf{Flux line}. COMSOL model of on-chip flux line to bias the dc SQUID and enable parametric interactions consisting of \textbf{(a)} a finite coplanar waveguide (CPW) with a lumped 50 $\Omega$ resistor shorting the center conductor to ground, transitioning to a coplanar stripline (CPS) via a balun, and \textbf{(b)} an inductive short in the form of a loop with mutual inductance to the dc SQUID.}
    \label{fig:flux_port}
\end{figure}

We model the flux port as a finite coplanar waveguide (CPW) with a 50 $\Omega$ lumped resistor shorting the center trace to ground. This CPW transitions to a differential excitation in a coplanar stripline (CPS) via a balun, as shown in Fig.~\ref{fig:flux_port}(a). The CPS then travels up the chip where it is inductively shorted in a loop with mutual inductance to the dc SQUID, as shown in Fig.~\ref{fig:flux_port}(b). This coupled line can therefore induce DC and AC circulating currents in the SQUID, facilitating parametric processes.

To model the effective SQUID inductance, we apply the following expression for an asymmetric dc SQUID:
\begin{equation}
    L_{\mathrm{SQUID}}(\Phi) = \frac{L_0}{\sqrt{\alpha^2 + (1 - \alpha^2) \cos^2(\pi \Phi / \Phi_0)}}
\end{equation}
where $\Phi$ is the external magnetic flux threading the loop, $\Phi_0$ is the magnetic flux quantum, $L_0$ is the unbiased SQUID inductance, and $\alpha$ is an asymmetry parameter of the junctions' critical currents, $I_{c,1} = I_c(1-\alpha)$ and $I_{c,2} = (1+\alpha)$. In the finite-element simulation, the dc SQUID is modeled as a loop with two lumped inductors representing the junctions. As the lumped inductors are in parallel, each is assigned the inductance $2 \times L_{\mathrm{SQUID}}$ in the simulation.

\subsection{Other simulation parameters}
In the model for the device, the values of unbiased dc SQUID inductance $L_0$ and transmon junction inductance $L_J$ are extracted from room-temperature resistance measurements. We tuned the dielectric constant of sapphire in the finite-element model in an effort to match the measured qubit anharmonicity (see next subsection) until it reached the lower bound of the acceptable range for sapphire ($\epsilon_r$ = 9.3). Subject to variations from machining and fabrication tolerances, the simulated geometry and external flux bias otherwise match the implemented device parameters. The relevant simulated device parameter inputs are summarized in Table~\ref{tbl:EM_sim_inputs}.
\begin{table}
\begin{tabular}{ |c||c| }
    \hline
    Substrate $\epsilon_r$ & 9.3\\
    \hline
    Transmon $L_J$ & 6.3808 nH\\
    \hline
    SQUID $L_0$ & 0.2023 nH\\
    \hline
    SQUID $\alpha$ & 0.1\\
    \hline
\end{tabular}
\caption{EM simulation inputs}
\label{tbl:EM_sim_inputs}
\end{table}

\section{Experimental diagram}
\label{apdx:experimental_diagram}
Below is a diagram of the experimental setup utilized in this work. We send attenuated drives at the buffer and qubit frequencies to the charge-coupled port of the readout network. For the three beam-splitter interactions (AB, BC, and AC), we send attenuated signals to the flux line of the device, as well as tee in a DC bias at the mixing chamber, allowing the pumps to turn on three-wave mixing processes. Signals traveling from the device pass through a low-noise microwave measurement chain enabled by a TWPA. Notably, we place external filters to protect leakage at frequencies outside the buffer mode for the signal line, and at frequencies above the pump band for the flux line. Drives outside the signal line's bandpass Purcell filter are still permitted to reach the device through the coupled port of a directional coupler placed between the filter and the device. The finite directivity of this coupler ($\sim$25 dB), which could be further degraded due to possible impedance mismatches, likely creates the Fano interference features that make an accurate fit of the scattering parameters difficult to achieve (necessitating the catch-and-release calibrations detailed in Appendix~\ref{apdx:c_r_eom}). For the preliminary scattering parameter fits which produce the data shown in Figs.~\ref{fig:dev_overview}(c) and \ref{fig:linewidths_v_bias}, we replace the Purcell bandpass filter and directional coupler within the device shield with a single-stage circulator outside the device shield.

\twocolumngrid
\clearpage

%


\end{document}